\documentclass{aa}
\usepackage{graphicx}
\usepackage[flushleft]{threeparttable}
\usepackage{amsmath}

\def\la{\lower.5ex\hbox{$\; \buildrel < \over \sim \;$}}
\def\ga{\lower.5ex\hbox{$\; \buildrel > \over \sim \;$}}

\begin{document}      

   \title{ESO\,137-001 -- a jellyfish galaxy model}

   \author{B.~Vollmer\inst{1}, M.~Sun\inst{2}, P.~Jachym\inst{3}, M.~Fossati\inst{4}, A.~Boselli\inst{5}}

   \offprints{B.~Vollmer, e-mail: Bernd.Vollmer@astro.unistra.fr}

   \institute{Universit\'e de Strasbourg, CNRS, Observatoire Astronomique de Strasbourg, UMR 7550, 67000 Strasbourg, France \and
     Physics Department, University of Alabama in Huntsville, Huntsville, AL 35899, USA \and
     Astronomical Institute of the Czech Academy of Sciences, Bocni II 1401, 141 00 Prague, Czech Republic \and
     Dipartimento di Fisica G. Occhialini, Universit`a degli Studi di Milano-Bicocca, Piazza della Scienza 3, I-20126 Milano, Italy \and
     Aix-Marseille Univ., CNRS, CNES, LAM, Marseille, France
              }

   \date{Received / Accepted}

   \authorrunning{Vollmer et al.}

\abstract{
  Ram pressure stripping of the spiral galaxy ESO~137-001 within the highly dynamical intracluster medium (ICM) of the Norma cluster lead to spectacular extraplanar
  CO, optical, H$\alpha$, UV, and X-ray emission. The H$\alpha$ and X-ray tails extend up to $80$~kpc from the galactic disk.
  Dynamical simulations of the ram pressure stripping event are presented to investigate the physics of the stripped gas and its ability to from stars,
  to predict H{\sc i} maps, and to constrain the orbit of ESO~137-001 within the Norma cluster. Special care was taken for the stripping of the diffuse gas.
  In a new approach, we analytically estimate the mixing between the intracluster and interstellar media. Different temporal ram pressure profiles and the ICM-ISM mixing rate were tested.
  Three preferred models show most of the observed multi-wavelength characteristic of ESO~137-001.
  Our highest-ranked model best reproduces the CO emission distribution, velocity for distances $\la 20$~kpc from the galactic disk, and the available NUV observations.
  The second and third preferred models reproduce best the available X-ray and H$\alpha$ observations of the gas tail including the H$\alpha$ velocity field.
  The angle between the direction of the galaxy's motion and the galactic disk is between $60^{\circ}$ and $75^{\circ}$.
  Ram pressure stripping thus occurs more face-on. The existence of a two-tail structures is a common feature in our
  models. It is due to the combined action of ram pressure and rotation together with the projection of the galaxy on the sky.
  Our modelling of the H$\alpha$ emission caused by ionization through thermal conduction is consistent with observations.
  H{\sc i} emission distributions for the different models are predicted.
  Based on the 3D velocity vector derived from our dynamical model we derive a galaxy orbit, which is close to unbound.
  We argue that compared to an orbit in an unperturbed spherical ICM ram pressure is enhanced by a factor of $\sim 2.5$.
  This increase can be obtained in two ways: an increase of the ICM density or a moving ICM opposite to the motion of the galaxy within the cluster.
  In a strongly perturbed galaxy cluster as the Norma cluster with an off-center ICM distribution the two possibilities
  are probable and plausible.
  }

\maketitle
\nolinenumbers

\section{Introduction \label{sec:intro}}

Galaxy clusters represent ideal laboratories for galaxy evolution. Within a cluster environment galaxies can undergo different
interactions: with the gravitational potential of the cluster, with other cluster galaxies, or with the intracluster medium.
Whereas gravitational interaction affect the stellar and gaseous content of a galaxy, the interaction with the hot cluster
atmosphere (ram pressure stripping) only affects the gas. If a galaxy is on a rather radial orbit within the cluster its
velocity increases when approaching the cluster center (Boselli \& Gavazzi 2006). At the same time the ambient gas density of the intracluster medium (ICM)
increases toward the cluster center. Thus, ram pressure, which is ICM density times the square of the galaxy velocity, can increase
dramatically when the galaxy reaches its closest distance to the cluster center. 

Spiral galaxies that underwent or are undergoing ram pressure stripping 
show a truncated gas disk together with a symmetric stellar disk (e.g., VLA Imaging of Virgo in Atomic gas, VIVA, Chung et al. 2009).
If the interaction is ongoing, a gas tail mainly detected in H{\sc i} is present (e.g., Chung et al. 2007).
The gas truncation radius is set by the galaxy's closest passage to the cluster center via the
criterion introduced by Gunn \& Gott (1972):
\begin{equation}
\rho_{\rm ICM} v_{\rm gal}^2 = \pi \Sigma v_{\rm rot}^2/R \ ,
\label{eq:rps}
\end{equation}
where $\rho_{\rm ICM}$ is the ICM density, $v_{\rm gal}$ the galaxy velocity, $\Sigma$ the
surface density of the interstellar medium (ISM), $v_{\rm rot}$ the rotation velocity of the galaxy,
and $R$ the stripping radius. If peak ram pressure occurred more than $400$-$500$~Myr ago,
the gas tails have disappeared and the truncated gas disk has become symmetric again (Vollmer 2009).

Under extreme conditions, ram-pressure stripped galaxies often show extraplanar, one-sided optical and UV emission 
as well as important tails of ionized gas (e.g., Yagi et al. 2010, 2017; Poggianti et al. 2017). Because of their optical appearance, these
objects are called jellyfish galaxies. 
A significant fraction of the H$\alpha$ emission in these tails can be due to photoionization by massive stars born in situ in the 
tails (Poggianti et al. 2019). Extraplanar molecular gas is often found in jellyfish galaxies (Jachym et al. 2014, 2019;
Moretti et al. 2018), from which the ionizing stars are formed.
The Virgo cluster spiral dwarf galaxy IC~3418 also shows a one-sided filamentary optical and UV tail (Hester et al. 2010).
Kenney et al. (2014) called the extraplanar H{\sc ii} regions with a tail of young stars detected in UV fireballs.

\begin{figure}
  \centering
  \resizebox{\hsize}{!}{\includegraphics{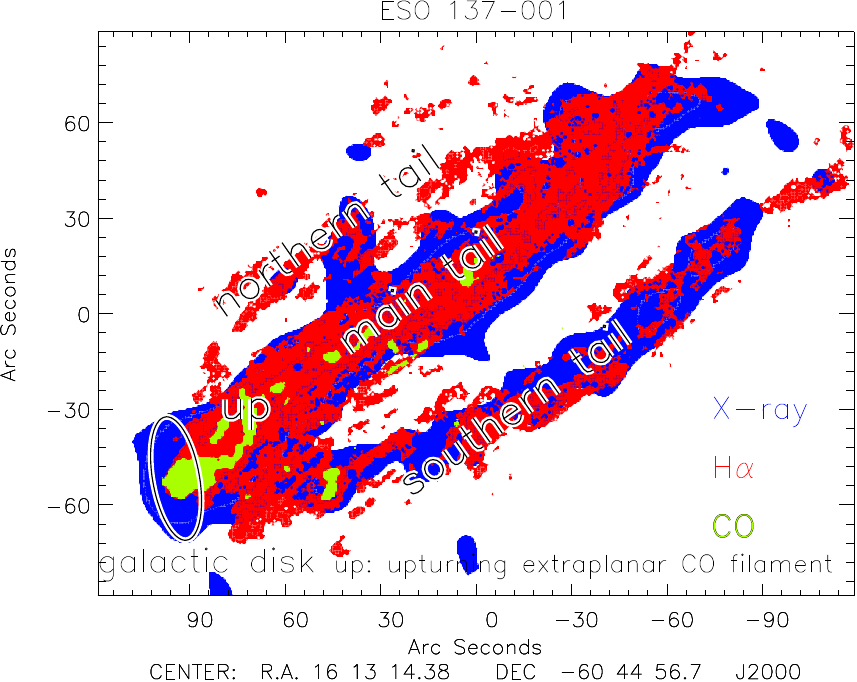}}
  \caption{ESO~137-001. Blue: X-rays (Sun et al. 2010), red: H$\alpha$ (Sun et al. 2022), green: CO (Jachym et al. 2019).
    The main features of the system are labeled. It appears that the X-ray stripping front is ahead of  the H$\alpha$ stripping front because of the
    large spatial smoothing scale of the X-ray image.
    Actually, the X-ray and H$\alpha$ stripping fronts are at the same position as shown in Fig.~2 of Luo et al. (2023).
    For a more detailed image see Fig.~\ref{fig:observations}.
  \label{fig:esoplot_sketch}}
\end{figure}

ESO~137-001, which is located in the Norma cluster is one of the nearest jellyfish galaxies ($D=70$~Mpc).  
Sun et al. (2006) discovered a bright $80$~kpc-long X-ray tail to the northwest of the galaxy pointing away from the cluster center.
With a deeper Chandra exposure, Sun et al. (2010) discovered a fainter secondary X-ray tail to the south of the bright tail.
The X-ray tail was also detected in H$\alpha$ emission by Sun et al. (2007), Fumagalli et al. (2014), and Fossati et al. (2016).
New deep MUSE observations lead to the detection
of a faint H$\alpha$ tail to the north of the bright X-ray tail (Sun et al. 2022, Luo et al. 2023).
Jachym et al. (2019) found a rich distribution of mostly compact CO regions extending to nearly $60$~kpc in length and $25$~kpc in width.
In total, about $10^9$~M$_{\odot}$  of molecular gas were detected with ALMA in the tail, assuming the standard Galactic CO-to-H 2 conversion factor.
The $80$~kpc gas tail of ESO~137-001 thus has a
multi-phase nature where the dense molecular gas is co-located with the warm and hot ionized gas (Fig.~\ref{fig:esoplot_sketch}).

ESO~137-001 is located at a projected distance of about $200$~kpc from the cluster center. Its line-of-sight (LOS) 
velocity ($4680 \pm 71$~km\,s$^{-1}$ ; Woudt et al. 2004) is close to the average cluster velocity ($4871 \pm 54$~km\,s$^{-1}$ ; Woudt et al. 2008).
This suggests that the galaxy orbit lies nearly within the plane of the sky. This is consistent with the large size of the tail.
The Norma cluster is located close to the center of the Great Attractor, at the crossing of a web of ﬁlaments of galaxies (called the
Norma wall; Woudt et al. 2008). The cluster is strongly elongated along the Norma wall, indicating an ongoing merger.

B\"ohringer et al. (1996) presented ROSAT data of the Norma cluster. The distribution of the X-ray emission of the ICM is elongated along the
southeast--northwest direction. When a spherical X-ray emission distribution was subtracted from the image, an elongated southeastern structure and
a northwestern ridge reminiscent of a large-scale shock became apparent. Moreover,
the high-surface-brightness X-ray emission of the ICM detected by ROSAT and XMM-Newton is displaced toward the northwest of the central cD galaxy ESO~137-006
(see Fig.~2 of Sun et al. 2010). It thus appears that ESO~137-001 evolves in an asymmetric and highly dynamical ICM.

We present dynamical simulations of the ram pressure stripping event to investigate the physics of the stripped gas and its ability to from stars,
to predict H{\sc i} maps, and to constrain the orbit of ESO~137-001 within the Norma cluster.
This article is structured in the following way: the dynamical model is presented in Sect.~\ref{sec:model} with an emphasis on the stripping and
mixing of the diffuse gas (Sect.~\ref{sec:iong}) and the tested temporal ram pressure profiles (Sect.~\ref{sec:param}).
The already existing CO, H$\alpha$, and X-ray observations are briefly described in Sect.~\ref{sec:observations} followed by the results of our
simulations (Sect.~\ref{sec:results}). The influence of the ICM-ISM mixing rate and the hot gas stripping efficiency on the simulation
results are shown in Sects.~\ref{sec:mixing} and \ref{sec:efficiency}. The existence of a third gas tail in the simulations is
discussed in Sect.~\ref{sec:northerntail}. The multi-phase gas masses, the dense molecular gas and CO emission, and the UV emission and star formation
are inspected in Sects.~\ref{sec:gasmasses}, \ref{sec:moleculargas}, and \ref{sec:UVstar}. Predicted H{\sc i} maps and the H$\alpha$--X-ray correlation
of the stripped gas are presented in Sects.~\ref{sec:HI} and \ref{sec:haxcorrelation}. A possible orbit of ESO~137-001 within the Norma
cluster is proposed in Sect.~\ref{sec:Norma}. Finally, we give our conclusions in Sect.~\ref{sec:conclusions}.

\section{The dynamical model\label{sec:model}}

For a proper treatment of the different gas phases during a ram pressure stripping event three-dimensional
high-resolution hydrodynamic simulations are the first choice (e.g., Tonnesen \& Bryan 2021, Lee et al. 2022).
However, these simulations are complex and very much time-consuming. We think that, as a first attempt, simplified simulations with
well-controlled sub-grid physics can help to understand the physics of the ram-pressure stripped ISM and its ability to form stars.

We used the N-body code described in Vollmer et al. (2001), which consists of two components: (i) a non-collisional component that simulates
the stellar bulge, stellar disk, and the dark halo and (ii) a collisional component that simulates the ISM as an ensemble of gas clouds
(Sect.~\ref{sec:halostarsgas}). Each particle of the collisional component or sticky particle corresponds to a gas cloud.
A scheme for star formation was implemented, where stars are formed during cloud collisions and then evolve as non-collisional particles
(Sect.~\ref{sec:sfr}).
The effect of ram pressure is simulated as an additional force acting on gas clouds, which are not
protected from the ram pressure wind by other clouds. Simulations with $19$ different ram pressure profiles were calculated (Sect.~\ref{sec:param}).

Since our code is not able to treat diffuse gas of low density and  high volume filling factor in a consistent way,
we assume that warm gas clouds become diffuse if their densities fall below a critical density and if
they are stripped out of the galactic plane, i.e. the stellar density drops below a given limit.
When the clouds become diffuse their surface density $\Sigma$ decreases and the acceleration by ram pressure $p_{\rm ram}$ increases
($p_{\rm ram} \propto \Sigma^{-1}$; Sect.~\ref{sec:iong}).

Furthermore, the stripped warm gas gradually mixes with the ambient ICM. The mixing rate is given by an analytical model of a radiative
turbulent mixing layer (Eq.~\ref{eq:mdotmix}; Fielding et al. 2020). Once the mixed mass is equal to the total cloud mass, the cloud temperature is set to
$10^7$~K. At the same time the stripping efficiency is increased by a factor of $10$ (or three in a second set of simulations) because of a decreased surface density
of the heated clouds with respect to the warm clouds. Only for the calculation of the
model X-ray and H$\alpha$ maps the hot and warm gas mass fractions are taken into account (Sect.~\ref{sec:extract}).
Finally, the equation of motion of the hot gas clouds was modified by adding the acceleration caused by the gas pressure gradient
$a=- \nabla p_{\rm ISM}/\rho_{\rm ISM}$ using a Smoothed-particle hydrodynamics (SPH) formalism.

\subsection{Halo, stars, and gas \label{sec:halostarsgas}} 

\begin{figure}
  \centering
  \resizebox{\hsize}{!}{\includegraphics{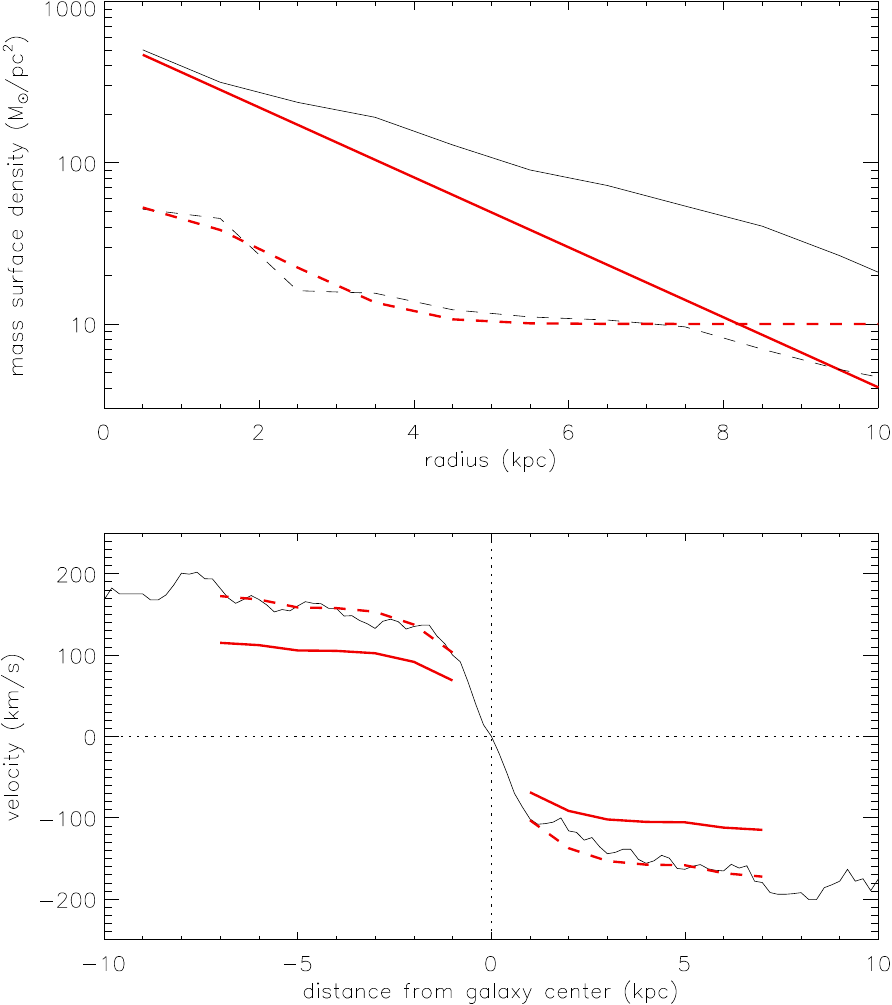}}
  \caption{Initial conditions. Upper panel: model stellar (black solid line) and gas (black dashed line) mass surface density profiles.
    Solid red line: exponential stellar mass surface density profile with a scale length of $2$~kpc. Dashed red line: analytical approximation to the gas
    surface density profile (see text).
    Lower panel: model (black solid line) and observed, asymmetric-drift-corrected (red line; Luo et al. 2023) rotation curve.
    Red dashed line: observed rotation curve multiplied by a factor of $1.5$.
  \label{fig:rotcurve}}
\end{figure}

The non--collisional component consists of 81\,920 particles, which simulate
the galactic halo, bulge, and disk.
The characteristics of the different galactic components are shown in
Table~\ref{tab:param}.
\begin{table}
      \caption{Total mass, number of collisionless particles $N$, particle mass $M$, and smoothing
        length $l$ for the different galactic components.}
         \label{tab:param}
      \[
         \begin{array}{lllll}
           \hline
           \noalign{\smallskip}
           {\rm component} & M_{\rm tot}\ ({\rm M}$$_{\odot}$$)& N & M\ ({\rm M}$$_{\odot}$$) & l\ ({\rm pc}) \\
           \hline
           {\rm halo} & 1.9 \times 10$$^{11}$$ & 32768 & $$5.9 \times 10^{6}$$ & 1200 \\
           {\rm bulge} & 6.7 \times 10$$^{9}$$ & 16384 & $$0.4 \times 10^{6}$$ & 180 \\
           {\rm disk} & 3.3 \times 10$$^{10}$$ & 32768 & $$1.0 \times 10^{6}$$ & 240 \\
           \noalign{\smallskip}
        \hline
        \end{array}
      \]
\end{table}
    Our model was not initially intended to reproduce ESO~137-001 in detail. Its stellar content is twice as massive as that of ESO~137-001.
      The resulting stellar surface density profile and the model rotation curve are presented in Fig.~\ref{fig:rotcurve}.
      For comparison we extracted a radial surface brightness profile from the HST H band image of ESO~137-001 and derived an exponential scale length of $2$~kpc.
      The profile was scaled such that the disk is maximum at a radius of two times the scale length (grey line in the upper panel
      of Fig.~\ref{fig:rotcurve}). The model disk has a $1.5$ times larger scale length and thus an about two times higher mass than observed. 
      The model rotation velocity is $\sim 180$~km\,s$^{-1}$ (lower panel of Fig.~\ref{fig:rotcurve}), and the rotation curve becomes flat at
      a radius of about 7~kpc. We also show the observed rotation curve (Luo et al. 2023) corrected for asymmetric drift using Eq.~9 of Iorio
      et al. (2017). For the calculation of the stellar velocity dispersion
      we used Eq.~B3 of Leroy et al. (2008) where we conservatively set the stellar scale length to $1$~kpc. With $L_{\rm K}=2.4 \times 10^{10}$~L$_{\odot}$
      this is at the lower end of the $L_{\rm K}$--$R_{\rm e,K}$ relation of disk galaxies (Courteau et al. 2007).
      As noted by Jachym et al. (2014), the corresponding $L_{\rm K}$--$v_{\rm rot}$ relation yields $v_{\rm rot}=110$--$120$~km\,s$^{-1}$ for ESO~137-001.
      This is consistent with the asymmetric-drift corrected rotation curve. 
      The model stellar rotation curve is about $50$\,\% higher than observed.

We adopted a model where the ISM is simulated as a collisional component,
that is, as discrete particles that each possess a mass and a radius and 
can have inelastic collisions (sticky particles).
The advantage of our approach is that ram pressure can be easily included as an additional
acceleration on particles that are not protected by other particles (see Vollmer et al. 2001).

The 20\,000 particles of the collisional component represent gas cloud complexes that
evolve in the gravitational potential of the galaxy.
The gas surface density profile is presented in the upper panel of Fig.~\ref{fig:rotcurve}.
It can be approximated up to a radius of $\sim 8$~kpc by $\Sigma=(1+4.5 \exp{(-r/2.2~{\rm kpc}})) \times 10$~M$_{\odot}$pc$^{-2}$.
The total assumed gas mass is $M_{\rm gas}^{\rm tot}=5.2 \times 10^{9}$~M$_{\odot}$,
which corresponds to the total neutral gas mass before stripping.
A radius is attributed to each particle, which depends on its mass assuming a constant surface density.
The normalization of the mass-size relation was taken from Vollmer et al. (2012a).
During each cloud-cloud collision, the overlapping parts of the clouds are calculated. 
Let $b$ be the impact parameter and $r_1$ and $r_2$ the radii of the larger and smaller clouds. 
If $r_1+r_2 > b > r_1-r_2$, the collision can result in fragmentation (high-speed encounter) or
mass exchange. If $b < r_1-r_2$, mass exchange or coalescence (low-speed encounter) can occur.
The outcome of the collision is simplified following Wiegel (1994). 
If the maximum number of gas particles or clouds ($20000$) is reached, only coalescent or mass-exchanging collisions are allowed.
In this way, a cloud mass distribution is naturally
produced. The energy loss by partially inelastic cloud-cloud collisions results in an effective gas viscosity in the disk.

As the galaxy moves through the ICM, its clouds are accelerated by
ram pressure $a=\rho_{\rm ICM}v_{\rm gal}^2/\Sigma$. In addition, the gas clouds are accelerated by the 
gradients of the gravitational potential $a=- \nabla \phi$.
Within the galaxy's inertial system, the galaxy's clouds
are exposed to a wind coming from a direction opposite to that of the galaxy's 
motion through the ICM. 
The temporal ram pressure profile has the form of a Lorentzian,
which is realistic for galaxies on highly eccentric orbits within the
Virgo cluster (Vollmer et al. 2001).
The effect of ram pressure on the clouds is simulated by an additional
force on the clouds in the wind direction. Only clouds that
are not protected from the wind by other clouds are affected.
This results in a finite penetration length of the ICM into the ISM.
Since the gas cannot develop instabilities, the influence of turbulence 
on the stripped gas is not included in the model. The mixing of the
ICM into the ISM is very crudely approximated by the finite
penetration length of the ICM into the ISM; in other words, up to this penetration
length, the clouds undergo an additional acceleration caused by ram pressure.

\subsection{Star formation \label{sec:sfr}}

We assume that the star formation rate (SFR) is proportional to the cloud collision rate.
At the end of each collision, a collisionless particle is created, which is added to the ensemble of particles.
Since the mass of the newly formed stars is small compared to the total stellar mass, the newly created collisionless particles have zero mass
(they are test particles) and the positions and velocities of the colliding clouds after the collision.
These particles are then evolved passively with the whole system. 
The information about the time of creation is attached to each newly created star particle.
We verified that the SFR of an unperturbed galaxy
is constant within $1$~Gyr. The simulations do not include stellar feedback. Clouds can lose kinetic energy
via partially inelastic collisions. This energy loss does not lead to a significant decrease of the velocity dispersion of the gas clouds
within an unperturbed galactic disk during $1$~Gyr.

In the following, we link the star formation recipe based on cloud--cloud collisions to the recipes based on the local and global gas densities.
Vollmer \& Beckert (2003) expanded on an analytical model for a galactic gas disk that considers the warm, cold, and molecular phases 
of the ISM as a single, turbulent gas. This gas is assumed to be in vertical hydrostatic equilibrium, with the midplane pressure balancing 
the weight of the gas and the stellar disk. The gas is taken to be clumpy, so that the local density is enhanced relative to the average 
density of the disk. Using this local density, the free-fall time of an individual gas cloud (i.e., the fastest timescale for star formation)
can be determined. The SFR is used to calculate the rate of energy injection by supernovae. This rate is related to the 
turbulent velocity dispersion and the driving scale of turbulence. These quantities in turn provide estimates of the clumpiness of gas 
in the disk (i.e., the contrast between the local and average density) and the rate at which viscosity moves matter inward.
Vollmer \& Leroy (2011) applied the model successfully to a sample of local spiral galaxies.
Within the framework of the model, the SFR per unit volume is given by
\begin{equation}
\dot{\rho}_*=\Phi_{\rm V} \rho\,t_{\rm ff,cl}^{-1} \ ,
\end{equation}
where $\Phi_{\rm V}$ and $t_{\rm ff,\ cl}^{-1}$ are the volume filling factor and free-fall time of self-gravitating gas clouds, respectively, and
$\rho$ is the global gas density. The volume filling factor is $\Phi_{\rm V}=n_{\rm cl}\,4 \pi / 3 \,r_{\rm cl}^3$, where
$n_{\rm cl}$ is the number density and $r_{\rm cl}$ the radius of the self-gravitating gas clouds. For a self-gravitating cloud, the free-fall time equals
the turbulent crossing time $t_{\rm ff,cl}=2\,r_{\rm cl}/v_{\rm turb,cl}$. With the collision time given by
$t_{\rm coll}=(n_{\rm cl}\,\pi\,r_{\rm cl}^2 v_{\rm turb})^{-1}$, this leads to $\dot{\rho}_*= 2/3\,\rho\,t_{\rm coll}^{-1}$.
Thus, the SFRs of the two recipes are formally equivalent.

In numerical simulations, the star formation recipe usually involves the gas density $\rho$ and the free-fall time 
$t_{\rm ff}=\sqrt{3\,\pi/(32\,G \rho)}$: $\dot{\rho}_* \propto \rho\, t_{\rm ff}^{-1} \propto \rho^{1.5}$.
We verified that our star formation recipe based on cloud-cloud collisions leads to the same exponent
of the gas density in a simulation of an isolated spiral galaxy.
As a consequence, our code reproduces the observed SFR-total gas surface 
density, SFR-molecular gas surface density, and SFR-stellar surface density relations (Vollmer et al. 2012b).

\subsection{Diffuse gas stripping and mixing \label{sec:iong}}

Our numerical code is not able to treat diffuse gas of low density and  high volume filling factor in a consistent way. For a realistic treatment, 3D hydrodynamical simulations
should be adopted. Nevertheless, we can mimic the action of ram pressure on diffuse gas by applying
very simple recipes based on the fact that the acceleration by ram pressure is inversely proportional to the gas
surface density, which in turn depends on the gas density for a gas cloud of constant mass.
For the sake of simplicity, we do not modify the radii of the diffuse gas clouds for the calculation of the cloud-cloud collisions.
The diffuse clouds are thus mostly ballistic particles under the influence of a ram-pressure induced acceleration.
We divide the gas in our simulations into a dense and diffuse phase according to the local density.
The gas and stellar densities are calculated via the $50$ nearest neighboring particles.

Following Vollmer et al. (2021) we assumed that the warm ($\sim 10^4$-$10^5$~K) gas clouds become diffuse (i.e., their sizes and volume filling factor
increase and their densities and column densities decrease)
if they are stripped out of the galactic plane and if their densities fall below the critical density
of $n_{\rm crit}^{\rm warm}=5 \times 10^{-3}$~cm$^{-3}$. We further assume that the first condition is fulfilled if the
stellar density at the location of the gas particle is lower than $\rho_*^{\rm crit}=2.5 \times 10^{-4}$~M$_{\odot}$pc$^{-3}$. 
For our model galaxy, this density is reached at a disk height of $\sim 3.5$-$4$~kpc.

Once the gas has a high volume filling factor (i.e., it becomes diffuse), its volume increases and the surface densities decreases.
The decrease in the surface density is taken into account by considering a gas cloud of a constant mass:
for $\rho_* <  \rho_*^{\rm crit}$, the cloud size is proportional to $n^{-1/3}$, the cloud surface density 
is proportional to $n^{2/3}$, and the acceleration caused by ram pressure $a=\rho_{\rm ICM}v_{\rm gal}^2/\Sigma$ is increased by
a factor of $(0.044~{\rm cm}^{-3}/n_{\rm ISM})^{2/3}$. Since it is assumed that the gas becomes diffuse, we set $n_{\rm ISM}$ to the global gas density.
This last normalization and the critical densities were chosen such that they led to acceptable results for several observed 
galaxies undergoing ram pressure stripping. The critical stellar density is motivated
by the fact that a significant change in the ISM properties only occurs once the ISM has entirely left the galactic disk. 
It turns out that this condition is necessary to avoid excessive gas stripping.

The initial temperature of the ISM is $T_{\rm ISM}=10^4$~K.
The hot ($>10^6$~K) diffuse gas is taken into account in the following way: 
we assume that once the stripped gas has left the galactic disk, it mixes with the ambient ICM.
If (i) the gas density falls below the critical
value of $n_{\rm crit}^{\rm hot}=5 \times 10^{-4}$~cm$^{-3}$, (ii) the stellar density is below $\rho_*^{\rm crit}$,
and (iii) the ISM temperature is below $9 \times 10^6$~K,
ICM-ISM mixing begins to take place. On the other hand, if the stellar density exceeds its critical value (i.e., the gas is located within or close
to the galactic disk), then the gas is assumed to cool rapidly and its temperature is set to $T_{\rm ISM}=10^4$~K.

In Vollmer et al. (2021) we assumed that mixing occurs instantaneously and raises the temperature of the mixed gas clouds to
\begin{equation}
\label{eq:ttt}
T=\frac{n_{\rm ICM}\,6 \times 10^7\ {\rm K}+\sqrt{n_{\rm ICM}\,n_{\rm ISM}}\,T_{\rm ISM}}{n_{\rm ICM}+\sqrt{n_{\rm ICM}\,n_{\rm ISM}}}\ ,
\end{equation}
where $n_{\rm ICM}$ is the density of the ICM and $n_{\rm ISM}$ is the density of the mixed ISM, which was continuously calculated for each particle.
With $n_{\rm ICM}=2 \times 10^{-3}$~cm$^{-3}$, $n_{\rm ISM}=0.1$~cm$^{-3}$, and $T_{\rm ISM}=10^4$~K the temperature of the mixed gas
is $T=0.9 \times 10^7~{\rm K}$, consistent with the observed X-ray tail temperature of $0.8$~keV (Sun et al. 2010).

In a new approach, we estimate the ICM-ISM mixing analytically based on the recipe of Fielding et al. (2020).
These authors considered a radiative turbulent mixing layer in which cold and hot gas in pressure and thermal
equilibrium move relative to each other. The Kelvin Helmholtz instability quickly develops turbulence that promotes
mixing and populates a rapidly cooling intermediate-temperature phase.
In quasi-steady state in the frame of the interface, radiative cooling losses are balanced by the advection of hot gas.
Hot gas ﬂows into the cooling layer at a speed $v_{\rm in}$. The mass transfer rate from hot to cold gas is
\begin{equation}
\label{eq:dotM}
\dot{M} \sim \rho_{\rm ICM}L^2 v_{\rm in}\ ,
\end{equation}
where $\rho_{\rm ICM}$ is the ICM density and $L$ the characteristic length of the mixing layer.
According to Eq.~6a of Fielding et al. (2020) the ratio between the inflow and the relative velocity between the hot and cold phases is
\begin{equation}
\label{eq:vinvrel}
\frac{v_{\rm in}}{v_{\rm rel}}=\big(\frac{\rho_{\rm ISM}}{\rho_{\rm ICM}}\big)^{\frac{3}{8}} \, \big(\frac{L}{v_{\rm rel}t_{\rm cool}}\big)^{\frac{1}{4}} \, \big(\frac{v_{\rm turb}}{v_{\rm rel}}\big)^{\frac{3}{4}}\ ,
\end{equation}
where the cooling time is $t_{\rm cool}=\alpha /\rho_{\rm ISM}$, $\rho_{\rm ISM}$ is the ISM density, and $v_{\rm turb}$ the turbulent velocity.
Fielding et al. (2020) estimated $\big(\frac{v_{\rm turb}}{v_{\rm rel}}\big) \sim 0.1$--$0.2$. We assume $\big(\frac{v_{\rm turb}}{v_{\rm rel}}\big) = 0.2$.

Combining Eqs.~\ref{eq:dotM} and \ref{eq:vinvrel} yields
\begin{equation}
  \dot{M} = \rho_{\rm ICM}^{\frac{5}{8}} \big(\frac{M_{\rm ISM}}{4/3 \pi}\big)^{\frac{3}{4}} \rho_{\rm ISM}^{-\frac{1}{8}} \alpha^{-\frac{1}{4}}v_{\rm rel}^{\frac{3}{4}} \big(\frac{v_{\rm turb}}{v_{\rm rel}}\big)^{\frac{3}{4}} \ .
  \label{eq:mdotmix}
\end{equation}
We call $\dot{M}$ the ICM-ISM mixing rate.
With $n_{\rm ICM}=10^{-3}$~cm$^{-3}$, $\alpha=1.5 \times 10^7 n_{\rm ISM}^{-1}$~yr, a molecular weight of $\mu_{\rm ICM}=0.6$,
and $v_{\rm rel}=3000$~km\,s$^{-1}$ we obtain
\begin{equation}
\label{eq:inflowmix}
\dot{M}=10^{-3} \big(\frac{M_{\rm ISM}}{10^5~{\rm M}_{\odot}}\big)^{\frac{3}{4}} \big(\frac{n_{\rm ISM}}{\rm cm^{-3}}\big)^{-\frac{1}{8}} \ {\rm M}_{\odot} {\rm yr}^{-1} \ ,
\end{equation}
where $M_{\rm ISM}$ is the mass of an ISM cloud.
The cloud mass $M_{\rm ISM}$ (in M$_{\odot}$) and the ISM density are calculated from the dynamical model at each timestep $\Delta t$.
The total mixed mass is then updated by $M_{\rm mix}=M_{\rm mix} + \dot{M} \times \Delta t$. 
Once the mixed mass is equal to the total cloud mass we set the cloud temperature to $10^7$~K which corresponds to the temperature given by Eq.~\ref{eq:ttt}.
For a cloud mass of $M_{\rm ISM}=10^5$~M$_{\odot}$ and a density of $n_{\rm ISM}=1$~cm$^{-3}$ the ICM-ISM mixing time is about $100$~Myr.

For the stripped hot gas clouds ($>10^6$~K), we modified the equation of motion by adding the acceleration caused by the gas pressure gradient
$a=- \nabla p_{\rm ISM}/\rho_{\rm ISM}$ using a Smoothed-particle hydrodynamics (SPH) formalism. 
For the sake of simplicity, an isothermal stripped ISM with a sound speed of $c_{\rm s}=235$~km\,s$^{-1}$, which corresponds to a temperature of
$2.4 \times 10^6$~K, is assumed for this purpose.
We thus do not solve an explicit energy equation and the calculation of hydrodynamic effects is approximate.
Its main purpose is to keep the hot diffuse ISM from clumping.

For the calculation of the acceleration caused by ram pressure $a=\rho_{\rm ICM}v_{\rm gal}^2/\Sigma$, the acceleration is further increased
by a heuristic factor of ten if the ISM temperature exceeds $10^4$~K and the stellar density is below $\rho_*=2.5 \times 10^{-4}$~M$_{\odot}$pc$^{-3}$.
Pressure equilibrium and a temperature ratio between the hot ($\sim 10^7$~K) and the cool ($\sim 10^4$~K) gas imply a density ratio of
$n_{\rm hot}/n_{\rm cool}=10^{-3}$. The ratio of the surface densities $\Sigma \propto \rho^{\frac{2}{3}}$ is then $\Sigma_{\rm hot}/\Sigma_{\rm cool}=10^{-2}$.
The heuristic factor is expected to be smaller than the inverse of this ratio.
To investigate the effect of this heuristic factor, we recalculated all simulations with a factor of three instead of ten (Sect.~\ref{sec:efficiency}).
It turned out that the initial factor of ten used by Vollmer et al. (2021) gave the best results for ESO~137-001.

We note that without the inclusion of hydrodynamical effects, thin gas filaments cannot be produced by our simple model.  
It is assumed that the diffuse warm gas is heated and ionized by thermal conduction (see Sect.~\ref{sec:extract}). 
Thus, the diffuse gas can significantly emit in the H$\alpha$ line if its temperature is below $10^5$~K

\subsection{Parameters of the ram pressure stripping event \label{sec:param}}

\begin{figure}[!ht]
  \centering
  \resizebox{\hsize}{!}{\includegraphics{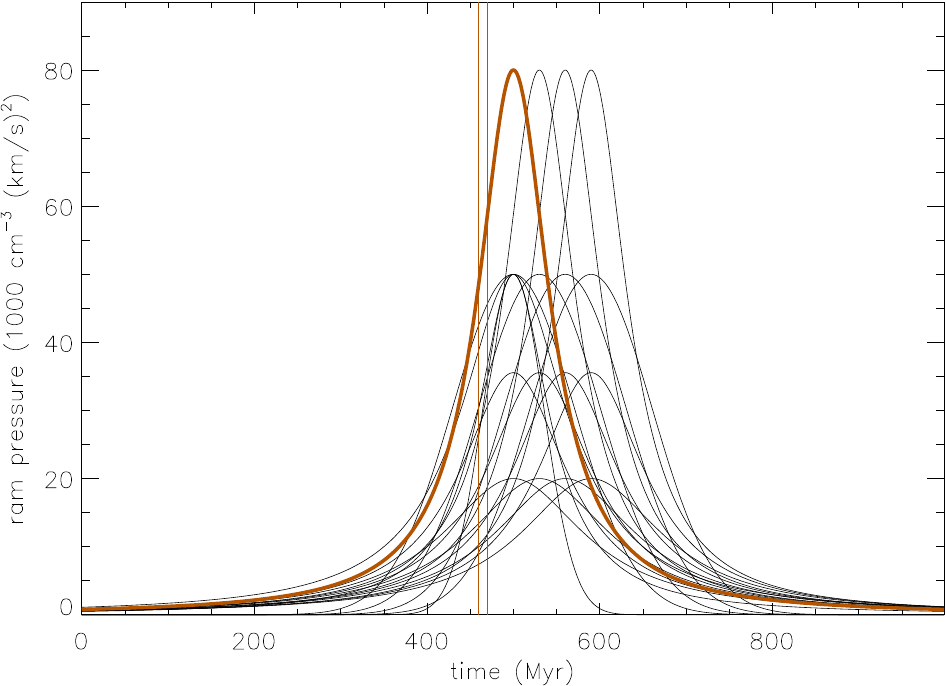}}
  \caption{Model ram pressure profiles (Table~\ref{tab:molent1}). The thick red line shows the highest-ranked model, and the vertical red lines show the
    range of timesteps of the highest-ranked model.
  \label{fig:profiles}}
\end{figure}

Following Vollmer et al. (2001), we used a Lorentzian profile for the time evolution of ram pressure stripping:
\begin{equation}
p_{\rm rp}=p_{\rm max} \frac{t_{\rm HW}^2}{(t-t_{\rm peak})^2+t_{\rm HW}^2}\ ,
\end{equation}
where $p_{\rm max}$ is the maximum ram pressure occurring at the galaxy's closest passage to the cluster center,
$t_{\rm HW}$ is the width of the profile, and $t_{\rm peak}=500$-$590$~Myr is the time of peak ram pressure.
The simulations were calculated from $t=0$ to $t=800$~Myr.
We set $p_{\rm max}=20000,\ 36000,\ 50000,\ 80000~{\rm cm}^{-3}({\rm km\,s}^{-1})^2$ and $t_{\rm HW}=50,\ 750,\ 100$~Myr.
Following Nehlig et al. (2016) and Vollmer et al. (2018), we investigated the influence of galactic structure (i.e., the position of spiral arms) on the
results of ram pressure stripping by varying the time of peak ram pressure between $t_{\rm peak}=530$~Myr and $590$~Myr.
For a different peak stripping the galactic spiral arms are not in the same place with respect to the leading edge of the interaction. 
We also used two different Gaussian profiles. The ram pressure profiles are specified in Table~\ref{tab:molent1} and shown in Fig.~\ref{fig:profiles}.
Each simulation took about four weeks on a single CPU.
\begin{table}[!ht]
      \caption{Model ram pressure stripping profiles.}
         \label{tab:molent1}
      \[
         \begin{array}{lrccc}
           \hline
           {\rm model} & {\rm profile} & {\rm amplitude} & {\rm width} & {\rm peak\ time} \\
             & & p_{\rm max} & t_{\rm HW} & t_{\rm peak} \\    
            & & ({\rm cm}^{-3}({\rm km\,s}^{-1})^2) & ({\rm Myr}) & ({\rm Myr}) \\
           \hline
           {\rm 1} & {\rm Lorentzian} & 20000 & 100 & 500 \\
           {\rm 1a} & {\rm Lorentzian} & 20000 & 100 & 530 \\
           {\rm 1b} & {\rm Lorentzian} & 20000 & 100 & 560 \\
           {\rm 1c} & {\rm Lorentzian} & 20000 & 100 & 590 \\
           {\rm 2} & {\rm Lorentzian} & 36000 & 75 & 500 \\
           {\rm 2a} & {\rm Lorentzian} & 36000 & 75 & 530 \\
           {\rm 2b} & {\rm Lorentzian} & 36000 & 75 & 560 \\
           {\rm 2c} & {\rm Lorentzian} & 36000 & 75 & 590 \\
           {\rm 3} & {\rm Lorentzian} & 50000 & 75 & 500 \\
           {\rm 4} & {\rm Lorentzian} & 50000 & 50 & 500 \\
           {\rm 5} & {\rm Lorentzian} & 80000 & 50 & 500 \\

           {\rm 5a} & {\rm Lorentzian} & 80000 & 50 & 530 \\
           {\rm 5b} & {\rm Lorentzian} & 80000 & 50 & 560 \\
           {\rm 5c} & {\rm Lorentzian} & 80000 & 50 & 590 \\
           {\rm 6} & {\rm Gaussian} & 50000 & 50 & 500 \\
           {\rm 7} & {\rm Gaussian} & 50000 & 100 & 500 \\
           {\rm 7a} & {\rm Gaussian} & 50000 & 100 & 530 \\
           {\rm 7b} & {\rm Gaussian} & 50000 & 100 & 560 \\
           {\rm 7c} & {\rm Gaussian} & 50000 & 100 & 590 \\
        \noalign{\smallskip}
        \hline
        \noalign{\smallskip}
        \hline
        \end{array}
      \]
\end{table}

The inclination angle of ESO~137-001 is $i \sim 65^{\circ}$ (Sun et al. 2007). If the galaxy is moving in the plane of the sky in the opposite direction of its gas tail
(close to the minor axis of the galaxy),
the angle between the disk plane and the ram pressure wind is also about $65^{\circ}$.
To investigate the influence of the angle between the disk plane and the ram pressure wind on the resulting gas distribution and
velocity field, we set this angle to ($50^{\circ}$, $60^{\circ}$, $75^{\circ}$). For the given parameter set we calculated 33 models.

In addition, we recalculated the model set (i) with a heuristic increase of the acceleration caused by ram pressure by a factor of three instead of ten
and with a (ii) three times higher and (iii) three times lower ICM-ISM mixing rate,
i.e. mass inflow rate within the ICM-ISM mixing layer (Eq.~\ref{eq:inflowmix}).
In total we calculated 99 different models. It turned out that the models with the inflow rate of Eq.~\ref{eq:inflowmix} gave the best results.

\subsection{Extraction of the observables \label{sec:extract}}

From the model calculations the following observables were extracted: H{\sc i}, CO, H$\alpha$, NUV, and X-ray emission. 
Following Vollmer et al. (2008), the separation between the atomic and molecular gas phases is based on the gas density:
\begin{equation}
\label{eq:phases}
\frac{M_{\rm mol}}{M_{\rm tot}}=min(4.2 \times \rho_{\rm tot}, 1) \ ,
\end{equation}
where $M_{\rm mol}$ and $M_{\rm tot}$ are the molecular and total gas mass of the cloud and $\rho_{\rm tot}$ is the total gas density in M$_{\odot}$pc$^{-3}$.
Elmegreen (1993) found the following relation between the molecular to atomic gas fraction and the gas pressure:
$M_{\rm mol}/M_{\rm HI} \propto P^{1.2}$. For a constant gas velocity dispersion the gas pressure is proportional to the gas density and
$M_{\rm mol}/M_{\rm tot}$ is about proportional to the gas density.
The atomic gas mass of the cloud then is $M_{\rm HI}=1-M_{\rm mol}/M_{\rm tot}$.

The calculation of the model X-ray emission map is based on the emission measure of the hot stripped ISM.
For a temperature of $T \sim 10^7$~K the X-ray emissivity approximately yields
\begin{equation}
\epsilon_{\nu} \sim 10^{-23} n_{\rm e}^2~{\rm erg}\,{\rm cm}^{-3}{\rm s}^{-1}\ ,
\end{equation}
(e.g., Fig.~34.2 of Draine 2011) where $n_{\rm e}$ is the electron density in cm$^{-3}$. It is assumed that the gas is fully ionized.
The X-ray luminosity of a hot gas particle or cloud is given by
\begin{equation}
L_{\rm X}=4\,\pi\,R^3 \epsilon_{\nu} = 5.3 \times 10^{35} \big(\frac{M_{\rm cl}}{1\,{\rm M}_{\odot}}\big) x_{\rm hot}\,\big(\frac{\rho_{\rm hot}}{1\,{\rm M}_{\odot}{\rm pc}^{-3}}\big)~{\rm erg\,s}^{-1}\ ,
\end{equation}
where $\rho_{\rm hot}$ is the average density of the hot gas and $x_{\rm hot}$ its overdensity.
To reproduce the observed X-ray luminosity of the stripped gas tail ($L_{0.5-2~{\rm keV}} \sim 10^{41}$~erg\,s$^{-1}$; Sun et al. 2010) by
the best-fit model we set $x_{\rm hot}=70$.
For the calculation of the hot gas mass the hot portion of a gas cloud is taken into account.
Hot gas within the galaxy, which is heated by supernova explosions, is not taken into account.
Therefore, the observed strong X-ray emission emitted by the galactic disk is not reproduced by the model.

The model H$\alpha$ images consist of two components: (i) the H{\sc ii} regions ionized by young massive stars and (ii) diffuse ionized gas that is
ionized by the stellar UV radiation, heat conduction (Cowie \& McKee 1977), or possibly by strong shocks induced by ram pressure stripping
(as, e.g. in the diffuse H$\alpha$ tail of NGC~4569; Boselli et al. 2016).
The first component is modeled by the distribution of stellar particles with ages less than $20$~Myr. 
This is the only component used for the images of the models without diffuse gas stripping.
For the diffuse component we assumed that the gas with temperatures lower than $10^6$~K is ionized by thermal conduction:
the mixing between the hot ICM and the warm stripped gas leads to a relatively dense ($\sim 10^{-2}$~cm$^{-3}$; Sun et al. 2010)
hot medium at a temperature of $\sim 10^7$~K (Sect.~\ref{sec:iong}). Thermal electrons from
this mixed ISM-ICM gas penetrate into the neutral warm stripped ISM clouds ionizing and heating them. Ultimately, this leads to the evaporation
of the ISM clouds (Cowie \& McKee 1977). The typical evaporation timescale for a cloud of $10^{21}$~cm$^{-2}$ is $\sim 100$~Myr (Vollmer et al. 2001).
In the case of a magnetic field configuration that inhibits heat flux (e.g., a tangled magnetic field), this evaporation time can increase significantly
(Cowie, McKee, \& Ostriker 1981) and might attain several $100$~Myr. Stripped clouds of warm neutral gas can thus survive within the stripped gas tail.
The influence of magnetic fields on thermal conduction in the turbulent stripped gas is discussed in Sect.~\ref{sec:magfield}.

For the determination of the emission measure of an evaporating gas cloud we followed Eq.~25 of McKee \& Cowie (1977):
\begin{equation}
\label{eq:em}
\frac{{\rm d}(EM)}{{\rm d}\ln T}=\big(\frac{n_{\rm ICM}}{n}\big)^2 \frac{{\rm d}N}{{\rm d}\ln T}=3 \sigma_0^{\frac{7}{8}}n_{\rm ICM}^2 \big(\frac{T}{T_{\rm ICM}}\big)^{\frac{3}{2}} R\ ,
\end{equation}  
where $EM$ is the emission measure of the conduction front, $T$ and $T_{\rm ICM}$ the temperature of the warm ISM and hot ICM, $N$ and $n$ are the column density and
density of the conduction front, and
$\sigma_0 \simeq \big( \frac{T_{\rm hot}}{1.54 \times 10^7~{\rm K}}\big)^2 n_{\rm ICM}^{-1} R_{\rm pc}^{-1}$ is the saturation parameter.
To also include the case of a classical evaporating cloud (Eq.~22 of McKee \& Cowie 1977), we set $\sigma_0=1$ if $\sigma_0 < 1$. 
The gas cloud radius is given by $R=(3\,M_{\rm cl}/(4\,\pi \rho))^{1/3}$.
Since the highest column densities occur at the transition between the inner classical and the saturated zone of the conduction front, we only used
$\frac{{\rm d}N}{{\rm d}\ln T}$ of the inner classical zone (Eq.~23 of McKee \& Cowie 1977).

For all gas particles with temperatures lower than $10^5$~K the H$\alpha$ luminosity was calculated with
\begin{equation}
\label{eq:lumem}
L_{{\rm H}\alpha}=EM \alpha_{{\rm H}_2}^{\rm eff} \pi R^2 h \nu_{{\rm H}\alpha}=2.7 \times 10^{36} \sigma_0^{\frac{7}{8}} n_{\rm e}^2 (x\,n)^{-1}\ {\rm erg\,s}^{-1},
\end{equation}  
where $\alpha_{\rm H_2}^{\rm eff}=  1.17 \times 10^{-13}$~cm$^{3}$s$^{-1}$ is the H$\alpha$ effective recombination coefficient, $h$ the Planck constant, $\nu_{{\rm H}\alpha}$
the frequency of the H$\alpha$ line, and $n$ and $x=10^3$ are the average density and the overdensity of the warm stripped ISM.
As for the X-ray emission, we set the local electron density of the hot gas to $70$ times the mean ICM density. The overdensity of the warm gas was chosen 
such that the observed H$\alpha$ luminosity of the gas tail ($L_{{\rm H}\alpha} \sim 3 \times 10^{40}$~erg\,s$^{-1}$) is approximately reproduced by our best-fit model.
For a justification of the chosen overdensities we refer to Sect.~\ref{sec:magfield}.

As explained in Sect.~\ref{sec:sfr}, the information about the time of creation is attached to each newly created star particle.
In this way, the H$\alpha$ emission distribution caused by H{\sc ii} regions can be modeled by the distribution of
star particles with ages younger than $20$~Myr. The UV emission of a star particle in the two GALEX bands
is modeled by the UV flux of single stellar population models produced from the STARBURST99 software (Leitherer et al. 1999). 
The age of the stellar population equals the time since the creation of the star particle.
The total UV distribution is then the extinction-free distribution of the UV emission of the newly created
star particles.

The model H{\sc i}, CO data cubes and the model NUV map were convolved with Gaussian kernels to the spatial resolutions
of the actual observations. The X-ray map of Sun et al. (2010) was adaptively smoothed. We decided to convolve our model X-ray map with
a Gaussian kernel with a half power width of $2.2$~kpc, which lead to results well comparable to the adaptively smoothed X-ray map.
The model distribution of the diffuse warm ionized gas was convolved with a Gaussian kernel with a half power width of $1$~kpc,
that of the H{\sc ii} regions with a kernel with a half power width of $0.6$~kpc. The total H$\alpha$ image was obtained by adding
the images of the diffuse warm ionized gas and the H{\sc ii} regions.
We then added the H{\sc ii} region component with a heuristic normalization to the model image of the diffuse warm ionized gas, which
worked best to insure that extraplanar H{\sc ii} regions are well visible within the emission of the diffuse ionized component.
In these images, the surface brightness of the disk emission with respect to the extraplanar emission is realistic in a qualitative,
but not quantitative, sense.
We used the following projection angles of ESO~137-001: inclination $i=66^{\circ}$, position angle $PA=10^{\circ}$. 

\section{Observations and model fitting \label{sec:observations}}

\begin{figure}[!ht]
  \centering
  \resizebox{\hsize}{!}{\includegraphics{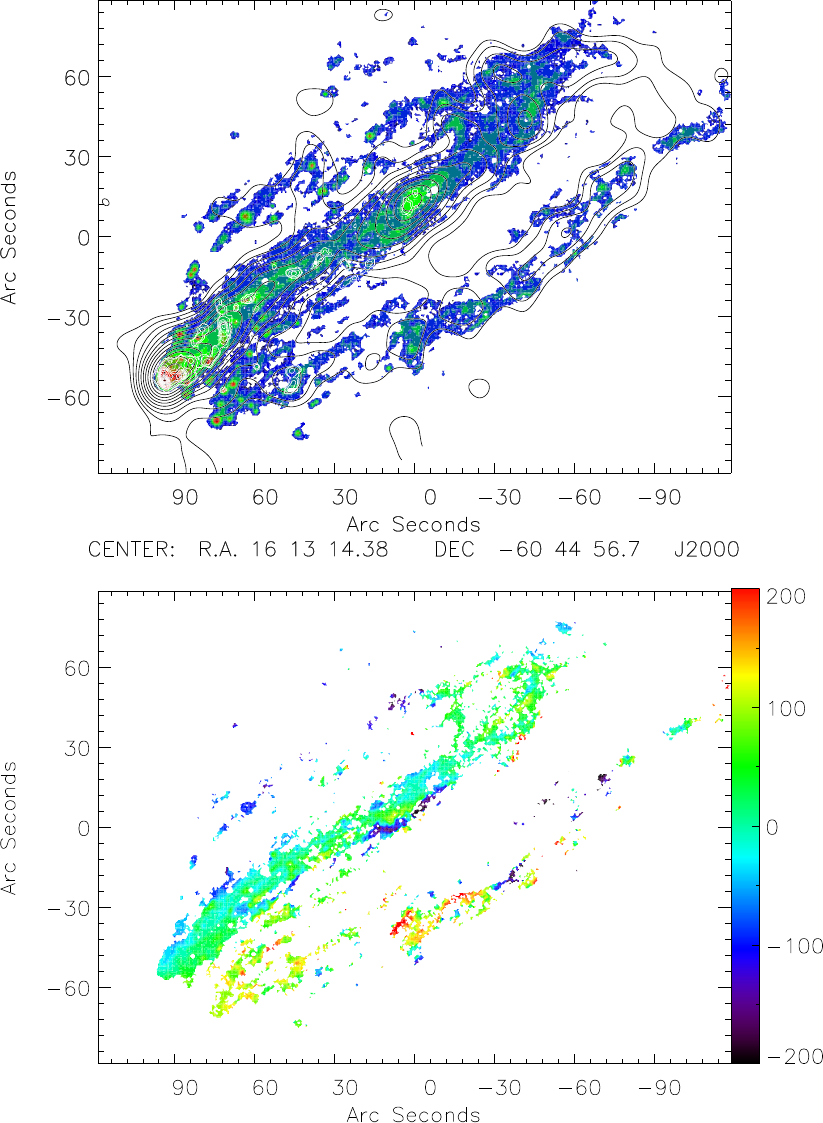}}
  \caption{Multi-wavelength observations of ESO~137-001. Upper panel: color: MUSE H$\alpha$ (Sun et al. 2022, Luo et al. 2023); black contours:
    Chandra X-ray (Sun et al. 2010); white contours: ALMA CO (Jachym et al. 2019).
    It appears that the X-ray stripping front is ahead of  the H$\alpha$ stripping front because of the large spatial smoothing scale of the X-ray image.
    Actually, the X-ray and H$\alpha$ stripping fronts are at the same position as shown in Fig.~2 of Luo et al. (2023). 
    Lower panel: MUSE H$\alpha$ velocity field (Sun et al. 2022).
    The color scale corresponds to the LOS velocities with respect to the systemic velocity in km\,s$^{-1}$.
    At a distance of $70$~Mpc $30''$ correspond to about $10$~kpc. The size of the images is $80$~kpc~$\times$~$60$~kpc.
  \label{fig:observations}}
\end{figure}

ESO~137-001 is one of the rare cases where CO, H$\alpha$, and X-ray data at decent spatial resolutions ($\la 1$~kpc) are available.
The X-ray observations were made by Chandra (Sun et al. 2010), the H$\alpha$ observations by MUSE (Sun et al. 2022), and the CO
observations by ALMA (Jachym et al. 2019). All these observation are displayed in Fig.~\ref{fig:observations}.

  There are several ways to quantitatively compare our simulations to the observations: global properties as integrated fluxes of the disk and tail regions
  at a given wavelength or length of the tail, brightness profiles or mean LOS velocity along the axis of the system (along the tail), or a
  direct comparison of the resolved emission distributions. Global properties do not depend on the morphology of
  the emission in a given region (e.g., the tail region). Profiles along given axes are averaged over the perpendicular direction and thus 
  depend on the global morphology along the axis. These quantities have the advantage that they mostly depend on the large-scale morphology and
  not much on particular realizations of a model, which might have different small-scale morphologies.
  In our case, different initial conditions of the galactic disk in terms of spiral arms and surface density profile will lead to a different
  morphology of the stripped gas tail (see, e.g., Vollmer et al. 2021).
  In the case of a limited number of simulations with only one initial condition, the comparison of
  global properties and profiles along the tail axis are the best choice. Since we made simulation with different ram pressure wind profiles
  and different initial conditions (Sect.~\ref{sec:param}), a direct comparison of the resolved model and observed emission distributions is appropriate.
  Because the emission distributions depend on the chosen projection (the inclination and position angles are given by the observations,
  the azimuthal angle has to be chosen), we produced model images with three different azimuthal angles.
  In addition, we allowed for a small possible rotation and a shrinking or expansion between the model and observed images.
  With about a hundred different simulations and three different projections,
  we are confident that it is possible to constrain the ram pressure profile and the time to peak ram pressure by a direct comparison of the resolved
  emission distributions.

For the search of the highest-ranked models we calculated the goodness of the fit for all timesteps of all models. This was also done for the velocity field.
We define the goodness as the sum of the absolute differences between the model and observed image pixels.
Before the calculation of the goodness of the model X-ray, H$\alpha$, and CO maps, the given model map was divided by the normalization factor $Q$ described in Eq.~16
of Vollmer et al. (2020). The factor $Q$ minimizes the $\chi^2$ between the model and observed maps.

We  clipped the model CO data at a surface density of $1$~M$_{\odot}$pc$^{-2}$ in a $10$~km\,s$^{-1}$ channel. If a detection in three adjacent channels is required
this yields a detection limit of $3$~M$_{\odot}$pc$^{-2}$, a value that is broadly consistent with the lowest contour of $6$~M$_{\odot}$pc$^{-2}$ in Fig.~4
of Jachym et al. (2019).

The clipping values of the model X-ray and H$\alpha$ surface brightness distributions were $10^{-6}$~erg\,cm$^{-2}$s$^{-1}$. 
The model H$\alpha$ clipping value corresponds to the MUSE $3\,\sigma$ H$\alpha$ surface brightness of $1.6 \times 10^{-18}$~erg\,s$^{-1}$cm$^{-2}$arcsec$^{-2}$
(Luo et al. 2023). The model X-ray clipping value corresponds to about one fourth of the Chandra $3\,\sigma$ X-ray surface brightness of
$\sim 6 \times 10^{-18}$~erg\,s$^{-1}$cm$^{-2}$arcsec$^{-2}$ for $\sim 16$~kpc scales (Sun et al. 2010).

For the comparison of model and observed the H$\alpha$ velocity fields we calculated the goodnesses for the clipped and unclipped model H$\alpha$ data.
The visual inspection of the best-fit models showed that the comparison based on the unclipped model H$\alpha$ data lead to the best result,
that is model H$\alpha$ velocity fields, which reproduce the main characteristics of the observed  H$\alpha$ velocity field.
The H$\alpha$ and CO maps were aligned to the X-ray image (same center and same pixel size) before the calculation of the goodnesses.

Because of the unknown CO-H$_2$ conversion factor especially in the tail of ESO~137-001 we decided not to use the CO surface brightness maps but to produce binary maps of
the model and observed CO maps. A pixel of the observed and model map was set to unity if the flux exceeds three times the rms and to zero otherwise.
In this way the fitting procedure favored models whose qualitative morphologies resemble that of the observed CO distribution.
In parallel, we also calculated the goodnesses using the CO surface brightness maps for the models with the nominal ICM-ISM mixing rate and stripping efficiency.
It appeared that in this case model timesteps smaller than $-50$~Myr from peak ram pressure are preferred when the goodness is based on the CO,
X-ray, H$\alpha$ emission, and the H$\alpha$ velocity field. However, the observed bifurcated structure of the tail was not well recovered by these models.
Reassuringly, our highest ranked model (Fig.~\ref{fig:eso137-001all_5687}) was consistently among the preferred models for both types of goodnesses.
The goodness distributions are smooth (Fig.~\ref{fig:goodnessdistribution}) with a change of slope after the first $\sim 100$ highest-ranked models.
For the sake of efficiency, we decided to inspect and present the $50$ highest-ranked models.
The visual inspection showed that the goodnesses based on the CO binary maps led to the best results in terms of model H$\alpha$ and X-ray maps as well
as the H$\alpha$ velocity field. In particular, the emission of the model tails was recovered in the model CO images in a satisfactory way.

Since we only calculated a restricted number of simulations, we took into account (i) a small possible rotation between the model and observed images
from $-15^{\circ}$ to $15^{\circ}$ in steps of $5^{\circ}$ and (ii) a shrinking or expansion by a factor of $0.8$ to $1.2$ in steps of $0.1$.
These modifications were applied to the observed images. In addition, we added and subtracted $30^{\circ}$ to and from the azimuthal projection angle.

Whereas the highest-ranked model at a single wavelength corresponds to the minimum goodness, the comparison of goodness at different
wavelengths is not straight forward. We decided to rank the models at the different wavelengths to calculate the sum of the ranks at
the different wavelengths. We define the highest-ranked model as the model with the smallest value of the sum of the ranks.
The highest-ranked models were determined for (i) the combination of the different wavelengths, (ii) this combination plus the
H$\alpha$ velocity field, and (iii) only the H$\alpha$ velocity field.
Of course, a valuable model should reproduce the spatial distribution of the gas phases and their velocity fields if available.
The 50 highest-ranked models of cases (i) to (iii) (Tables~\ref{tab:molent} to \ref{tab:molent2}) were inspected by eye,
with only a small portion presented in the form of images in this article.

\section{Results\label{sec:results}}

The first 50 models with the smallest total ranks are presented in the upper part of Table~\ref{tab:molent} to \ref{tab:molent2} for 
the comparison based on (i) the CO binary, H$\alpha$, X-ray images and the H$\alpha$ velocity field and (ii) the CO binary image and the H$\alpha$ velocity field,
and (iii) the CO binary , H$\alpha$ and X-ray images.
After the visual inspection of all highest-ranked models, we chose three pre-peak models, for which there is a good resemblance between the
model and observed multi-wavelength images. The selection process is described in Appendix~\ref{sec:selection}.

Gaussian models (models~6 and 7) are excluded by our selection process.
  When the model selection is only based on the CO binary , H$\alpha$ and X-ray maps (models with $\Delta t < 0$ in Table~\ref{tab:molent2}),
  models~3 (high ram pressure and intermediate width, galaxy orbit of intermediate eccentricity) and 5
  (highest ram pressure and smallest width or highly eccentric galaxy orbit) are preferred.
  When the model selection is based on the CO binary map and the H$\alpha$ velocity field
  (models with $\Delta t < 0$ in Table~\ref{tab:molent1a}), the models with the lowest peak pressure and the largest width are preferred (model~1; less eccentric
  galaxy orbit).
  In addition, three models with the highest peak ram pressure and smallest width are present among the $50$ highest-ranked models.
   When the model selection is based on the CO binary, H$\alpha$ and X-ray maps and the H$\alpha$ velocity field (models with $\Delta t < 0$ in Table~\ref{tab:molent}),
  the models with the highest peak pressure and the smallest width are preferred (model~5).
  Model~5 with a highly eccentric galaxy orbit is the overall preferred model because it is among the $50$ highest-ranked models
  based on all three selection criteria. Shifting the time of peak ram pressure (models~5a and 5b) leads to a different morphology of
  the tail while the tail length and width are generally conserved.

\begin{figure}[!ht]
  \centering
  \resizebox{\hsize}{!}{\includegraphics{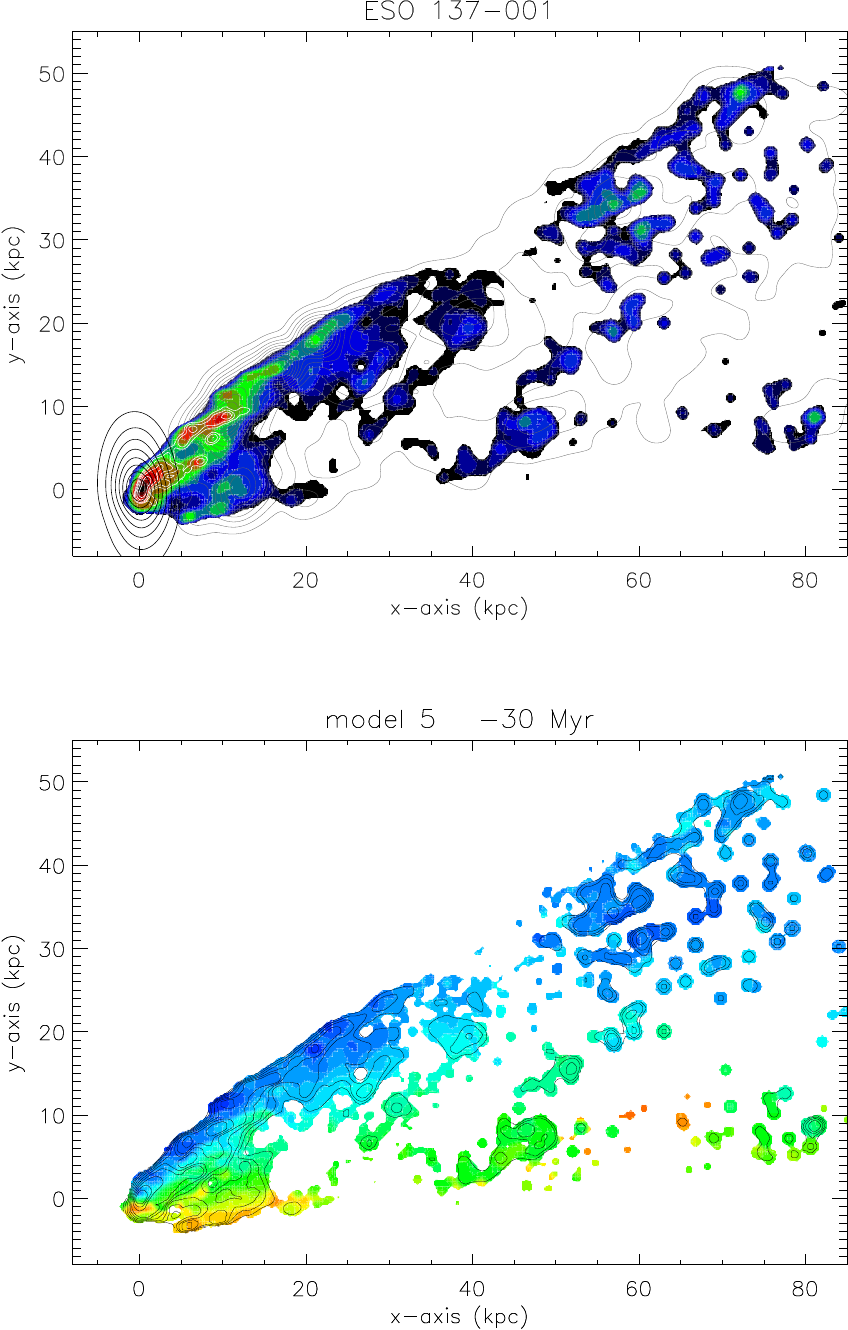}}
  \put(-220,360){\large \bf highest-ranked model~A}
  \put(-60,177){$\Theta=60^{\circ}$}
  \caption{Highest-ranked model~A of ESO~137-001. Upper panel: color: H$\alpha$; dark gray contours:
    X-ray; white contours: CO ; black contours: stellar content. The relative contours are same the as in Fig.~\ref{fig:observations}:
    H$\alpha$: logarithmic transfer function with $15$ contour levels; X-ray and CO: square root transfer function with $15$ contour levels;  
    Lower panel: H$\alpha$ velocity field. The colors are the same as in Fig.~\ref{fig:observations}. The corresponding time evolution 
    can be found here: (observations\_modelA.gif). The time evolution of the mass surface densities of the different gas phases can be found here:
    (gasphases\_modelA.gif).
  \label{fig:eso137-001all_5687}}
\end{figure}

It turned out that the observed extraplanar CO emission within $10$~kpc of the
galactic disk with the upturning extraplanar CO filament 
(Fig.~\ref{fig:esoplot_sketch} and upper panel of Fig.~4 of Jachym et al. 2019) was very hard to reproduce.
These features are only present in model~5 at $\Delta t \sim -20$-$-40$~Myr, which are amongst the highest-ranked pre-peak models of the comparison
based on the CO/H$\alpha$/X-ray emission distributions and the H$\alpha$ velocity field (Table~\ref{tab:molent}).
We call this model the highest-ranked model~A (Figs.~\ref{fig:eso137-001all_5687} and \ref{fig:zusammen_co}).

The parameters of the highest-ranked model~A are presented in Table~\ref{tab:molent}.
The CO, H$\alpha$, and X-ray emission distribution together with the H$\alpha$ velocity field are presented in Fig.~\ref{fig:eso137-001all_5687}.

\begin{figure}[!ht]
  \centering
  \resizebox{\hsize}{!}{\includegraphics{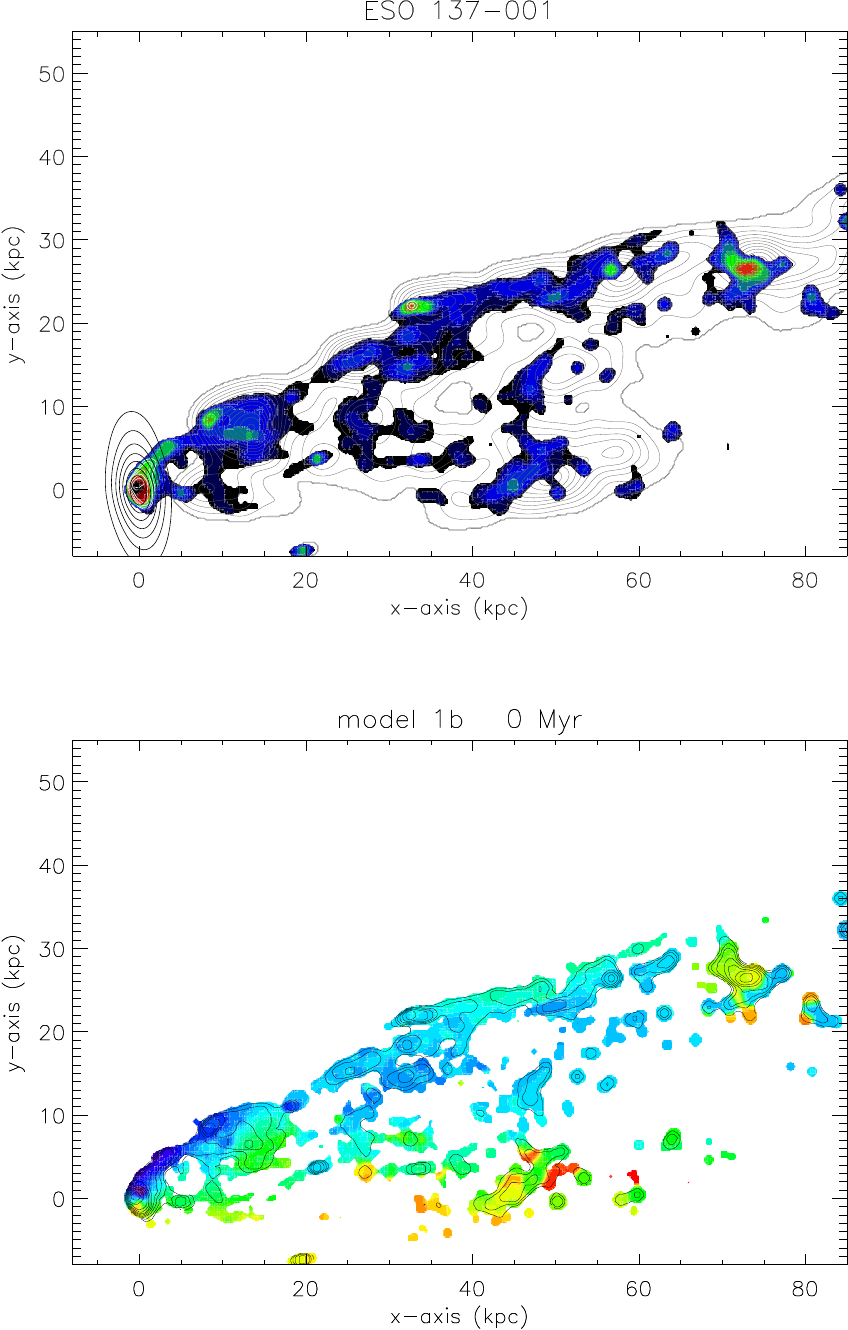}}
  \put(-220,360){\large \bf preferred model~B}
  \put(-60,177){$\Theta=75^{\circ}$}
  \caption{Preferred model~B of ESO~137-001. Upper panel: color: H$\alpha$; dark gray contours:
    X-ray; white contours: CO ; black contours: stellar content. Lower panel: H$\alpha$ velocity field. See Fig.~\ref{fig:eso137-001all_5687} for the
    description of the contours and colors. The corresponding time evolution 
    can be found here: (observations\_modelB.gif). The time evolution of the mass surface densities of the different gas phases can be found here:
    (gasphases\_modelB.gif).
  \label{fig:eso137-001all_471mom1_mixing1}}
\end{figure}

\begin{figure}[!ht]
  \centering
  \resizebox{\hsize}{!}{\includegraphics{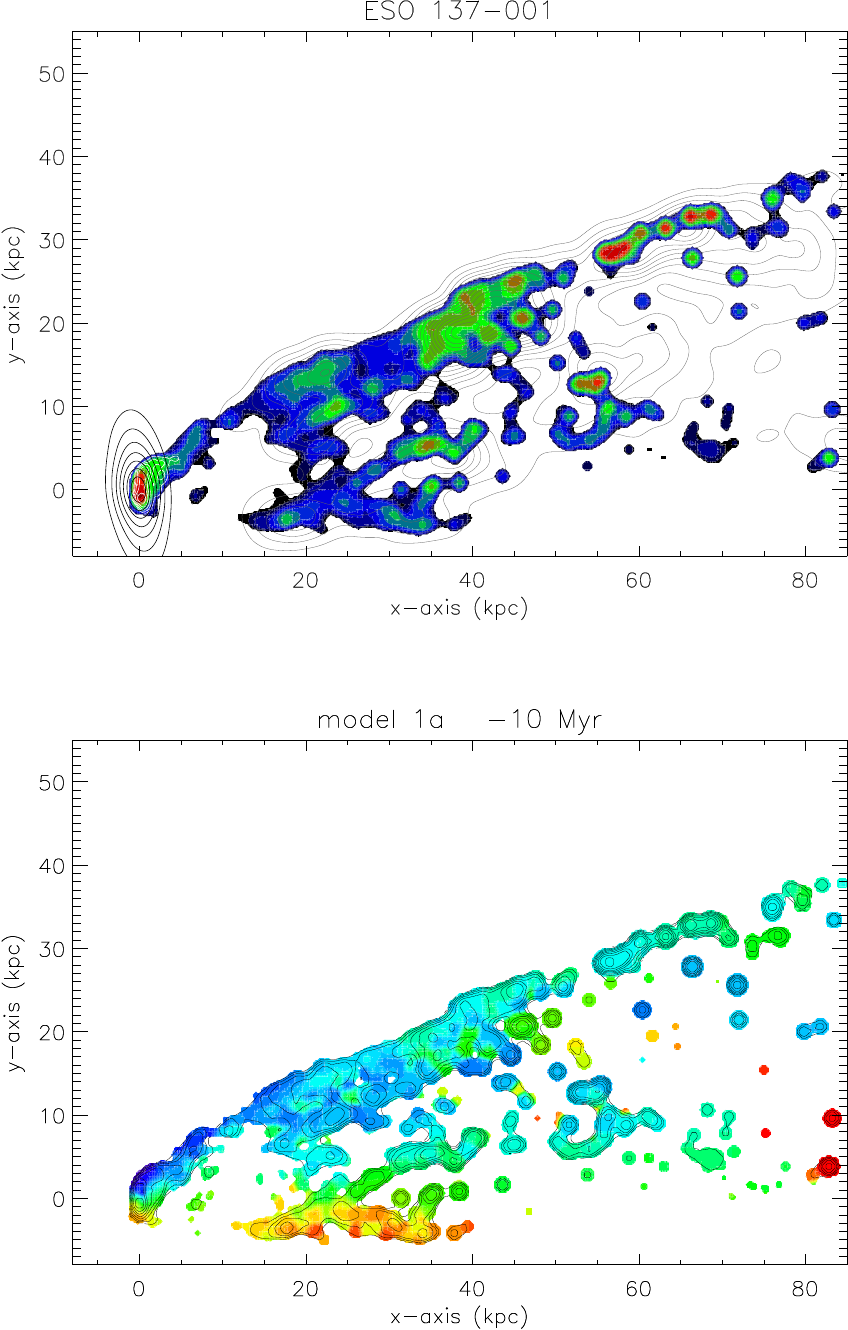}}
  \put(-220,360){\large \bf preferred model~C}
  \put(-60,177){$\Theta=75^{\circ}$}
  \caption{Preferred model~C of ESO~137-001. Upper panel: color: H$\alpha$; dark gray contours:
    X-ray; white contours: CO ; black contours: stellar content. Lower panel: H$\alpha$ velocity field. See Fig.~\ref{fig:eso137-001all_5687} for the
    description of the contours and colors. The corresponding time evolution 
    can be found here: (observations\_modelC.gif). The time evolution of the mass surface densities of the different gas phases can be found here:
    (gasphases\_modelC.gif).
  \label{fig:eso137-001all_472mom1_mixing1}}
\end{figure}

The peak-ram pressure model~1b is the highest-ranked model only based on the velocity field with $\Delta t \le 0$~Myr.
We call this model the preferred model~B (Fig.~\ref{fig:eso137-001all_471mom1_mixing1}; Table~\ref{tab:molent1a}).
Moreover, we visually identified model~1a with $\Delta t = -10$~Myr as preferred model~C (Fig.~\ref{fig:eso137-001all_472mom1_mixing1}; Table~\ref{tab:molent1a}).

In all models, A, B and C, the main tail and the southern secondary tail are present. As observed, the main tails of models A and C are significantly brighter than
the secondary tails. In model~B the two tails have about the same surface brightness.
The extents of the X-ray and H$\alpha$ tails are reasonably reproduced by the models. 
Whereas the lengths of the secondary tails are well reproduced by the models, the sizes of model main tails are
about $30$\,\% larger than the observed sizes. The observed broadening to the north of the main H$\alpha$ tail at distances $>40$~kpc
is only present in model~A.
Overall, the morphology of the observed X-ray emission distribution is better reproduced
by models~A and C than that of model~B. 
Especially, the observed local X-ray maximum $\sim 37$~kpc from the galaxy center is only reproduced by
model~C. Furthermore, the H$\alpha$ velocity field at distances $> 30$~kpc is better reproduced by models~B and C. This is expected
because they were selected based on the comparison of the velocity fields.
The observed northern faint H$\alpha$ tail is absent in all models, except for the slight broadening to the north of the main
H$\alpha$ tail at distances $>30$~kpc in the highest-ranked model~A.

The observed CO distribution is better reproduced by the highest-ranked model~A.
We observe two filamentary structures in the model CO emission distribution of the main tail: a northern convex and
southern concave filament. The latter corresponds to the upbending extraplanar CO filament (Fig.~\ref{fig:esoplot_sketch}).
Whereas the southern filament is stronger than northern one in the ALMA observations, the model shows
two filaments of similar column densities (see Sect.~\ref{sec:moleculargas}).

We conclude that the X-ray and H$\alpha$ observations are best reproduced by model~C, whereas the CO
observations are better reproduced by model~A. The existence of a two-tail structures is a common feature in our
models. It is due to the combined action of ram pressure and rotation together with the projection of the galaxy on the sky.
Magnetic fields might enhance the appearance of a tail that is bifurcated in the plane of the sky (Ruszkowski et al. 2014) but they are not included in our model.
Too strong magnetic field might suppress thermal conduction, which is needed to explain the observed H$\alpha$ tail emission (see Sect.~\ref{sec:magfield}). 

\subsection{The influence of the ICM-ISM mixing rate \label{sec:mixing}}

We recalculated all models of Tables~\ref{tab:molent} to \ref{tab:molent2} with a three times higher and lower ICM-ISM mixing rate.
The ranking results for the comparisons (i) to (iii) are presented in Tables~\ref{tab:moremolent} to \ref{tab:lessmolent2}.

As before, we focus on near ram-pressure peak models.
For a higher mixing rate model~1c is preferred at $\Delta t=-10$-$10$~Myr for comparisons (i) and (iii) (Tables~\ref{tab:moremolent} and \ref{tab:moremolent2}).
Based on the the comparison between the CO binary maps and the $H\alpha$ velocity field
models~1, 1a, and 1b at $\Delta t=-50$-$-10$~Myr are preferred (Table~\ref{tab:moremolent1}).
All preferred models show strong X-ray main tails, but faint or absent outer H$\alpha$ tails. 
The H$\alpha$ extents of these model tails with a higher mixing rate are significantly smaller than those
of the models with the nominal mixing rate. The reason for this behavior is the higher stripping efficiency (Sect.~\ref{sec:iong}) of mixed hot gas
together with the faster mixing. The outer parts of the tails in the model with the nominal mixing rate are already pushed to larger
distances out of the field of view in the model with the three times higher mixing rate. In all preferred models the southern
gas tail is barely visible in H$\alpha$ emission. 

The $50$ highest-ranked models with a three times lower mixing rate are presented in Tables~\ref{tab:lessmolent} to \ref{tab:lessmolent2}.
For a lower mixing rate the H$\alpha$ and X-ray morphologies of the model gas tails are very different from the observed morphologies
(Figs.~\ref{fig:moreeso137-001all} to \ref{fig:lesseso137-001allnovel}).
Model emission H$\alpha$ and X-ray is mostly seen up to distances of $\sim 20$~kpc from the galaxy center. In addition, rare patchy emission regions with
sizes of $\sim 10$~kpc are present at larger distances. The gas distribution in the tail is much smoother with less overdensities than in the models
with the nominal or a three times higher mixing rate. Because of the applied sensitivity limits only a small amount of emission is present
in the model X-ray and H$\alpha$ maps of the models with a lower mixing rate. None of the preferred model based on the comparisons (i) to (iii) reproduce
the available observations.

We conclude that the highest-ranked models with the nominal ICM-ISM mixing rate reproduce observations significantly better than the
models with a three times higher or lower ICM-ISM mixing rate.

\subsection{The influence of the hot gas stripping efficiency \label{sec:efficiency}}

As stated in Sect.~\ref{sec:iong} the calculation of the acceleration caused by ram pressure is $a=\rho_{\rm ICM}v_{\rm gal}^2/\Sigma$. This acceleration was further
increased by a heuristic factor of ten if the ISM temperature exceeds $10^4$~K. To investigate the influence of this enhanced stripping efficiency of
the hot gas on the model results, we recalculated all models of Tables~\ref{tab:molent} to \ref{tab:molent2} with a three times lower stripping
efficiency of the hot gas. The ranking results for the comparisons (i) to (iii) are presented in Tables~\ref{tab:xpushmolent} to \ref{tab:xpushmolent2}.
The highest-ranked models with $\Delta t \le 0$~Myr are model~2 for comparison (i), models~1, 1a, and 1b for comparison (ii), and models~2 and 3
for comparison (iii). 

The models with a lower stripping efficiency typically show stronger outer X-ray tails. This is expected because the gas in the outer tail is
less accelerated than in the model with the nominal stripping efficiency and thus stays denser.
Only the third and fourth highest-ranked model based on comparison (ii) (Fig.~\ref{fig:xpusheso137-001allonlyvel})
show two separate tails in X-ray and H$\alpha$ emission with sizes larger than $30$~kpc.

\begin{figure*}[!ht]
  \centering
  \resizebox{\hsize}{!}{\includegraphics{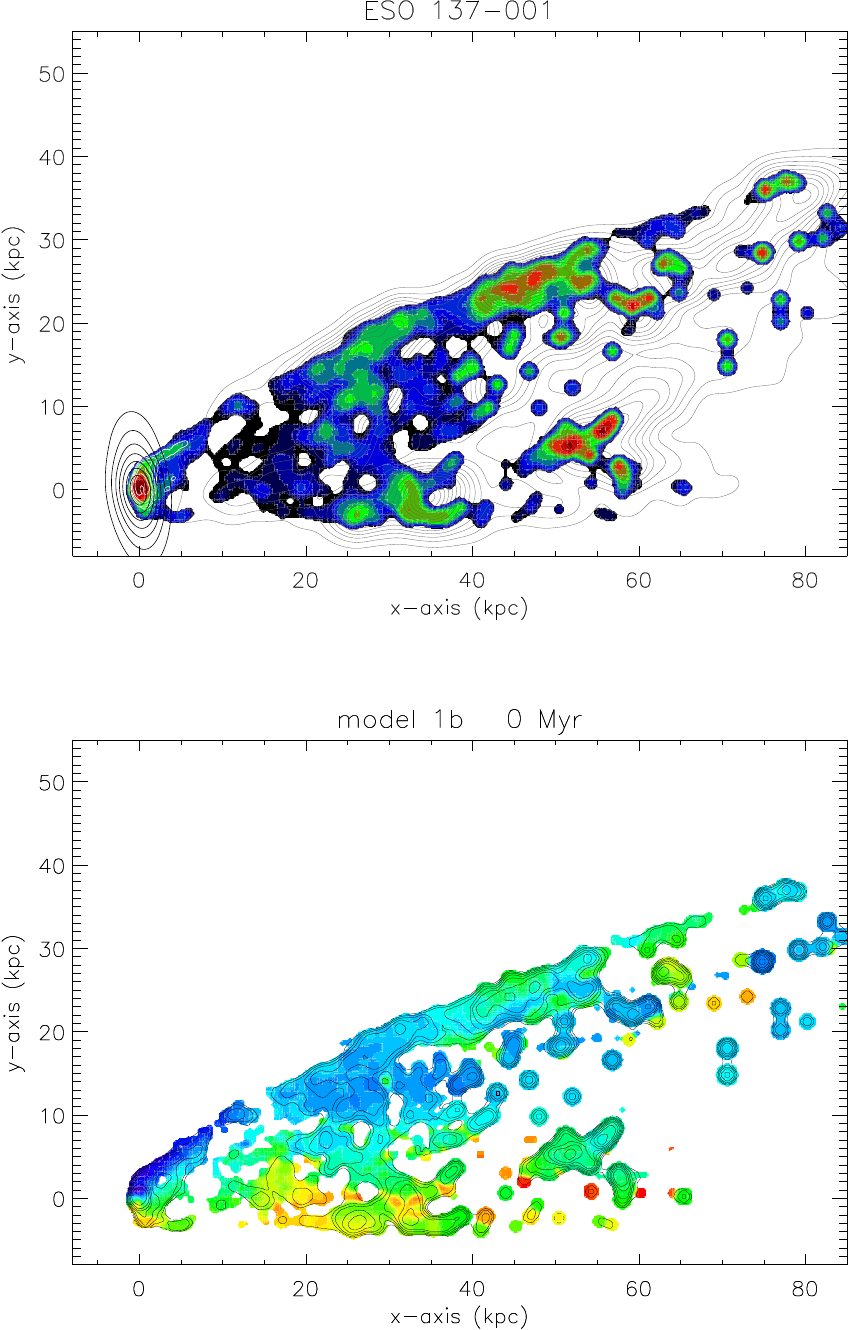}\includegraphics{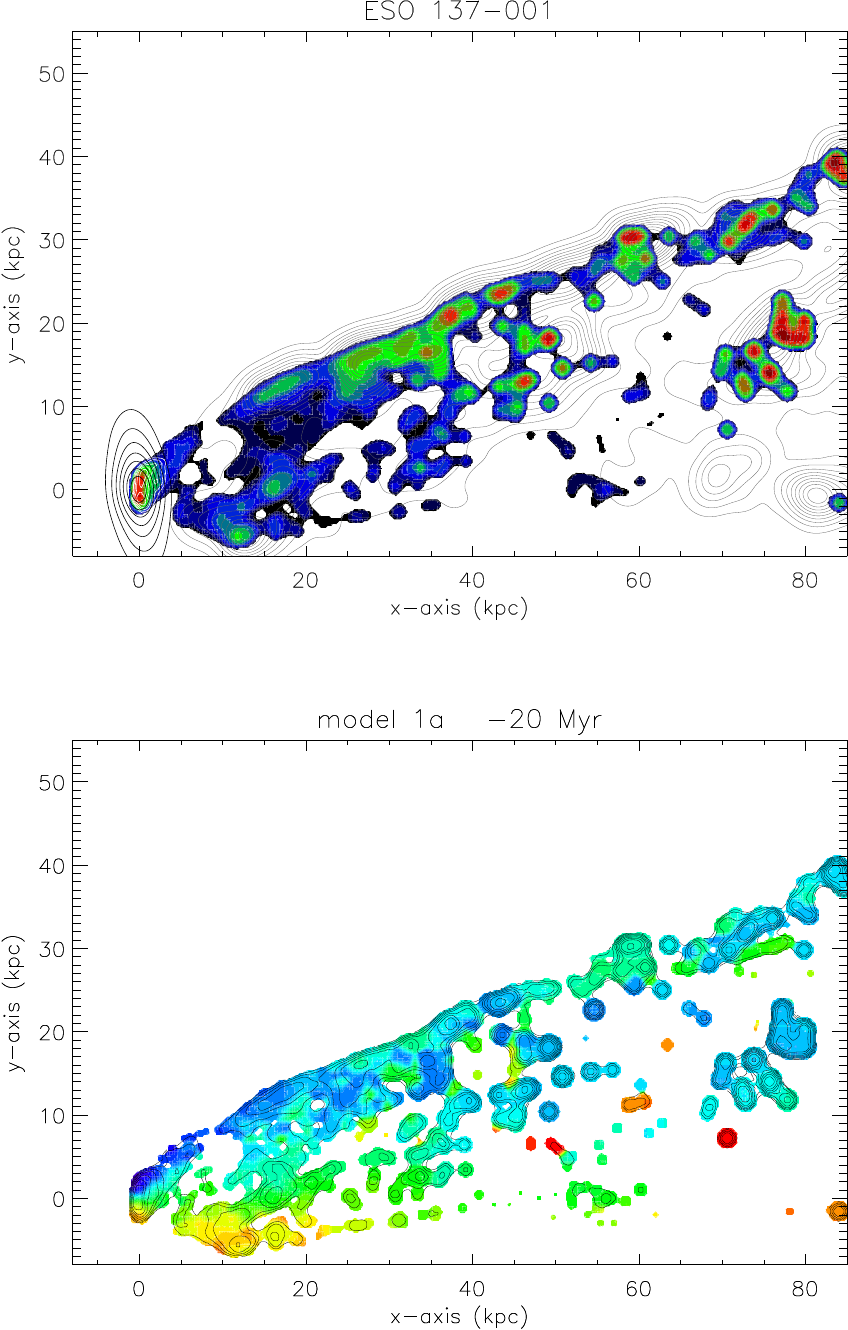}}
  \caption{Left panel: model corresponding to the preferred model~B of Fig.~\ref{fig:eso137-001all_471mom1_mixing1}
    with a three times lower stripping efficiency with respect to the nominal stripping efficiency.
    Right panel: model close to the preferred model~C ($\Delta t=-20$~Myr instead of $\Delta t=-10$~Myr) of Fig.~\ref{fig:eso137-001all_472mom1_mixing1}.
    Upper panel: color: H$\alpha$; dark gray contours:
    X-ray; white contours: CO ; black contours: stellar content. Lower panel: H$\alpha$ velocity field. See Fig.~\ref{fig:eso137-001all_5687} for the
    description of the contours and colors.
  \label{fig:eso137-001all_mixing1_Xpushonlyvel_04}}
\end{figure*}

For a direct comparison with the models of the nominal stripping efficiency, we show the preferred model~B with a
three times lower stripping efficiency in (Fig.~\ref{fig:eso137-001all_mixing1_Xpushonlyvel_04}).
The gas truncation radius within the disk is larger than that of the preferred model~B.
Moreover, the X-ray and H$\alpha$ tails are significantly stronger than those of the preferred model~B.
The emission of the northern and southern model tails are enhanced by about the same factor.
Especially the model H$\alpha$ emission of the southern tail is much stronger than it is observed.  
As expected, the velocity field is closer to the observed H$\alpha$ velocity field than that of the highest-ranked model~A.

We conclude that overall the CO/H$\alpha$/X-ray highest-ranked models with a three times lower stripping efficiency
reproduce the available observations less well than the highest-ranked models with the nominal stripping efficiency.

\section{Discussion \label{sec:discussion}}

After the detailed comparison between the models and observations based on different data ((i) CO/X-ray/H$\alpha$ emission and H$\alpha$ velocity field),
(ii) H$\alpha$ velocity field, (iii) CO/X-ray/H$\alpha$ emission) we are left with three models, which reproduce the available observations 
in a satisfactory way: the highest-ranked model~A (Fig.~\ref{fig:eso137-001all_5687}), the preferred model~B (Fig.~\ref{fig:eso137-001all_471mom1_mixing1})
and the preferred model~C (Fig.~\ref{fig:eso137-001all_472mom1_mixing1}). All these models have an ICM-ISM mixing rate, which is consistent with
theoretical expectations (Sect.~\ref{sec:iong}).

Model~A has an angle between the galactic disk and the ram pressure wind of $\Theta=60^{\circ}$, whereas models~B to E have $\Theta=75^{\circ}$.
The LOS components of the model galaxy velocity unit vector of model~A is tiny ($0.07$-$0.09$) and small of models~B and C ($0.33$).
Whereas the time to peak ram pressure of model~A is $\Delta t=-40$-$-20$~Myr (before peak ram pressure), it is $\Delta t=0$ for model~B and
$\Delta t=-30$-$-10$~Myr for model~C.
If we require that the galaxy is observed before peak ram pressure as deduced from its location within the Norma cluster and the direction
of its gas tail (Sun et al. 2010), we prefer models~A and C. If we also allow for peak ram-pressure, model~B is also a preferred model.

In the following we will investigate the existence of a third gas tail in the simulation (Sect.~\ref{sec:northerntail}) and
inspect the stripped multi-phase model gas masses (Sect.~\ref{sec:gasmasses}),
the model CO emission distribution and velocity field (Sect.~\ref{sec:moleculargas}), the model UV emission and star formation distributions (Sect.~\ref{sec:UVstar}),
and the H$\alpha$--X-ray correlation of the stripped gas tail (Sect.~\ref{sec:haxcorrelation}).
Moreover, we will give a possible orbit of ESO~137-001 within the Norma cluster (Sect.~\ref{sec:Norma}).

\subsection{The faint northern gas tail \label{sec:northerntail}}

\begin{figure}
  \centering
  \resizebox{\hsize}{!}{\includegraphics{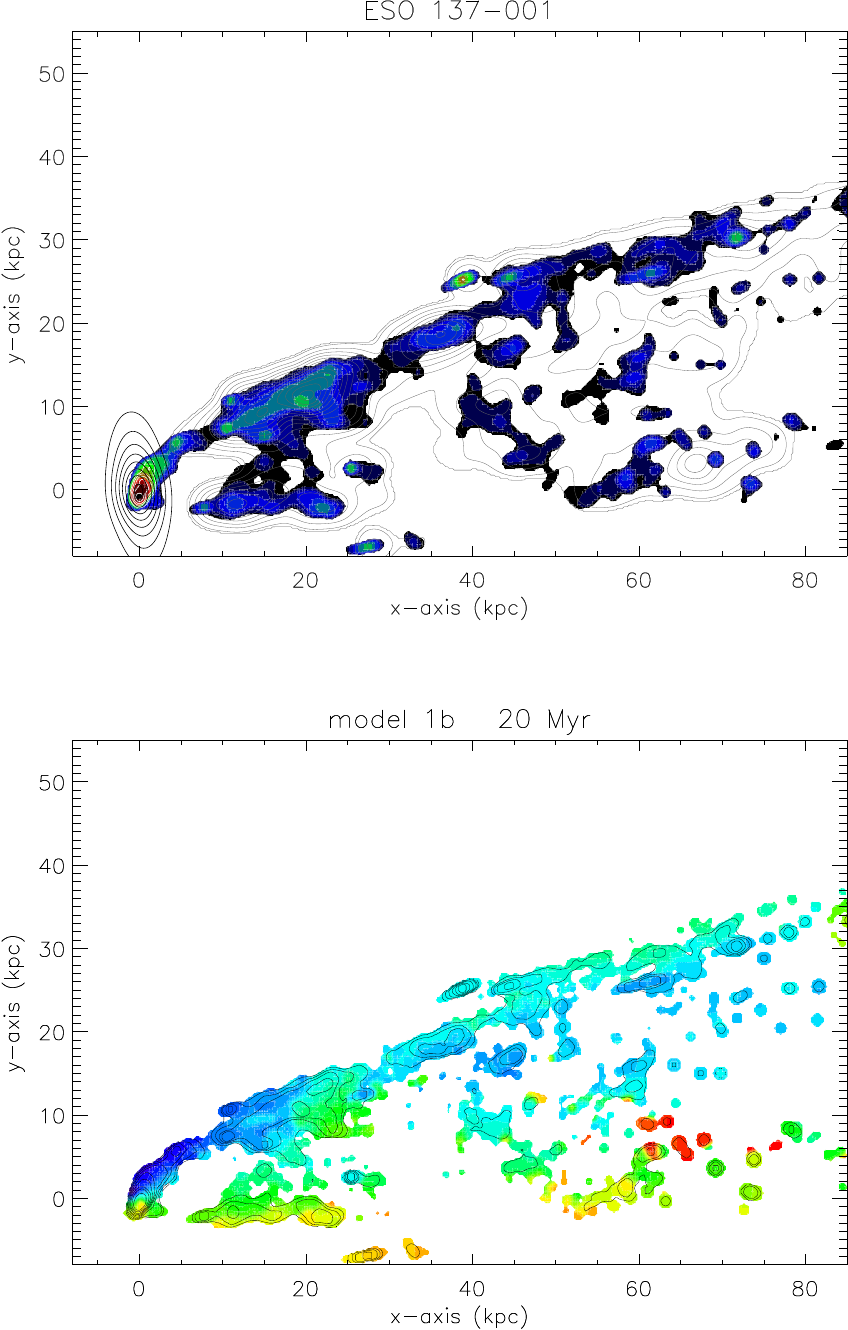}}
  \put(-150,330){detached H$\alpha$}
  \put(-125,235){detached H$\alpha$}
  \caption{ESO~137-001 model showing detached elongated H$\alpha$ emission regions in the north and south of the two main tails.
  \label{fig:eso137-001all_mixing1onlyvel_29}}
\end{figure}

As discussed in Sect.~\ref{sec:results} the faint northern gas tail (Fig.~\ref{fig:esoplot_sketch}) is absent in our preferred models.
However, such faint tail structures are present in our simulations but they are rare for the highest-ranked models.
We found such structures  in model~1b at timesteps $0~{\rm Myr} \le \Delta t \le 40$~Myr within the highest-ranked models when only the
H$\alpha$ velocity field is taken into account (Table~\ref{tab:molent1a}). The timestep $\Delta t=20$~Myr is presented in Fig.~\ref{fig:eso137-001all_mixing1onlyvel_29}.
This model corresponds to model~B with a $20$~Myr later time and a somewhat different projection.
Two detached elongated H$\alpha$ emission regions in the north and south of the two main tails are visible at $(40,\,25)$~kpc and $(25,\,-6)$~kpc
(upper panel of Fig.~\ref{fig:eso137-001all_mixing1onlyvel_29}).
They are made of gas that resists the ram pressure wind because of its high density. This gas is constantly ablated by the
ram pressure wind. The existence of such structures depends on the density structure of the stripped gas, which itself depends on the
initial gas distribution and the temporal ram pressure profile.
We conclude that our model is in principle able to produce structures as the observed faint northern gas tail, which is mainly detected in H$\alpha$ emission.

\subsection{Multi-phase gas masses \label{sec:gasmasses}}

To provide a more quantitative comparison with observations, we extracted the tail gas masses within the regions of the model maps
($90$~kpc$\times 60$~kpc; Table~\ref{tab:multiphase}). We did this for the cold neutral medium (CNM), the warm neutral medium (WNM), the warm ionized medium (WIM),
and the hot ionized medium (HIM). For the mass of the warm ionized medium of a gas particle we set
\begin{equation}
  M_{\rm WIM}=3 \sigma_0 70\, \rho_{\rm hot} V_{\rm warm} = 3 \sigma_0 70\, \rho_{\rm hot} M_{\rm cl} (10^3 \rho_{\rm warm})^{-1}\ ,
\end{equation}
where $\rho_{\rm hot, warm}$ are the average density of the hot and warm stripped gas.

\begin{figure}
  \centering
  \resizebox{\hsize}{!}{\includegraphics{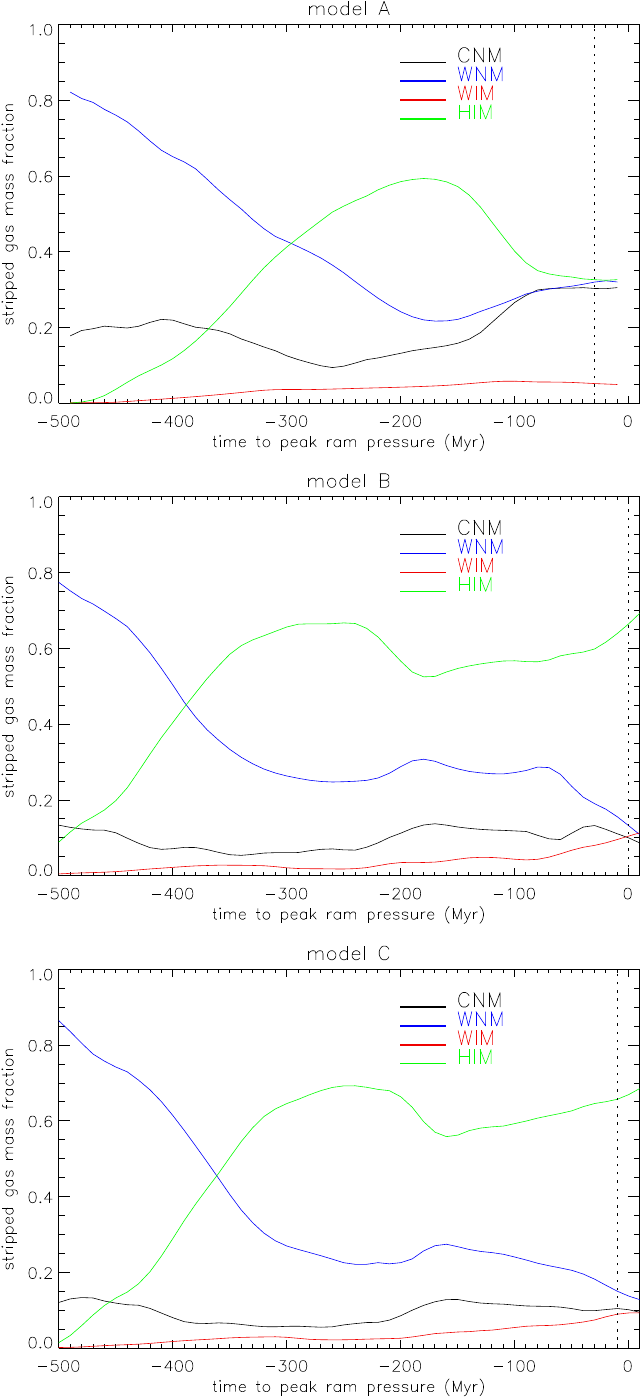}}
  \caption{Fractions of stripped ($2$~kpc$\le x \le 80$~kpc) gas in the different phases. The time $t=0$~Myr corresponds to peak ram pressure.
  \label{fig:strmasses}}
\end{figure}

The fractions of stripped ($2$~kpc$\le x \le 80$~kpc) gas in the different phases (CNM, WNM, WIM, and HIM) as a function of time are presented in Fig.~\ref{fig:strmasses}.
The WNM observed in the H{\sc i} line represents $\sim 80$\% of the stripped gas at the beginning of the all three simulations. The WNM mass fraction then
decreases to $\sim 20$\,\% after $200$-$300$~Myr. Whereas it increases slightly towards peak ram pressure in model~A, it decreases to $\sim 10$\,\% at peak
ram pressure in models~B and C. The CNM fraction varies between $10$\,\% and $20$\,\% in all three models. Only in model~A it increases to $\sim 30$\,\% at
peak ram pressure. The WIM fraction monotonically increases with time in all three simulations. It reaches a maximum of $\sim 5$\,\% in model~A and
$\sim 10$\,\% in models~B and C. The HIM fraction increases to $60$\,\% after $200$-$300$~Myr. It then decreases to $\sim 30$\,\% at peak ram pressure
in model~A and stays approximately constant in models~B and C. Near peak ram pressure all three gas phases are equally present in the stripped gas in model~A,
whereas the HIM dominates the mass budget in models~B and C. 

For the CNM Jachym et al. (2019) derived a mass of $M_{\rm CNM} \sim 9 \times 10^8$~M$_{\odot}$ within the disk and $M_{\rm CNM} \sim 10^9$~M$_{\odot}$
in the tail region assuming a Galactic CO-H$_2$ conversion factor.
Sun et al. (2007) estimated the mass of the WIM. Assuming a Galactic volume filling factor of $0.2$ (Boulares \& Cox 1990) yields
$M_{\rm WIM} \sim 2 \times 10^8$~M$_{\odot}$. Sun et (2006, 2010) derived a HIM mass of $M_{\rm HIM} \sim 10^9$~M$_{\odot}$ in the tail of ESO~137-001.
\begin{table*}[!ht]
      \caption{Model gas masses in the gas tail of ESO~137-001.}
         \label{tab:multiphase}
      \[
         \begin{tabular}{lcccccc}
           \hline
            & disk$^{\rm (a)}$ CNM & tail$^{\rm (a)}$ CNM & tail WNM$^{\rm (b)}$ & tail WIM & tail HIM & SFR$_{\rm disk}$/SFR$_{\rm tail}$ \\
            & ($10^8$~M$_{\odot}$) & ($10^8$~M$_{\odot}$) & ($10^8$~M$_{\odot}$)& ($10^8$~M$_{\odot}$)& ($10^8$~M$_{\odot}$) & \\
           \hline
           observed & $7$ & $9$ & -- & $2$ & $10$ & $2.0$ \\
           model~A & $5.8$ & $4.3$ & $5.0$ & $0.9$ & $4.1$ & $1.2$ \\
           model~B & $7.8$ & $0.4$ & $2.8$ & $0.8$ & $2.5$ & $6.9$ \\
           model~C & $7.0$ & $1.2$ & $2.7$ & $2.2$ & $5.9$ & $5.4$ \\
           \hline
        \end{tabular}
         \]
         \begin{list}{}{}
         \item $^{\rm (a)}$ the division between tail and disk region is at $x=2$~kpc.
         \item $^{\rm (b)}$ H{\sc i} contained in the maps of Fig.~\ref{fig:zusammen2_hi}. 
      \end{list}
\end{table*}

The model HIM masses vary between $2.5$ an $5.9 \times 10^8$~M$_{\odot}$. This is about half of the HIM mass derived from X-ray observations.
The model WIM masses, which critically depend on the assumed volume filling factor of the WIM, range between $0.8$ an $2.2 \times 10^8$~M$_{\odot}$.
This is quite close to the value derived from H$\alpha$ observations.
The WNM masses vary by about a factor of two between the different models ($2.7$-$5.0 \times 10^8$~M$_{\odot}$).
The mass distribution of the different
models is broadest for the CNM in the tail region\footnote{We define the tail region as regions with $x > 2$~kpc.} ($0.4$-$4.3 \times 10^8$~M$_{\odot}$). 
The CNM disk masses of all three models are comparable to the observed disk mass based on a Galactic CO-H$_2$ conversion factor.
The CNM mass of the tail of model~A is about half of the observed mass. The tail CNM masses of models~B and C are a factor of $11$ and
$4$ times smaller than that of model~A, respectively.

Waldron et al. (2023) give a star formation rate of $1.2$~M$_{\odot}$yr$^{-1}$ in the disk of ESO~137-001.
With a molecular gas mass within galactic disk of $7 \times 10^8$~M$_{\odot}$ (Jachym et al. 2019) the star formation efficiency (SFE)
is $17 \times 10^{-10}$~yr$^{-1}$. This is significantly higher than the average SFE in disks of local
spiral galaxies ($5.3 \pm 2.5 \times 10^{-10}$~yr$^{-1}$) and as high as the SFE in the centers of NGC~4736 and NGC~3351 (Leroy et al. 2008).

Leroy et al. (2013) stated that molecular gas in the central regions of spiral galaxies leads to more CO emission, appears more excited, and
forms stars more rapidly than molecular gas further out in the disks. Sandstrom et al. (2013) found that most galaxies exhibit a lower
conversion factor in the central kpc by factor of about two below the galaxy mean, on average. Moreover, the CO-H$_2$ conversion factor in the
central part of the Galaxy is about four times lower than that in the disk (Bolatto et al. 2013).
According to Sandstrom et al. (2013), NGC~4736 and NGC~3351 show a CO-H$_2$ conversion factor in the central kpc, which is more than five times
lower than the galactic value. With a two times lower conversion factor than Galactic the CNM mass within the galactic disk of ESO~137-001 would be
$\sim 4.5 \times 10^8$~M$_{\odot}$ consistent with the molecular gas mass of model~A. The SFE then is $8 \times 10^{-9}$~yr$^{-1}$.

The molecular gas mass in the tail region is $9 \times 10^8$~M$_{\odot}$ based on a Galactic CO-H$_2$ conversion factor.
Comparison of ﬂuxes of the ALMA+ACA observations with previous single-dish (APEX) observations (Jachym et al. 2019) indicated that, in
addition to the compact CO features, there is a substantial component (up to a factor of $3$ to $4$) of extended (scales $>6$~kpc) molecular gas in the tail.
Since a CO-H$_2$ conversion factor significantly lower than Galactic is expected in the tail, the missing ALMA flux might be compensated by the assumed
Galactic CO-H$_2$ conversion factor.
The star formation rate associated with this gas is about $0.5$~M$_{\odot}$yr$^{-1}$. Thus, whereas the CO flux in the disk region is as high as
that of the tail, the associated star formation is about twice that of the tail. Since we integrated the entire star formation activity of our models
without a sensitivity cutoff, our model SFR in the tail region represents an upper limit. The ratio SFR$_{\rm disk}$/SFR$_{\rm tail}$ is thus a lower limit.
When we take this limitation into account, model~A best reproduces the SFR found by Waldron et al. (2023).

We conclude that models~A best reproduces the observed CO emission and SFR fractions between the disk and tail regions.

\subsection{The dense molecular gas and CO emission \label{sec:moleculargas}}

\begin{figure*}[!ht]
  \centering
  \sidecaption
  \resizebox{12cm}{!}{\includegraphics{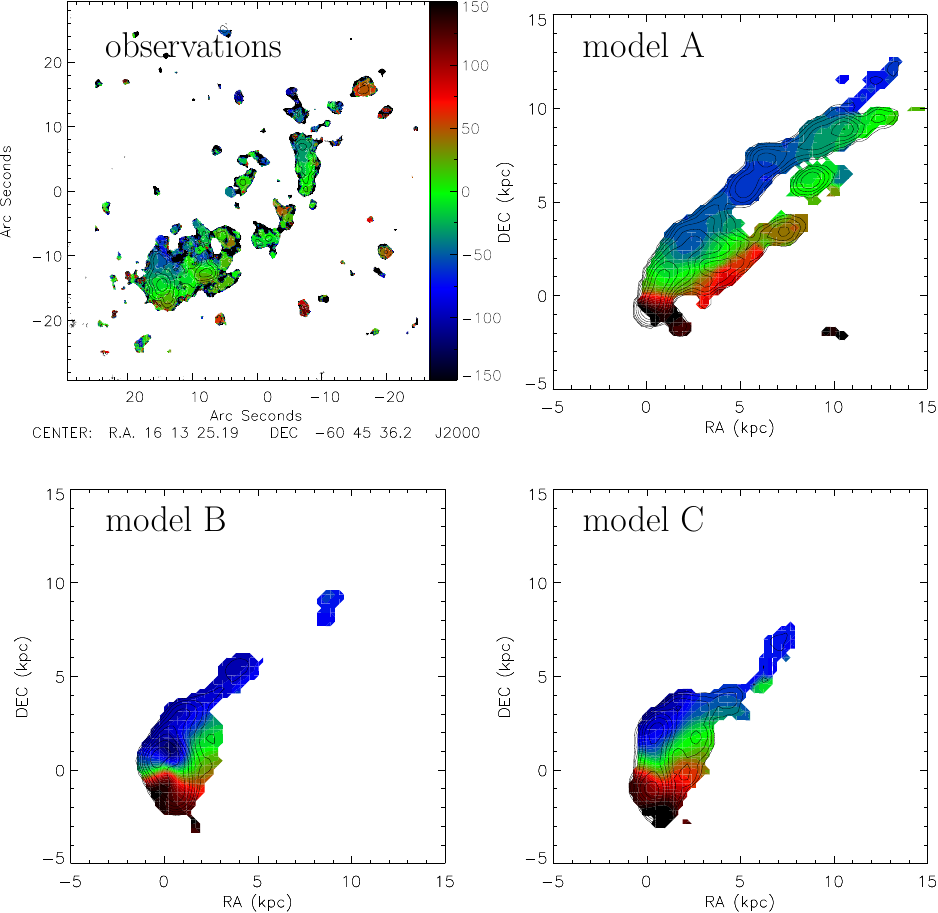}}
  \caption{CO emission distribution (contours) on CO velocity field. Upper left: ALMA observations. The color levels are the same in all panels.
  \label{fig:zusammen_co}}
\end{figure*}

The observed  (Jachym et al. 2019) and model CO velocity fields together with the CO emission distribution are presented in Fig.~\ref{fig:zusammen_co}.
The CO disk emission shows a north-south velocity gradient due to galactic rotation. The observed velocity amplitude is about $50$~km\,s$^{-1}$
(see also Fig.~4 of Jachym et al. 2019). LOS velocities $> 50$~km\,s$^{-1}$ are observed in the H$\alpha$ velocity field of the
southern tail (Fig.~3 of Luo et al. 2023). The upturning extraplanar CO filament mainly has velocities close to the systemic velocity.
Only the outer northern part shows negative velocities $< -20$~km\,s$^{-1}$. 

As already mentioned in Sect.~\ref{sec:results} the morphology of the model dense gas tail is different from the observed one:
model~A shows a northern dense model gas filament with an extent of about $15$~kpc. This filament is almost absent in the ALMA observations.
The observed upturning southern filament is best reproduced by models~A. Model~C also shows a southern upturning filament but its
extent is smaller than $5$~kpc.

The observed velocity field of the model~A southern filament
is consistent with observations for distances $> 5$~kpc from the galactic disk. At smaller distances the model velocities
are significantly higher than the observed ones. Higher model velocities compared to observations are expected as the model rotation velocities are
a factor of $1.5$ higher than observed (Fig.~\ref{fig:rotcurve}). In addition, the model gas extent within the disk plane is larger than observed
leading to higher LOS velocities due to the radially increasing rotation curve.
The observed negative velocities at the outer northern tip of the upturning filament is reproduced by model~A.

We conclude that model~A best reproduces the CO emission distribution and velocity for distances $\la 15$~kpc from the galactic disk.

\subsection{UV emission and star formation \label{sec:UVstar}}

\begin{figure*}[!ht]
  \sidecaption
  \centering
  \resizebox{12cm}{!}{\includegraphics{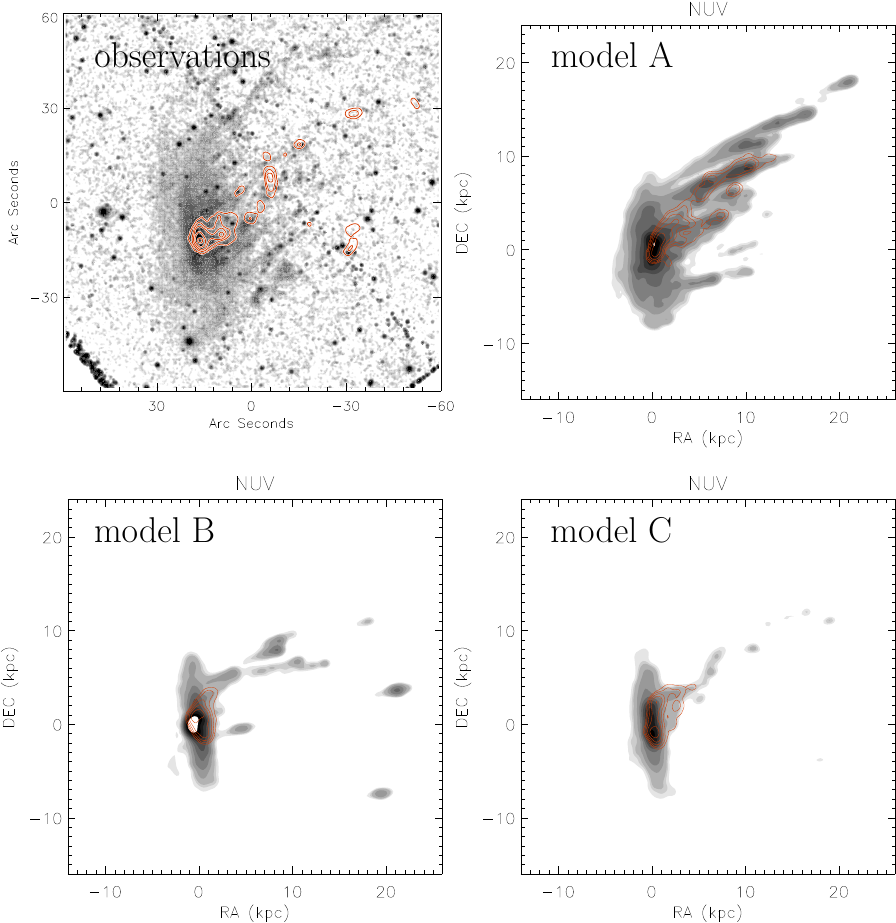}}
  \caption{CO emission (red contours) on NUV emission distribution. Upper left: ALMA and HST observations.
  \label{fig:zusammen1_nuv}}
\end{figure*}

The observed HST F275W NUV emission distribution (see also Fig.~2 of Waldron et al. 2023) together with the CO emission distribution are presented
in Fig.~\ref{fig:zusammen1_nuv}.
The HST image clearly shows extraplanar NUV emission to the west of the galactic disk in the direction of the gas tail.
This diffuse emission extends about $40'' \sim 13.6$~kpc to the west, except in the north where the extent is $\sim 25$~kpc.
The model with the strongest extraplanar NUV emission is model~A, which
qualitatively reproduces the observations. The observed most northern NUV filament is not present in model~A.
The observed southern extension of the NUV disk is present in model~A albeit less extended.
The extraplanar NUV emission distributions of models~B and C are much less prominent than their observed counterpart.

We conclude that model~A best reproduces the available NUV observations.

\subsection{H{\sc i} emission \label{sec:HI}}

\begin{figure*}[!ht]
  \centering
  \resizebox{\hsize}{!}{\includegraphics{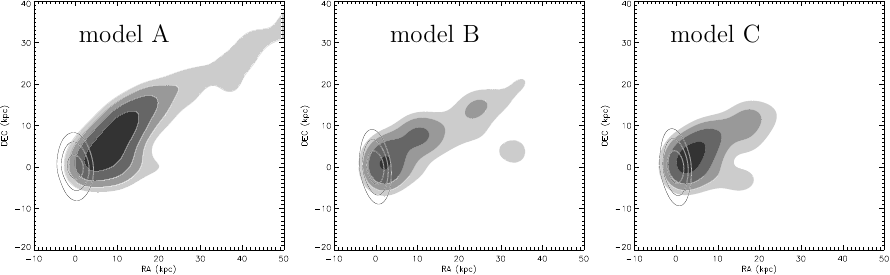}}
  \caption{Stellar disk (contours) on H{\sc i} emission distribution. The contour levels are $(1.5, 3, 6, 12, 24) \times 10^{19}$~cm$^{-2}$.
  \label{fig:zusammen2_hi}}
\end{figure*}

The atomic hydrogen mass distribution can be determined by taking into account the total gas mass, the molecular fraction (Eq.~\ref{eq:phases}),
and the amount of mixed gas (Sect.~\ref{sec:iong}).
We convolved the model H{\sc i} maps to a spatial resolution of $19''$. The resulting model H{\sc i} moment~0 maps are presented in
Fig.~\ref{fig:zusammen2_hi}. All model maps show an extended off-center maximum, which is elongated toward the northwest.
The extent of the high-surface-density gas significantly varies between the models. The smallest extent ($<20$~kpc) is observed in model~C,
the largest extent ($\sim 30$~kpc) in model~A. Model~B  shows and intermediate extent.
The extent of the low-surface-density tail is about proportional to that of the high-surface-density gas. The largest extent is present in
model~A ($\ga 50$~kpc) and the smallest extent in model~C.

\subsection{The H$\alpha$--X-ray correlation of the stripped gas \label{sec:haxcorrelation}} 

Recently, Sun et al. (2022) found a strikingly linear correlation between the H$\alpha$ and X-ray surface brightnesses of the diffuse stripped gas of
cluster spiral galaxies at $\sim 10-40$~kpc scales. ESO~137-001 is one of the prime examples in this work.
Lee et al. (2022) found in their 3D hydrodynamical simulations that the ICM-dominant tail gas shows flux ratio $F_{\rm X}/F_{{\rm H}\alpha} \sim 1$–$20$,
while the gas in the disk vicinity ($3$~kpc$ < z < 10$~kpc) exhibits a lower $F_{\rm X}/F_{{\rm H}\alpha}$ of $\sim 1$ (lower panel of their Fig.~15).

\begin{figure*}
  \sidecaption
  \centering
  \resizebox{12cm}{!}{\includegraphics{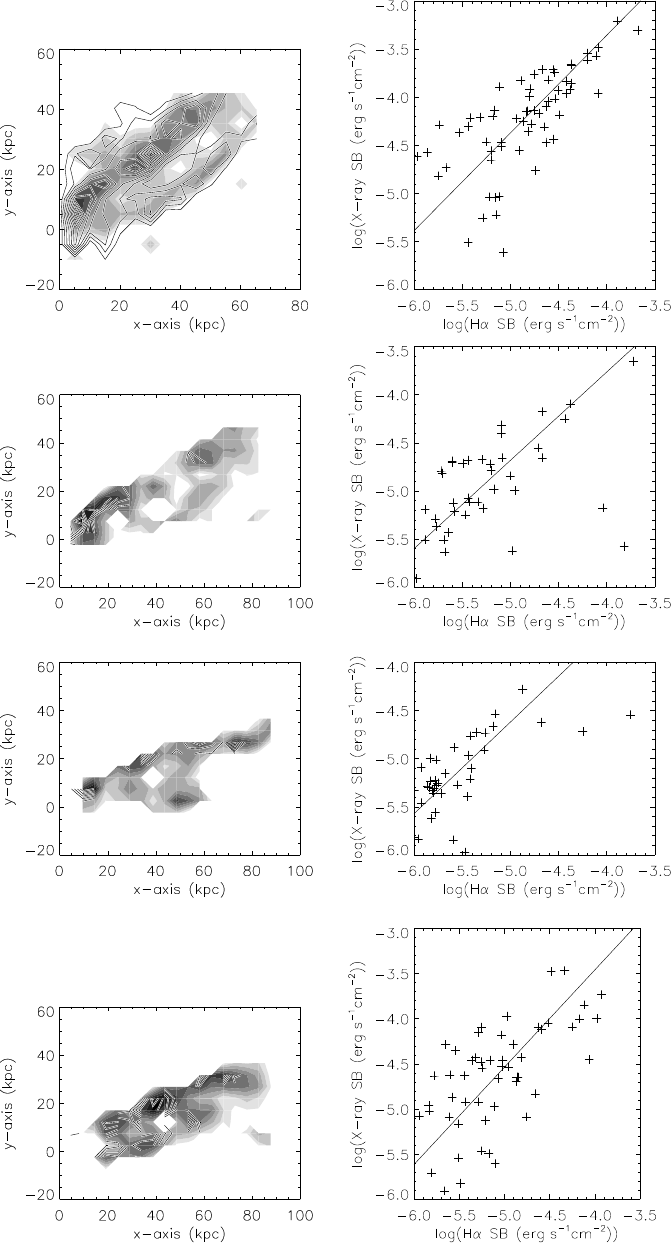}}
  \put(-305,595){\large \bf observations}
  \put(-305,420){\large \bf model~A}
  \put(-305,285){\large \bf model~B}
  \put(-305,110){\large \bf model~C}
  \caption{H$\alpha$--X-ray correlation (right panels). The solid lines correspond to an outlier-resistant linear regression.
    Left panels: grayscale: X-ray emission distribution; contour: H$\alpha$ emission distribution.
    The pixel size is $\sim 5$~kpc. 
  \label{fig:haxcorrelation_obs}}
\end{figure*}

Triggered by these results, we established the H$\alpha$--X-ray correlation for the observations and our models.
Sun et al. (2022) used uneven boxes to extract the mean surface brightnesses. For an objective comparison between observations and models
we extracted the mean surface brightnesses on an equidistant grid with a grid size of $5$~kpc.
To do so we first convolved the H$\alpha$ map with a Gaussian with a FWHM of $40$~pixels$=8"=2.7$~kpc.
The resulting map was regridded to match the X-ray data in terms of pixel size ($0.49"=0.17$~kpc) and image size.
The X-ray and H$\alpha$ maps were then regridded to a common pixel size of $14.9"=5$~kpc.
This choice of a $5$~kpc grid led to a linear slope for the observations (Fig.~\ref{fig:haxcorrelation_obs}), as it was derived by Sun et al. (2022).
The Spearman rank coefficient is $0.77$, meaning that the correlation is strong.

For the models the H$\alpha$ and X-ray maps were convolved with a Gaussian with a FWHM of $2.2$~kpc.
The H$\alpha$ map was regridded to match the X-ray data in terms of pixel size ($0.49"=0.17$~kpc)
and the X-ray and H$\alpha$ maps were regridded to a common pixel size of $14.9"=5$~kpc.
Finally, the X-ray and H$\alpha$ maps were clipped to yield the observed extents of the stripped gas tails.
We verified that the resulting correlations are not sensitive to the FWHM of the convolutions.

The ranges of the H$\alpha$ and X-ray surface brightnesses of all models are well comparable to the observed ranges. 
All three models show a clear correlation between the H$\alpha$ and X-ray surface brightnesses (Spearman rank coefficients of $0.53$,
$0.69$, and $0.62$ for models~A, B, and C, respectively). The correlations are weaker than the observed correlation.
The slopes of the log-log correlations are $0.92 \pm 0.11$, $0.95 \pm 0.10$, and $1.08 \pm 0.12$ for  models~A, B, and C, respectively.
The model slopes are thus consistent with the slope of the observed log-log correlation.

We conclude that our modelling of the H$\alpha$ emission caused by ionization through thermal conduction is consistent with
the results of Sun et al. (2022). We also tested a constant ionization fraction and an ionization fraction
caused by the equilibrium between X-ray ionization and recombination. Both recipes led to a much shallower slope of
the logarithmic H$\alpha$--X-ray correlation than observed. Thus, the ionization through thermal conduction sets the
slope of the correlation. The model correlation scatter is mainly set by the variation of the model gas density.
The main differences between the model and observed correlations are the higher H$\alpha$ and X-ray brightnesses of
the gas tail and the observed small scatter at high surface brightnesses. We can only speculate that the difference
is caused by our approximate description of ram pressure stripping of diffuse gas (Sect.~\ref{sec:iong}).

\subsection{Heat transfer via turbulence and thermal conduction in the presence of a magnetic field \label{sec:magfield}}

For the calculation of the emission measure of the stripped ISM due to thermal conduction (Eq.~\ref{eq:em}) the influence of the magnetic field was neglected.
We assume that at the interface between the hot ICM and the warm stripped ISM Kelvin-Helmholtz instabilities develop, which
lead to turbulent  mixing of the hot ICM with the warm ISM (see Sect.~\ref{sec:iong}). In this way turbulence is induced
into the warm partially ionized ISM. Li et al. (2023) found a turbulent energy injection scale of $L \sim 1$~kpc with an associated
velocity of $v_{\rm L} \sim 50$~km\,s$^{-1}$. Following Eq.~5.32 of Sarazin et al. (1988), the mean free path of electrons for $T=10^7$~K and
$n=0.3$~cm$^{-3}$ is about $\lambda \sim 1$~pc. If we further assume that the turbulence of the warm ISM is about sonic (Sonic Mach number
$M_{\rm S} \sim 1$) with a temperature of $10^4$~K dictated by the cooling curve ($v_{\rm turb} \sim 10$~km\,s$^{-1}$), the Alfv\'enic Mach number
is about $5$.

Following Lazarian (2006) the scale at which the magnetic ﬁeld gets dynamically important is $l_{\rm A} = L\,M_{\rm A}^{-3} \sim 8$~pc
and thus $l_{\rm A} > \lambda$. In this case Lazarian (2006) stated that heat conduction is decreased by a factor of three
with respect to the classical Spitzer value (Sect.~2.2 of Lazarian 2006). According to Eqs.~7, 21, and 25 of McKee \& Cowie (1977)
the emission measure is proportional to the square root of the conductivity. A reduction of the emission measure by a factor
of $\sqrt(3)$ due to the presence of magnetic fields is within the uncertainties of our calculations.
We note that the global heat transfer is dominated by turbulent advective heat transfer
(Fig.~1 of Lazarian 2006 with $M_{\rm S} > \lambda/(\alpha \, \beta \, L)$ with $\beta = 4$ and
$\alpha=(m_{\rm e}/m_{\rm p})^{\frac{1}{2}}$) and $M_{\rm A} =5 < (L/\lambda)^{\frac{1}{3}}= 10$) as assumed in Sect.~\ref{sec:iong}.

As a consistency check we can compare the density based on the observed mean emission measure to that based on the model overdensities of the warm and hot gas.
The extent of the ionized region $l$ of a stripped gas cloud can be estimated by setting the heat conduction timescale $t_{\rm heat}=l^2/(\lambda \, v_{\rm e})$ to
the turbulent timescale $t_{\rm turb}=L/(\sqrt(3)\,v_{\rm turb}) \sim 11$~Myr, where $v_{\rm e}=500$~km\,s$^{-1}$ is the velocity of the thermal electrons.
This yields $l=76$~pc and $44$~pc in the absence and presence of magnetic fields, respectively. An H$\alpha$ surface brightness of the tail
of $2 \times 10^{-17}$~erg\,cm$^{-2}$s$^{-1}$arcsec$^{-2}$ (Sun et al. 2022) corresponds to an emission measure of $EM \sim 0.6$~cm$^{-6}$pc, which leads to ionized gas densities of
$0.09$ and $0.12$~cm$^{-3}$, respectively. With an overdensity of $70$ for the hot gas and $1000$ for the warm gas (Sect.~\ref{sec:extract})
the warm gas is $\sim 14$ times  denser than the hot gas. With an observed  mean hot gas density of $10^{-2}$~cm$^{-3}$ (Sun et al. 2010) the warm gas density
is thus $0.14$~cm$^{-3}$, well comparable to the value based on the observed emission measure.

\subsection{ESO~137-001 within the Norma cluster \label{sec:Norma}}

We adopt the systemic velocity of ESO~137-001 of $4647$~km\,s$^{-1}$ from Luo et al. (2023). The cluster mean velocity is $4871$~km\,s$^{-1}$ (Woudt et al. 2008).
ESO~137-001 thus has a systemic velocity of $224$~km\,s$^{-1}$ with respect to the cluster mean velocity.
Assuming a LOS component of the unit galaxy velocity vector of $-0.07$ (model~5 of Table~\ref{tab:molent}) leads to a total galaxy velocity of $\sim 3200$~km\,s$^{-1}$. 
Its projected distance to the cluster center is $180$~kpc. Thus, the galaxy's smallest distance to the cluster center and thus the ram pressure
peak is expected to occur in about $50$~Myr. This is somewhat longer but comparable to the time to peak ram pressure of the highest-ranked model~A
($\Delta t = -40$-$-20$~Myr; Table~\ref{tab:molent}). The associated ram pressure is $49000$ and $59000$~cm$^{-3}$(km\,s$^{-1}$)$^2$.
With a total velocity of $3200$~km\,s$^{-1}$ the ICM density at the location of ESO~137-001 is $n_{\rm ICM}=4.8$-$5.8 \times 10^{-3}$~cm$^{-3}$.
This is about four times higher than the ICM density derived from X-ray observations (Sun et al. 2010).

\begin{figure}
  \centering
  \resizebox{7cm}{!}{\includegraphics{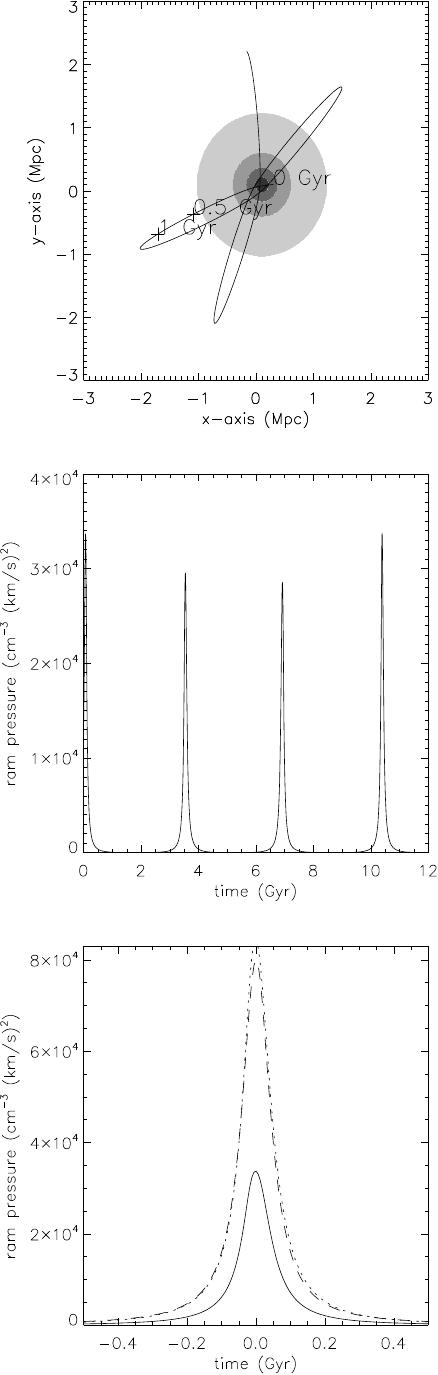}}
  \caption{A plausible orbit of ESO~137-001 within the Norma cluster. Upper panel: grayscale: ICM distribution.
    The timestep $t=0$~Myr corresponds to the current location of ESO~137-001. The orbit is extrapolated into the future.
    The  galaxy approaches the cluster center for the first time.
    Middle panel: ram pressure profile.
    Lower panel: solid line: last ram pressure stripping event of middle panel; dotted line: solid line multiplied by a factor of $2.5$;
    dashed line: ram pressure profile of the highest-ranked model~A (red line in Fig.~\ref{fig:profiles}).
  \label{fig:orbiteso137-001}}
\end{figure}

As the next step we calculated possible orbits of ESO~137-001 within the Norma cluster.
For the gravitational potential of the Norma cluster follow Jachym et al. (2014) by assuming an NFW halo density profile
with a Virial mass of $M_{\rm vir}=10^{15}$~M$_{\odot}$ and scaling radius $r_{\rm s}=346$~kpc.
For the ICM density distribution we used a $\beta$ profile
\begin{equation}
  \label{eq:beta}
  n_{\rm e} = n_{\rm e,0} \big(1+(r/r_{\rm c})^2\big)^{-3/2\,\beta}\,
\end{equation}
with $n_{\rm e,0}=2.4 \times 10^{-3}$~cm$^{-3}$, $r_{\rm c}=9.95'=200$~kpc, and $\beta=0.555$ (B\"ohringer et al. 1996).
The observed X-ray emission of the Norma cluster ICM is displaced from the cluster center (ESO~137-006) to the northwest by about $150$~kpc
(Fig.~1 of Jachym et al. 2014). We thus displaced the center of the distribution of Eq.~\ref{eq:beta} to $(100~{\rm kpc}, 100~{\rm kpc}, 0)$.
For the initial conditions we used the projected distance and the 3D velocity of ESO~137-001, which we derived from the highest-ranked model~A.
To simplify the model, we kept the orbit in the plane of the sky ($x$-$y$ plane).
$x_0=300$~kpc; $y_0=100$~kpc; $z_0=0$; $v_{x,0}=-3180$~km\,s$^{-1}$; $v_{y,0}=-280$~km\,s$^{-1}$; $v_{\rm z,0}=0$.
The resulting galaxy orbit is presented in the upper panel, the resulting ram pressure profile in the
middle panel of Fig.~\ref{fig:orbiteso137-001}.

This galaxy orbit is close to unbound and consistent with the orbital solutions found by Jachym et al. (2014). 
If the velocity of the galaxy is increased by a factor of $1.2$ the orbit becomes unbound. The initial total galaxy velocity for
the unbound orbit is $3840$~km\,s$^{-1}$ instead of $3200$~km\,s$^{-1}$ derived from the highest-ranked model~A.
The last ram pressure maximum of the upper panel is enlarged in the lower panel of Fig.~\ref{fig:orbiteso137-001} (solid line)
together with the ram pressure profile of the highest-ranked model~A (dashed line). The maximum ram pressure of the model orbit
is significantly smaller than that of model~A. Modifications of $y_0$ did not lead to a significantly higher peak ram pressure. 
On the other hand, when the last ram pressure profile is multiplied
by a factor of $2.5$ (dotted line) it is consistent with the ram pressure profile of the highest-ranked model~A.

This increase of the ram pressure profile can be obtained in two ways: (i) an increase of the ICM density or (ii)
a moving ICM opposite to the motion of the galaxy within the cluster as it occurs for NGC~4522 in the Virgo cluster
(Kenney et al. 2004, Vollmer et al. 2004). In the latter case the ICM velocity adds to the galaxy velocity
and ram pressure increases quadratically with velocity. An ICM velocity of $3000$~km\,s$^{-1}$ with respect to the
cluster mean velocity in the opposite direction to the motion of ESO~137-001 leads to an increase of
ram pressure by a factor of two. With a sound speed of $1263 \sqrt{kT /6~{\rm keV}}$~km\,s$^{-1}$ this corresponds to
a Mach number of $2.4$, comparable to that of the Bullet cluster (Markevitch et al. 2002).
In a strongly perturbed galaxy cluster as the Norma cluster with an off-center ICM distribution the two possibilities
(i) and (ii) are probable and plausible. If the moving ICM leads to a significant modification of the ram pressure
a Lorentzian profile as a function of time might not be expected.
In this context, Tonnesen et al. (2019) studied the influence of galaxy orbits within a cluster,
Roediger et al. (2014) the influence of instantaneous stripping  by means of eight “wind-tunnel” hydrodynamical simulations.
It is beyond the scope of this work to test different shapes of the ram pressure profile.

\section{Summary and conclusions \label{sec:conclusions}}

The Norma cluster galaxy ESO~137-001 is one of the rare ram-pressure stripped galaxies, for which deep, kpc-resolution observations are available
in CO line, H$\alpha$ line, and X-ray emission. Its stripped X-ray (Sun et al. 2010) and H$\alpha$ (Sun et al. 2022) tails are spectacular,
extending $\sim 80$~kpc behind the galactic disk. ESO~137-001 evolves in a highly dynamic ICM of the Norma cluster.
The cold molecular, warm ionized, and hot ionized gas coexist within the gas tail
as traced by the CO (Jachym et al. 2014, 2019), H$\alpha$, and X-ray emission. 
 
We made dynamical simulations of ESO~137-001 to constrain its 3D orbit within the Norma cluster and to investigate the physics of
the ISM-ICM turbulent mixing process. Our numerical code is not able to treat diffuse gas in a consistent way. For a realistic treatment,
3D hydrodynamical simulations should be adopted. Nevertheless, we can mimic the action of ram pressure on diffuse gas by applying
very simple recipes based on the fact that the acceleration by ram pressure is inversely proportional to the gas
surface density, which in turn depends on the gas density for a gas cloud of constant mass (Sect.~\ref{sec:iong}).
The hot ($>10^6$~K) diffuse gas is taken into account by assuming that once the stripped gas has left the galactic disk, it mixes with the ambient ICM.
In a new approach, we estimate the ICM-ISM mixing analytically (Eq.~\ref{eq:inflowmix}) based on the recipe of Fielding et al. (2020).

Following Vollmer et al. (2001), we used Lorentzian and Gaussian profiles for the time evolution of ram pressure stripping (Sect.~\ref{sec:param}).
Following Nehlig et al. (2016) and Vollmer et al. (2018), we investigated (i) the influence of galactic structure (i.e., the position of spiral arms) on the
results of ram pressure stripping by varying the time of peak ram pressure and (ii) the influence of the angle between the disk plane and the ram pressure wind
on the resulting gas distribution and velocity field.
Moreover, we recalculated the initial model set with different ICM-ISM mixing rates (Sect.~\ref{sec:mixing}) and stripping efficiencies
of the hot gas (Sect.~\ref{sec:efficiency}).

For the modelling of the warm diffuse ISM we assumed that the gas with temperatures lower than $10^6$~K is ionized by thermal conduction (Sect.~\ref{sec:extract}):
thermal electrons from the hot ICM penetrate into the neutral warm stripped ISM clouds ionizing and heating them. Ultimately, this leads to the evaporation
of the ISM clouds (Cowie \& McKee 1977). 

For the search of the highest-ranked models we calculated the goodness of the fit for all timesteps of all models (Sect.~\ref{sec:observations}).
This was also done for the velocity field.
Whereas the highest-ranked model at a single wavelength corresponds to the minimum goodness, the comparison of goodness at different
wavelengths is not straight forward. We decided to rank the models at the different wavelengths to calculate the sum of the ranks at
the different wavelengths. We define the highest-ranked model as the model with the smallest value of the sum of the ranks.

Based on the detailed comparison between our dynamical models and the multi-wavelength observation we conclude that
\begin{itemize}
\item
  the highest-ranked models with the nominal ICM-ISM mixing rate (Eq.~\ref{eq:inflowmix}) reproduce observations significantly better than the
  models with a three times higher or lower ICM-ISM mixing rate (Sect.~\ref{sec:results});
\item
  it was not possible to reproduce all observational characteristic with a single model. The X-ray and H$\alpha$ observations are better reproduced by the
  preferred model~C, whereas the CO observations are better reproduced by the highest-ranked model~A (Sect.~\ref{sec:results});
\item
  the angle between the direction of the galaxy's motion and the galactic disk is between $60^{\circ}$ and $75^{\circ}$. Ram pressure stripping thus
  occurs more face-on (Sect.~\ref{sec:results});
\item
  the existence of a two-tail structures is a common feature in our
  models. It is due to the combined action of ram pressure and rotation together with the projection of the galaxy on the sky (Sect.~\ref{sec:results});
\item
  structures like the observed northern faint H$\alpha$ tail can in principle be reproduced by the model (Sect.~\ref{sec:northerntail}).
  Their existence depends on the density structure of the stripped gas, which itself depends on the initial gas distribution and the temporal ram pressure profile.
\item
  overall the CO/H$\alpha$/X-ray highest-ranked models with a three times lower stripping efficiency
  reproduce the available observations less well than the highest-ranked models with the nominal stripping efficiency (Sect.~\ref{sec:efficiency});
\item
  model~A best reproduces the observed CO emission and SFR fractions between the disk and tail regions (Sect.~\ref{sec:gasmasses});
\item
  model~A also best reproduces the CO emission distribution, velocity for distances $\la 20$~kpc from the galactic disk, and the available NUV observations
  (Sects.~\ref{sec:moleculargas} and \ref{sec:UVstar}).
\end{itemize}

The recently established linear correlation between the H$\alpha$ and X-ray surface brightnesses (Sun et al. 2022) is reproduced by all three models,
albeit with a weaker correlation strength (Sect.~\ref{sec:haxcorrelation}). Our modelling of the H$\alpha$ emission caused by ionization through thermal conduction
is thus consistent with observations: the thermal electrons of the mixed ICM-ISM at $T \sim 10^7$~K penetrate into and ionize the warm stripped ISM.

Turbulent mixing is essential for modeling X-ray emission, and heat conduction is essential for modeling H$\alpha$ emission.
For the efficiency of heat conduction, the knowledge of the magnetic field strength, which can be estimated via radio continuum observations,
is important. We think that future 3D MHD ram pressure stripping simulations should be able to resolve turbulent mixing of diffuse gas in the tail
(Tonnesen et al. 2014, Ruszkowski et al. 2014) and account for thermal conduction.

We predict the H{\sc i} emission distributions for the different models (Fig.~\ref{fig:zusammen2_hi}). The observed total H{\sc i} mass
will be a critical test and an important additional constraint for our models.
Based on the 3D velocity vector derived from our dynamical model we derive a galaxy orbit, which is close to unbound (Sect.~\ref{sec:Norma}). 
We argue that compared to an orbit in an unperturbed spherical ICM ram pressure is enhanced by a factor of $\sim 2.5$.
This increase can be obtained in two ways: (i) an increase of the ICM density or (ii) a moving ICM opposite to the motion of the galaxy within the cluster.
In a strongly perturbed galaxy cluster as the Norma cluster with an off-center ICM distribution the two possibilities
are probable and plausible.

\begin{acknowledgements}
  The authors would like to thank J.D.P.~Kenney for useful comments on the article and the anonymous referee for their constructive
  questions and suggestions.
\end{acknowledgements}

\newpage

\begin{appendix}

  \section{Goodness distribution}

  \begin{figure}
  \centering
  \resizebox{\hsize}{!}{\includegraphics{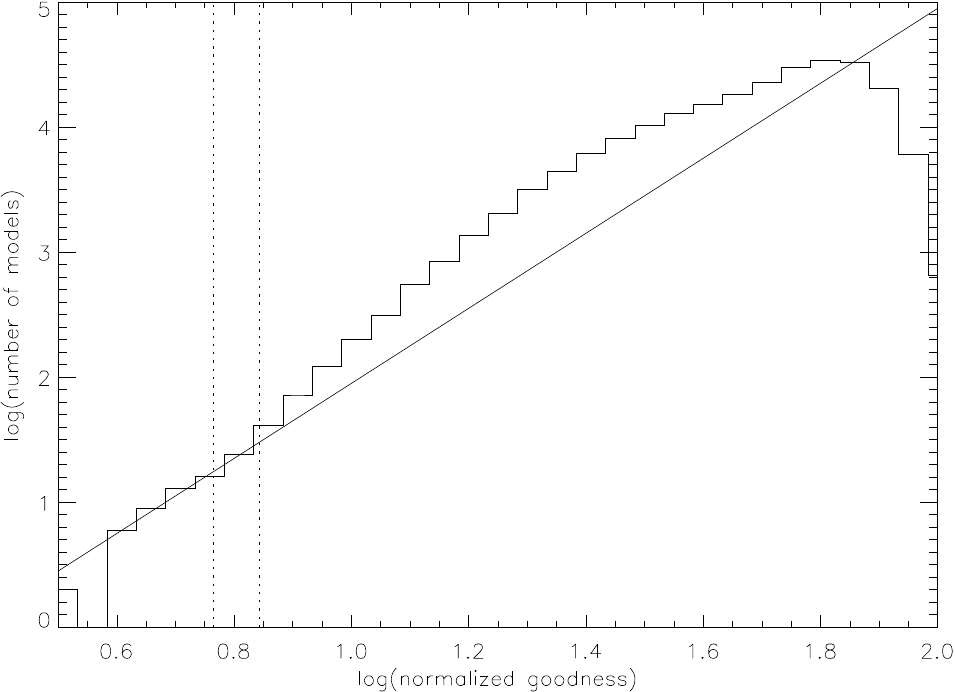}}
  \caption{Goodness distribution for the comparison based on the X-ray, H$\alpha$, and CO maps and the H$\alpha$ velocity field.
    The straight solid line represents a fit to the distribution of the first $100$ highest-ranked models.
    The dotted vertical lines correspond to the first $50$ and $100$ highest-ranked models, respectively.
  \label{fig:goodnessdistribution}}
\end{figure}
  
\section{Selection of the preferred models \label{sec:selection}}

\begin{figure*}[!ht]
  \centering
  \resizebox{14cm}{!}{\includegraphics{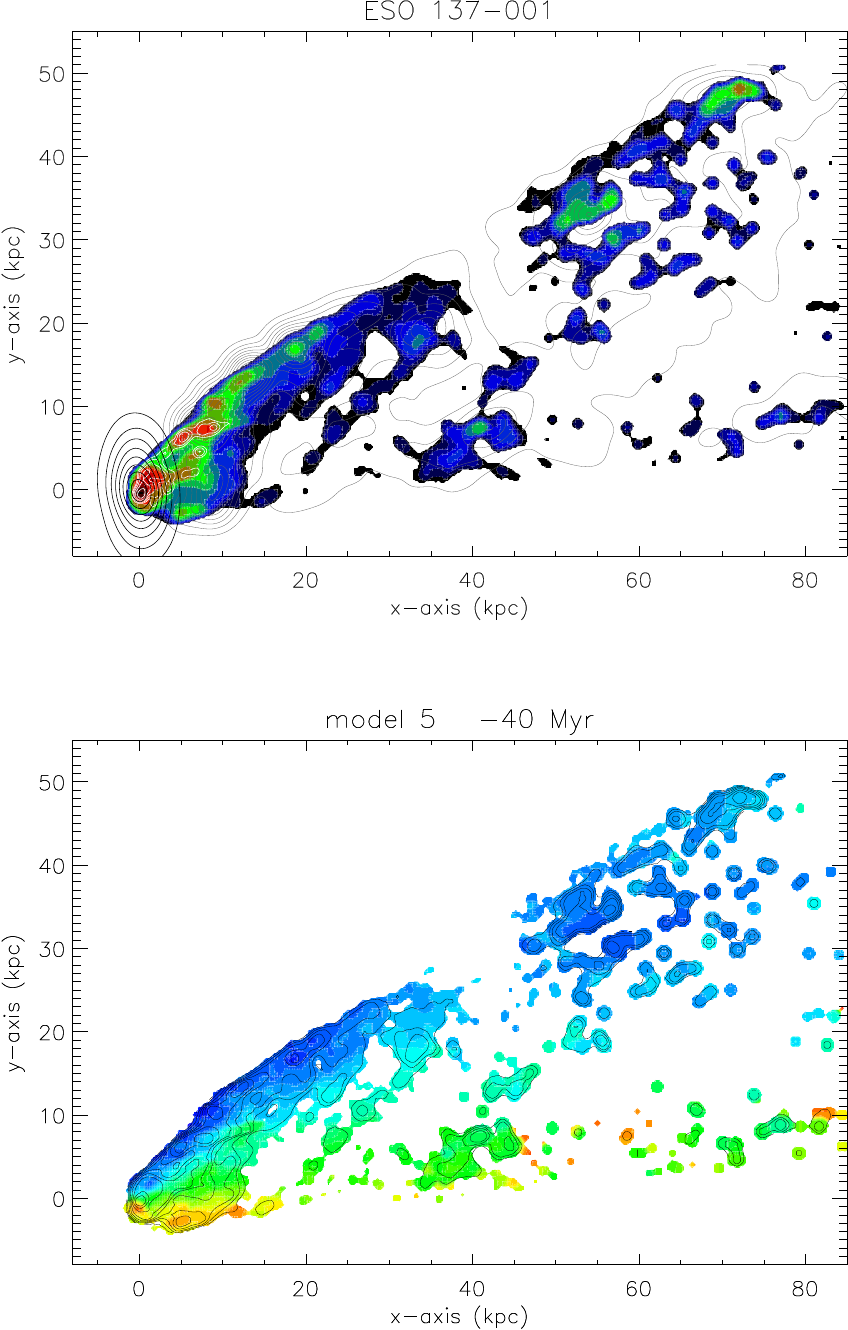}\includegraphics{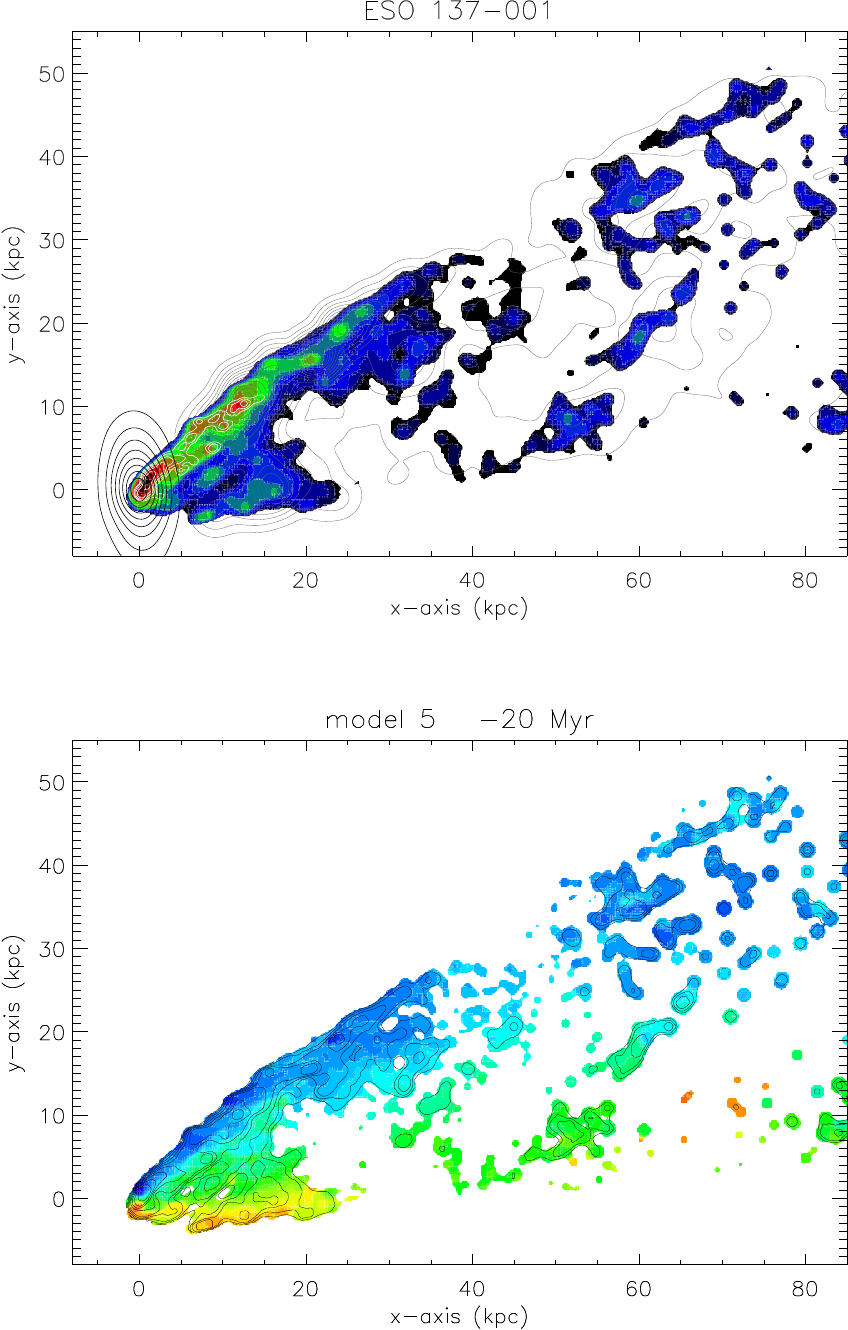}}
  \resizebox{14cm}{!}{\includegraphics{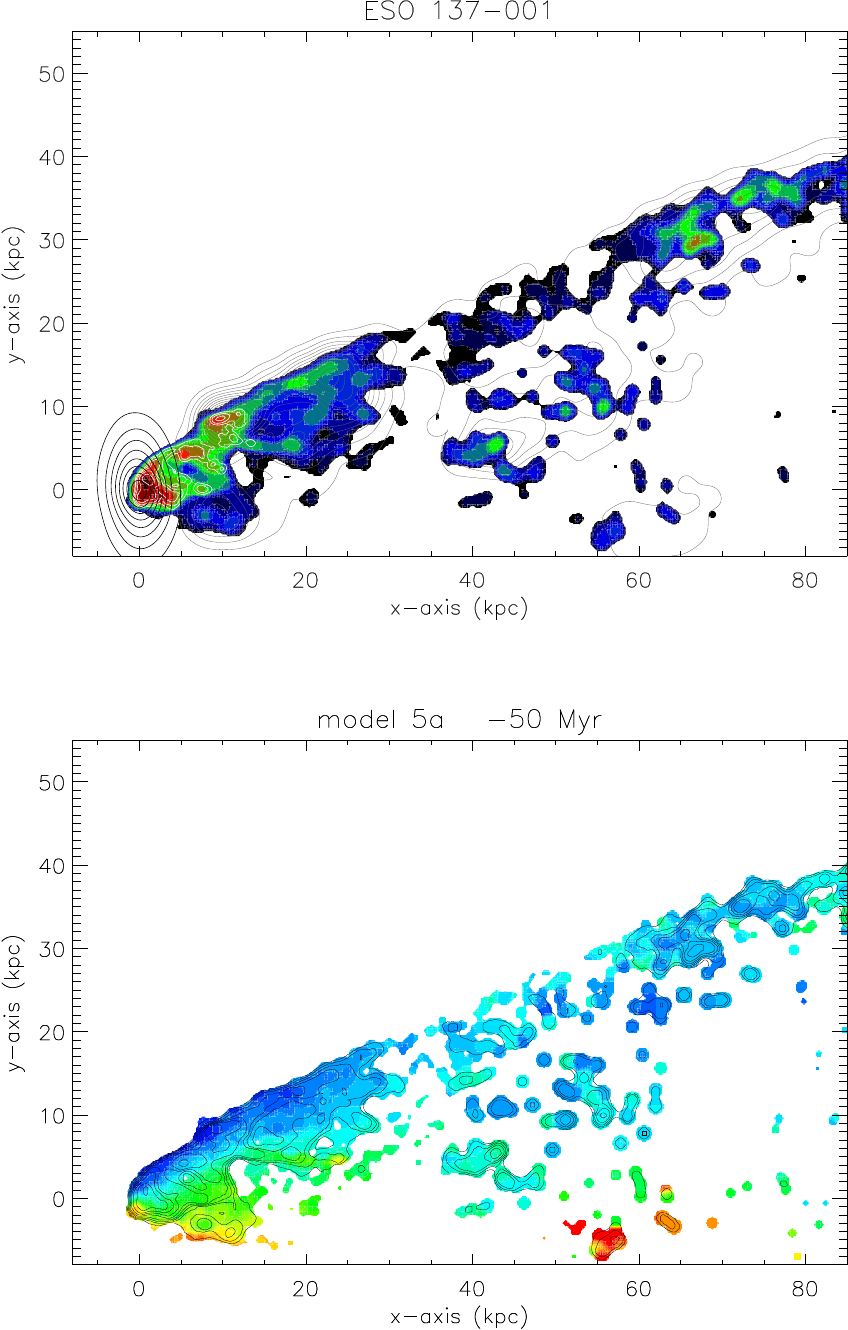}\includegraphics{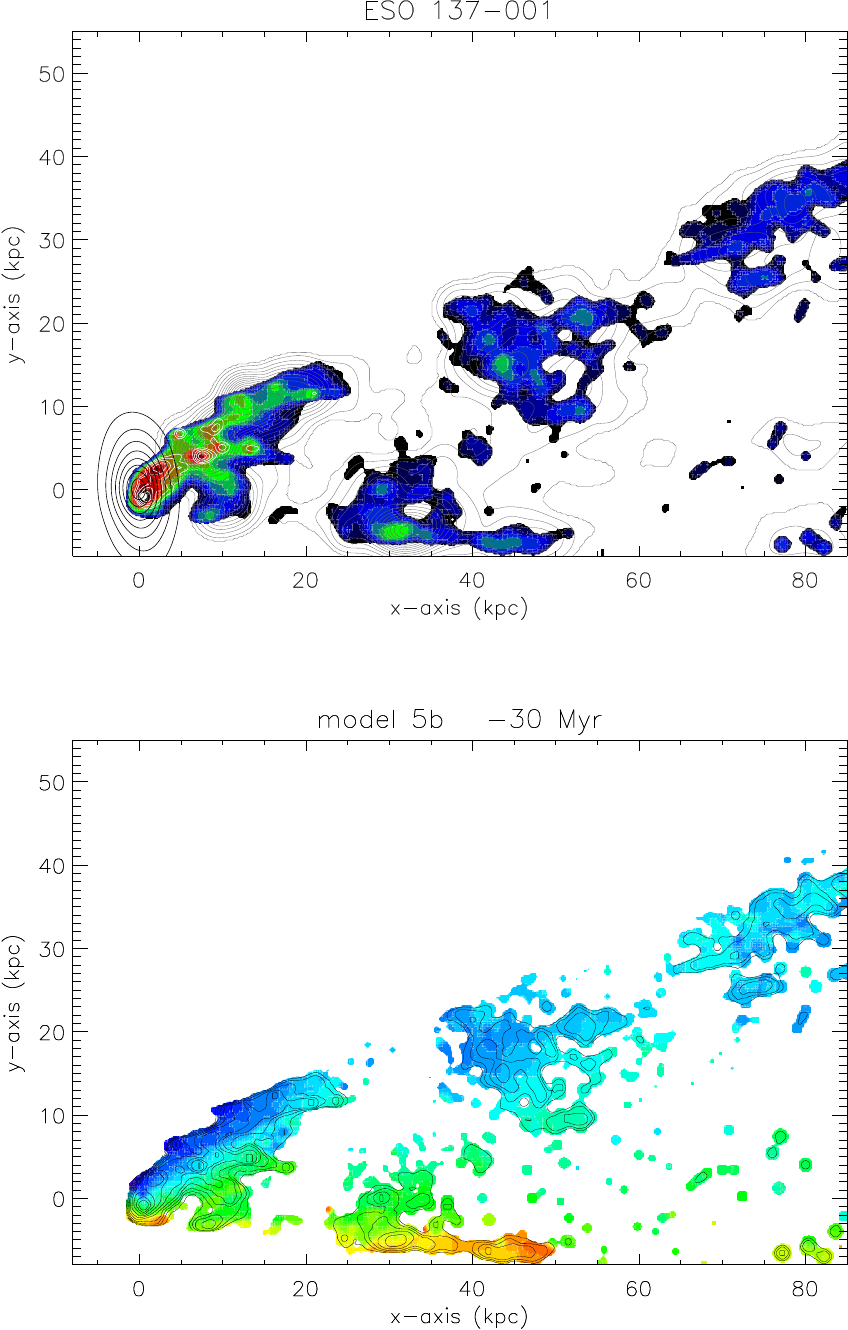}}
  \caption{Highest-ranked ram pressure pre-peak models of ESO~137-001 based on CO, H$\alpha$, X-ray, and H$\alpha$ velocity field.
    Upper panel: color: H$\alpha$; dark gray contours:
    X-ray; white contours: CO ; black contours: stellar content. Lower panel: H$\alpha$ velocity field. The colors are the same as in Fig.~\ref{fig:observations}.
  \label{fig:eso137-001all}}
\end{figure*}

\begin{figure*}[!ht]
  \centering
  \resizebox{14cm}{!}{\includegraphics{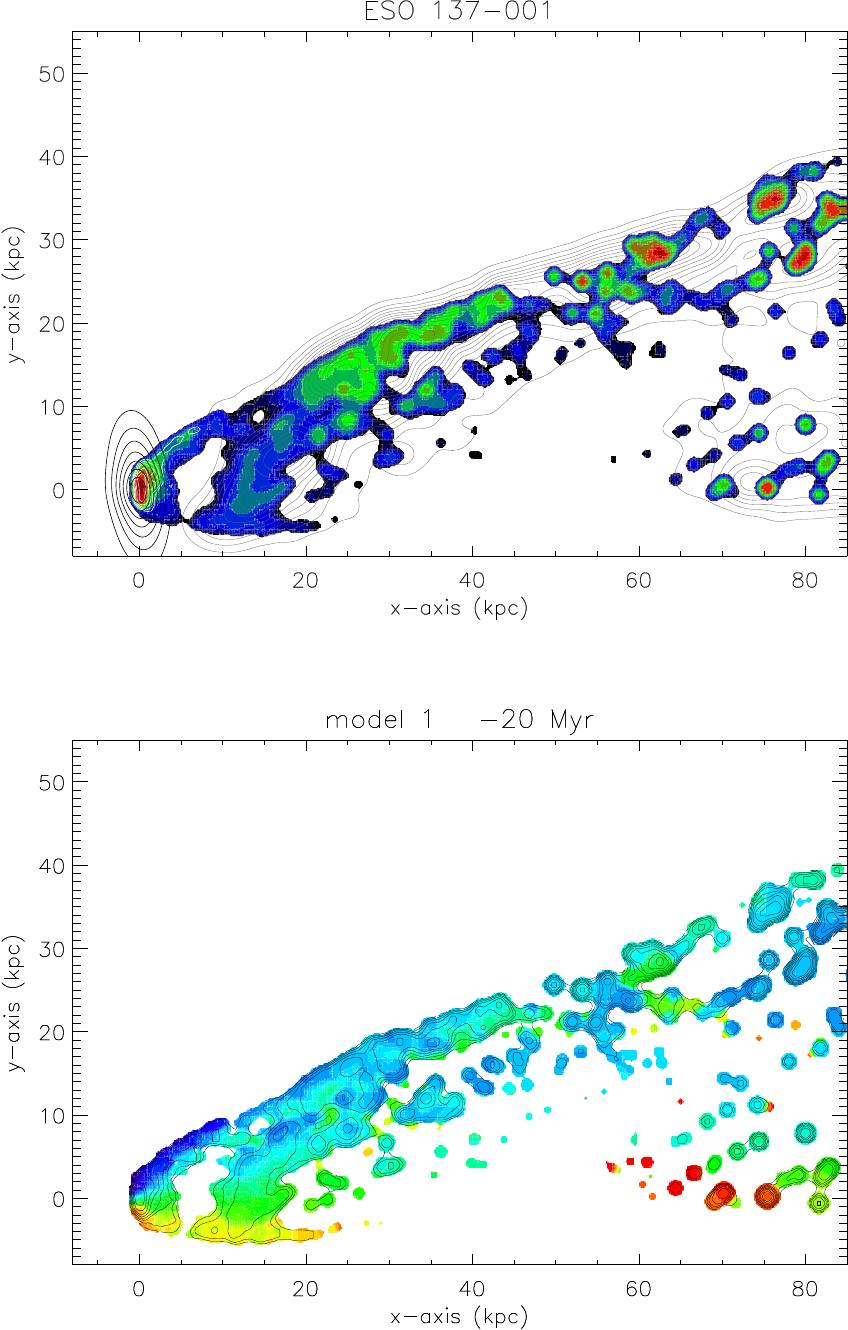}\includegraphics{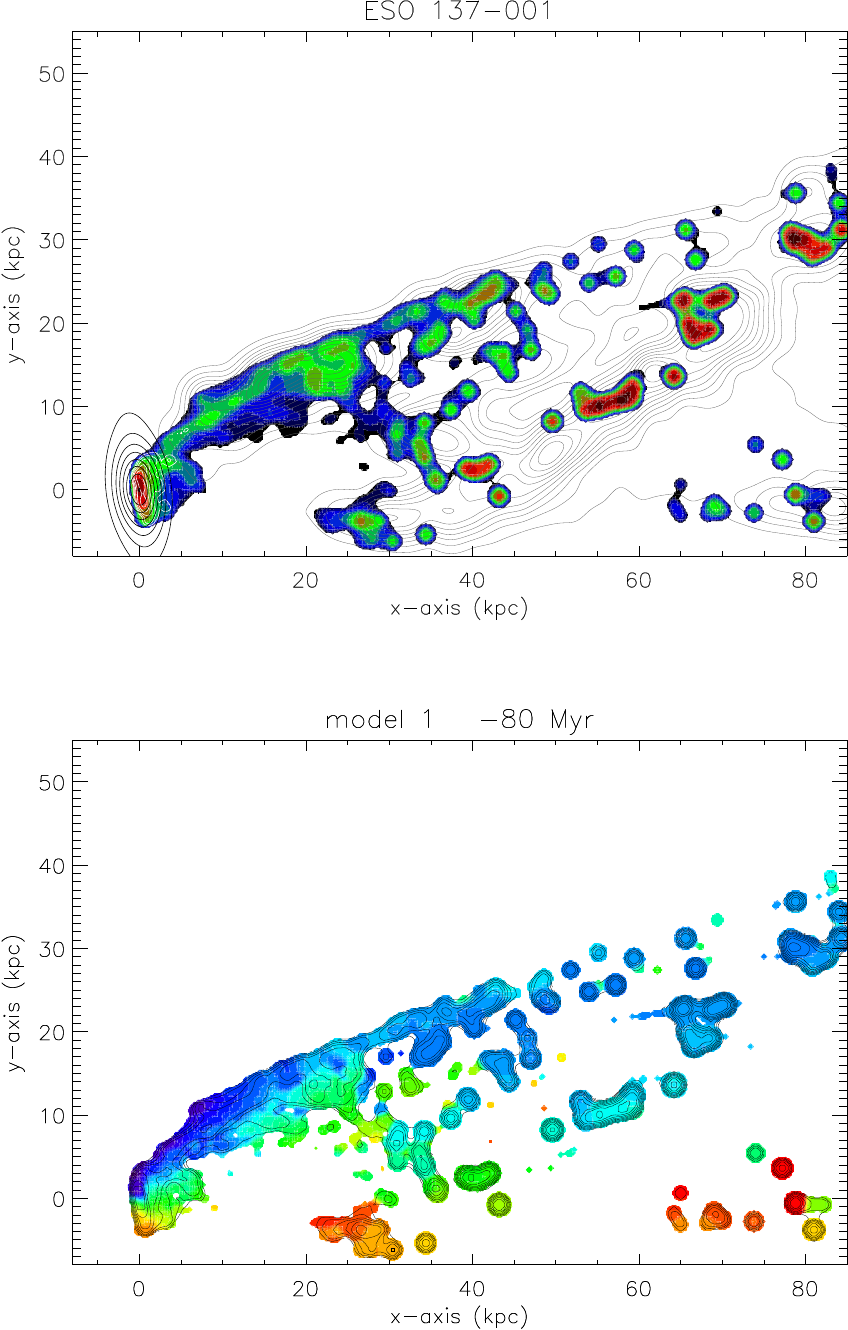}}
  \resizebox{14cm}{!}{\includegraphics{eso137-001all_mixing1_onlyvel10.pdf}\includegraphics{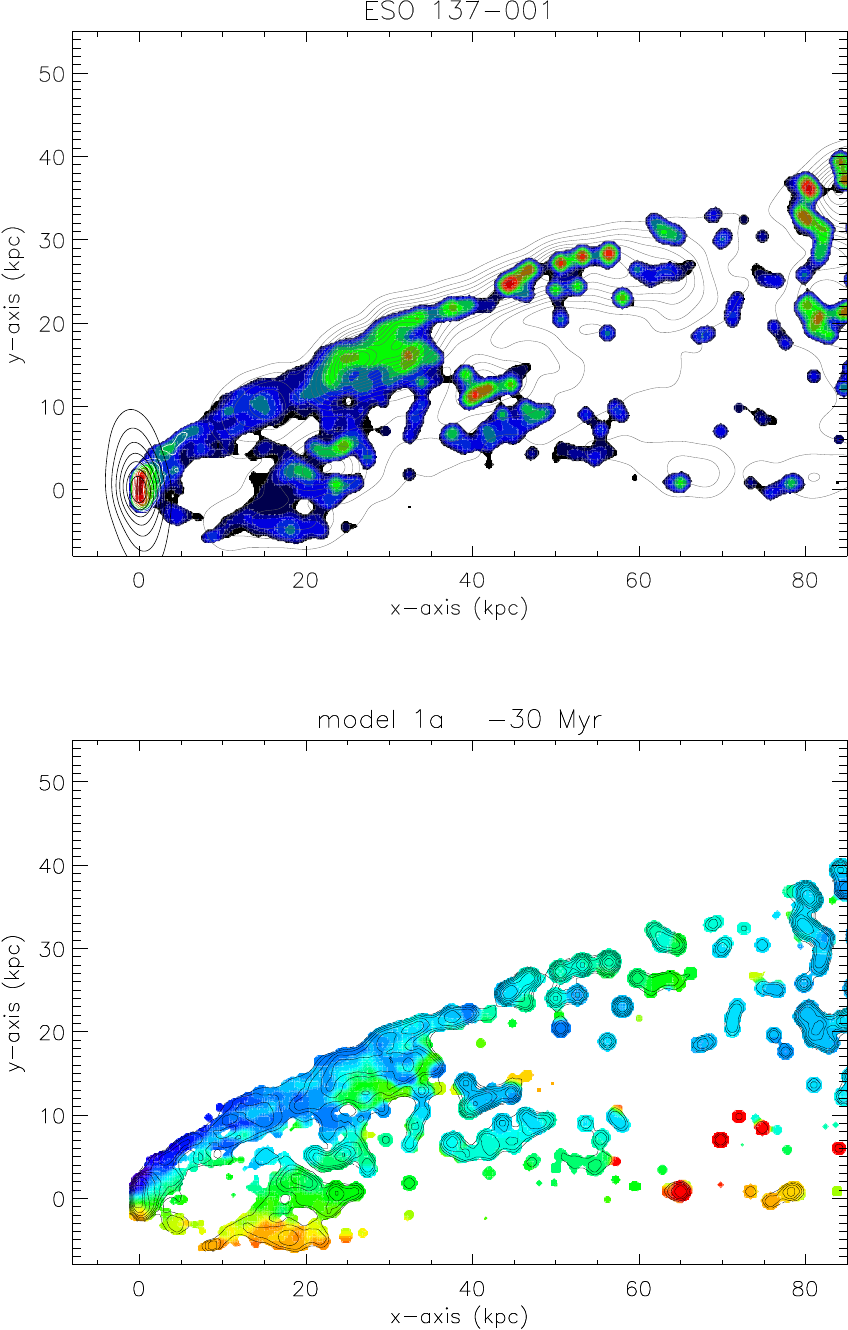}}
  \caption{Highest-ranked models of ESO~137-001 based on the H$\alpha$ velocity field. Upper panel: color: H$\alpha$; dark gray contours:
    X-ray; white contours: CO ; black contours: stellar content. Lower panel: H$\alpha$ velocity field. The colors are the same as in Fig.~\ref{fig:observations}.
  \label{fig:eso137-001allonlyvel}}
\end{figure*}

\begin{figure*}[!ht]
  \centering
  \resizebox{14cm}{!}{\includegraphics{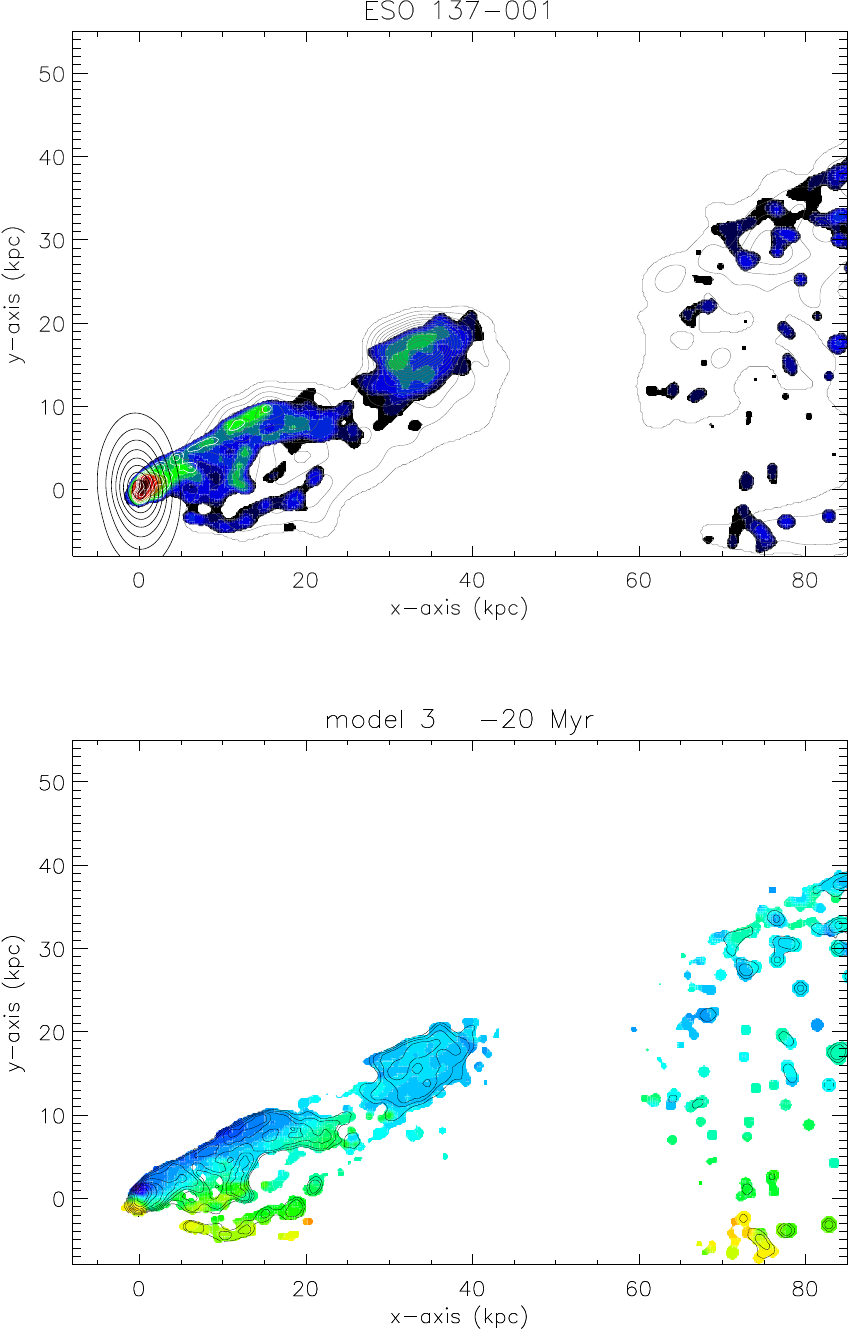}\includegraphics{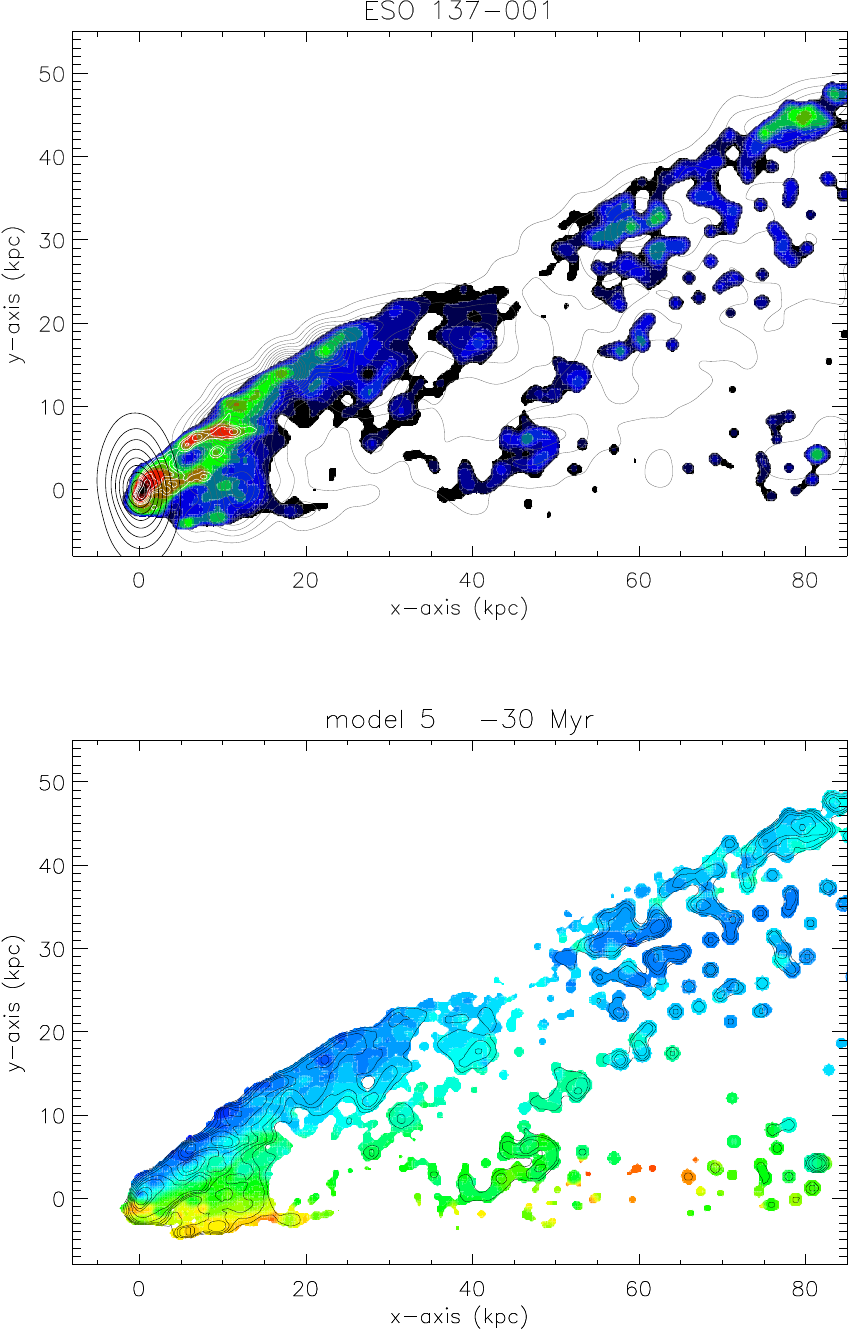}}
  \resizebox{14cm}{!}{\includegraphics{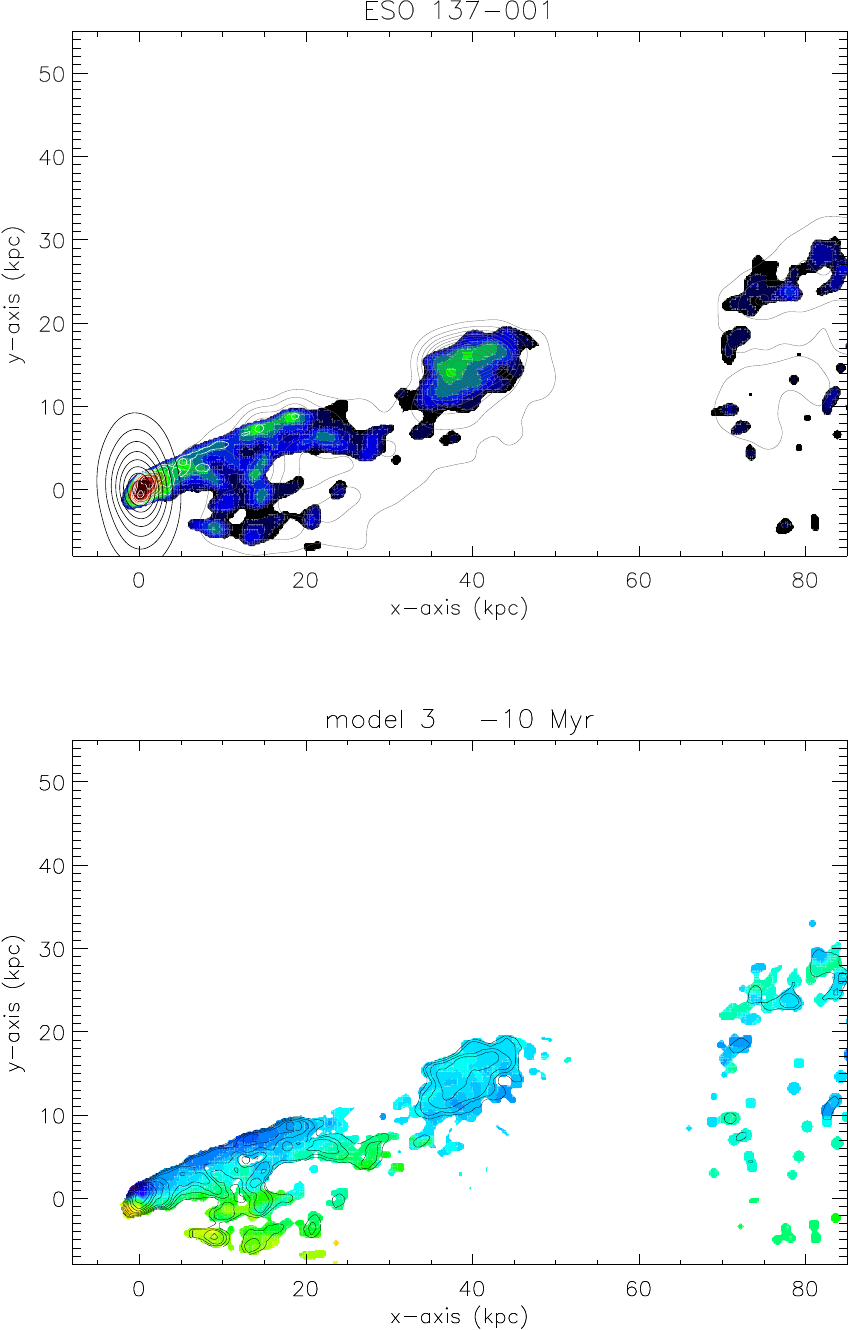}\includegraphics{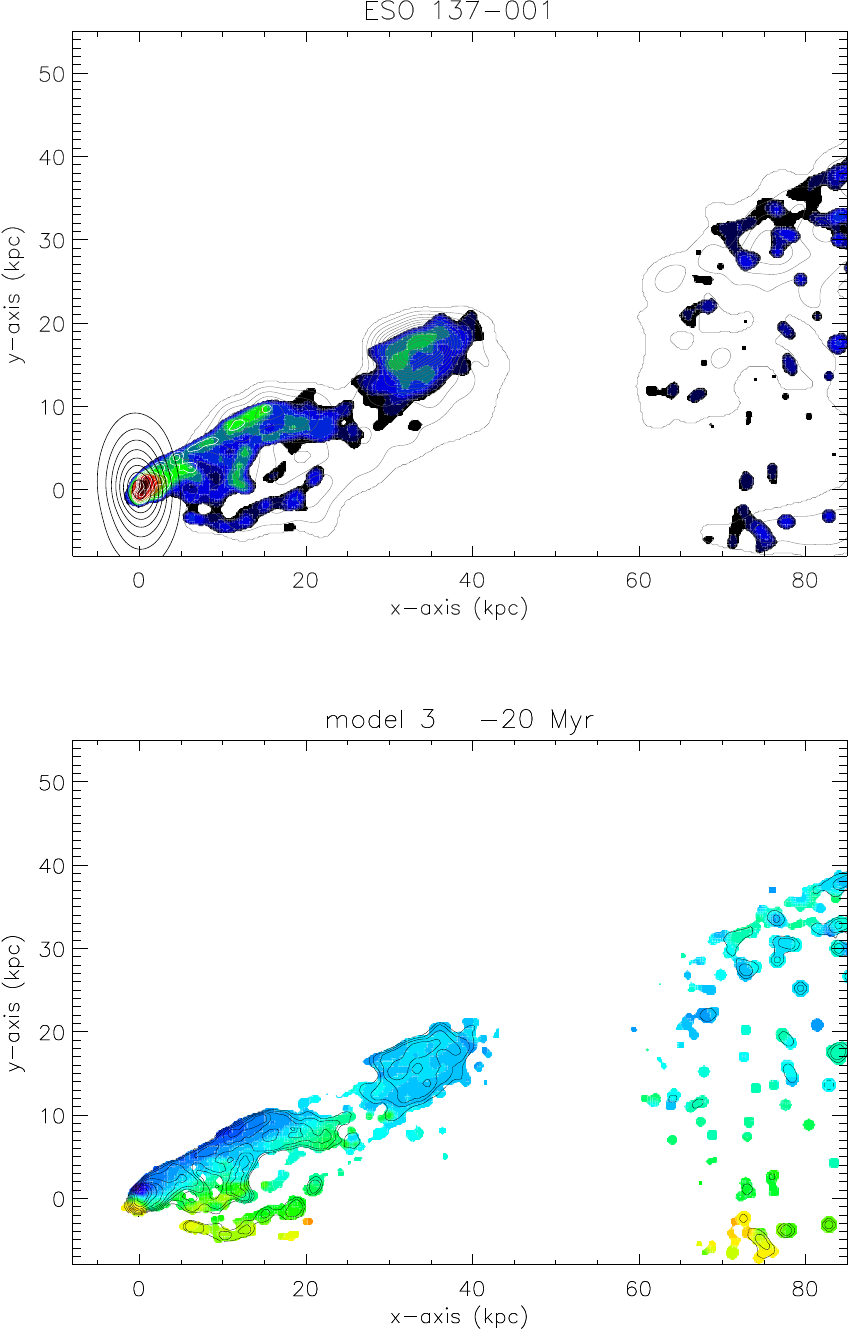}}
  \caption{Highest-ranked ram pressure pre-peak models of ESO~137-001 based on CO, H$\alpha$, and X-ray. Upper panel: color: H$\alpha$; dark gray contours:
    X-ray; white contours: CO ; black contours: stellar content. Lower panel: H$\alpha$ velocity field. The colors are the same as in Fig.~\ref{fig:observations}.
  \label{fig:eso137-001allnovel}}
\end{figure*}

Models~1 and 5 are the preferred models in all cases (i) to (ii). In case (iii) model~3 is also preferred.
The location of ESO~137-001 within the Norma cluster and the direction of its gas tail indicate that this galaxy is observed before the occurrence of
peak ram pressure (Fig.~1 of Sun et al. 2007). Since we only tested a limited number of temporal ram pressure profiles, we also consider models at
peak of ram pressure ($\le 0$~Myr).
In the following, we therefore focus on pre-peak models. Whereas models~1 and 5 are preferred when the comparison is only based on
the H$\alpha$ velocity field (Table~\ref{tab:molent1}), models~3 and 5 are preferred when the comparison is based on the CO/H$\alpha$/X-ray emission
distributions (Table~\ref{tab:molent2}). When the emission distributions and the velocity fields are combined
(Table~\ref{tab:molent}), the pre-peak models~5 are preferred.
All 50 highest-ranked model images and velocity fields were inspected by eye. To give an idea about the visual aspect of our models,
we show the first four highest-ranked pre-peak models in Figs.~\ref{fig:eso137-001all} to \ref{fig:eso137-001allnovel} for cases (i) to (iii).

In case (i) (Fig.~\ref{fig:eso137-001all}) the highest-ranked model~5 with $\Theta=60^{\circ}$ at $\Delta t=-40$-$-20$~Myr reproduces best the observed
CO emission distribution. The observed H$\alpha$ emission distribution of the main tail is reproduced by the model in a satisfactory way whereas the
model southern tail has an est-west component at distance of $60$ to $80$~kpc that is not observed.
Moreover, the north-western half of the southern model H$\alpha$ and X-ray tail
is closer to the main tail than it is observed. 

In all models of Fig.~\ref{fig:eso137-001all} the north-south gradient of the velocity field of the main tail is dominated by negative velocities
whereas the observed main tail is dominated by velocities close to systemic. There are negative velocities in the observed main tail but
they are located at or beyond the northern edge of the main tail. The observed velocity field of the southern tail is reasonably reproduced
by all models of Fig.~\ref{fig:eso137-001all}. Only the observed rare highly positive velocities (red) are not present in these models.

These highly positive velocities are present in three of the four models of case (ii) (Fig~\ref{fig:eso137-001allonlyvel}). In addition, the
observed close to systemic velocities in the main tail at distances $\ga 40$~kpc from the galaxy center are present in all models of
Fig~\ref{fig:eso137-001allonlyvel}. On the other hand, the model southern H$\alpha$ and X-ray tails are significantly shorter for three out of four models
than those of the models of case (i) (Fig.~\ref{fig:eso137-001all}) and the observed morphology of the CO tail is not well reproduced.

The X-ray and H$\alpha$ emission distributions of model~1a with $\Theta=75^{\circ}$ at 
$\Delta t=-10$~Myr better reproduces the observed southern tail structure than model~5. However, the observed CO emission distribution of the main
tail is less well reproduced by this model. 

The highest-ranked pre-peak models~5 of case (iii) are the same as for case (i). None of the pre-peak models~3 of case (iii) (Fig~\ref{fig:eso137-001allnovel})
reproduce the available observations of the gas tail in a satisfactory way.
Only models~1 and 5 are present within the twenty highest-ranked pre-peak models of comparisons (i) to (iii) (Tables~\ref{tab:molent} to \ref{tab:molent2}).

Whereas the comparison based on the velocity field alone favors an angle of $75^{\circ}$ between the galactic disk and the ram pressure wind,
the inclusion of the CO, H$\alpha$, and X-ray maps in the comparison (Table~\ref{tab:molent}) favors an angle of $60^{\circ}$.
It also favors lower absolute LOS components of the unit galaxy velocity vector.
\begin{table*}[!ht]
      \caption{Highest-ranked models based on CO, H$\alpha$, X-ray, and H$\alpha$ velocity field.}
         \label{tab:molent}
      \[
         \begin{tabular}{lcrccrcl}
           \hline
           model & $\Theta^{\rm (a)}$ & $\Delta t^{\rm (b)}$ & $az^{\rm (c)}$ & velocity vector & PA rotation & expansion/ & \\
            & (degrees) & (Myr) & (degrees) & & (degrees) & shrinking \\
           \hline  
             1a & 60 & 50 & 152 & -(0.95,0.32,0.07) &  5 & 1.1 & \\
             1b & 60 & 50 & 142 & -(0.97,0.24,0.09) &  0 & 1.1 & \\
             1b & 60 & 100 & 142 & -(0.97,0.24,0.09) &  0 & 0.9 & \\
             1b & 60 & 100 & 152 & -(0.95,0.32,0.07) &  10 & 1.2 & \\
             1a & 60 & 40 & 152 & -(0.95,0.32,0.07) &  5 & 1.2 & \\ 
             1b & 60 & 110 & 152 & -(0.95,0.32,0.07) &  10 & 1.1 & \\
             1b & 60 & 100 & 132 & -(0.98,0.15,0.10) &  10 & 1.0 & \\
             1a & 60 & 20 & 132 & -(0.98,0.15,0.10) &  10 & 1.2 & \\
             1b & 60 & 110 & 152 & -(0.95,0.32,0.07) &  10 & 1.0 & \\
             1a & 60 & 40 & 152 & -(0.95,0.32,0.07) &  5 & 1.1 & \\
             1b & 60 & 100 & 152 & -(0.95,0.32,0.07) &  10 & 1.1 & \\
             1b & 60 & 110 & 132 & -(0.98,0.15,0.10) &  10 & 1.0 & \\
             5 & 60 & -40 & 152 & -(0.95,0.32,0.07) &  5 & 1.2 & highest-ranked model A \\
             5 & 60 & -20 & 152 & -(0.95,0.32,0.07) &  5 & 1.1 & highest-ranked model A \\
             1a & 60 & 60 & 152 & -(0.95,0.32,0.07) &  10 & 1.0 & \\
             1b & 60 & 110 & 142 & -(0.97,0.24,0.09) &  10 & 1.0 & \\
             1b & 60 & 100 & 152 & -(0.95,0.32,0.07) &  10 & 1.0 & \\
             1b & 60 & 90 & 142 & -(0.97,0.24,0.09) &  10 & 0.9 & \\
             5a & 60 & -50 & 132 & -(0.98,0.15,0.10) &  10 & 1.2 & \\
             1a & 60 & 120 & 142 & -(0.97,0.24,0.09) &  5 & 1.1 & \\
             5 & 60 & -40 & 152 & -(0.95,0.32,0.07) &  5 & 1.0 & highest-ranked model A \\
             1b & 60 & 110 & 142 & -(0.97,0.24,0.09) &  10 & 0.9 & \\
             1b & 60 & 100 & 132 & -(0.98,0.15,0.10) &  10 & 1.1 & \\
             1b & 60 & 110 & 152 & -(0.95,0.32,0.07) &  10 & 1.2 & \\
             1b & 60 & 90 & 142 & -(0.97,0.24,0.09) &  0 & 0.9 & \\
             1a & 50 & 80 & 172 & -(0.89,0.46,0.01) &  -10 & 0.9 & \\
             1b & 60 & 30 & 142 & -(0.97,0.24,0.09) &  0 & 1.1 & \\
             5 & 60 & -30 & 152 & -(0.95,0.32,0.07) &  5 & 1.0 & highest-ranked model A (Fig.~\ref{fig:eso137-001all_5687})\\
             5b & 60 & -30 & 132 & -(0.98,0.15,0.10) &  0 & 1.1 & \\
             1b & 60 & 110 & 132 & -(0.98,0.15,0.10) &  10 & 1.1 & \\
             1b & 60 & 120 & 142 & -(0.97,0.24,0.09) &  5 & 1.0 & \\
             5a & 60 & -20 & 132 & -(0.98,0.15,0.10) &  10 & 0.9 & \\
             1b & 60 & 110 & 142 & -(0.97,0.24,0.09) &  5 & 1.0 & \\
             1b & 60 & 120 & 142 & -(0.97,0.24,0.09) &  10 & 1.0 & \\
             1a & 60 & 120 & 142 & -(0.97,0.24,0.09) &  5 & 1.0 & \\
             1b & 60 & 110 & 142 & -(0.97,0.24,0.09) &  5 & 0.9 & \\
             1a & 60 & 50 & 152 & -(0.95,0.32,0.07) &  10 & 1.0 & \\
             1a & 60 & 70 & 152 & -(0.95,0.32,0.07) &  10 & 1.1 & \\
             1a & 60 & 20 & 142 & -(0.97,0.24,0.09) &  10 & 1.2 & \\
             1b & 60 & 120 & 142 & -(0.97,0.24,0.09) &  5 & 1.1 & \\
             1b & 60 & 100 & 132 & -(0.98,0.15,0.10) &  10 & 0.9 & \\
             1a & 60 & 30 & 152 & -(0.95,0.32,0.07) &  5 & 1.2 & \\
             1b & 60 & 100 & 142 & -(0.97,0.24,0.09) &  10 & 0.9 & \\
             1b & 60 & 60 & 142 & -(0.97,0.24,0.09) &  0 & 1.1 & \\
             1b & 60 & 110 & 142 & -(0.97,0.24,0.09) &  5 & 0.8 & \\
             1a & 50 & 90 & 152 & -(0.95,0.31,0.08) &  -10 & 1.0 & \\
             3 & 60 & -30 & 132 & -(0.98,0.15,0.10) &  10 & 0.8 & \\
             1b & 60 & 100 & 152 & -(0.95,0.32,0.07) &  10 & 0.9 & \\
             5 & 60 & -40 & 152 & -(0.95,0.32,0.07) &  5 & 1.1 & highest-ranked model A \\
             3 & 60 & -30 & 132 & -(0.98,0.15,0.10) &  10 & 0.9 & \\
           \hline
        \end{tabular}
         \]
          \begin{list}{}{}
          \item $^{\rm (a)}$ angle between the direction of the galaxy's motion and the galactic disk
          \item $^{\rm (b)}$ time from peak ram pressure
          \item $^{\rm (c)}$ azimuthal viewing angle 
      \end{list}
\end{table*}

\begin{table*}[!ht]
      \caption{Highest-ranked models based on CO and the H$\alpha$ velocity field.}
         \label{tab:molent1a}
      \[
         \begin{tabular}{lcrccrcl}
           \hline
          model & $\Theta^{\rm (a)}$ & $\Delta t^{\rm (b)}$ & $az^{\rm (c)}$ & velocity vector & PA rotation & expansion/ & \\
            & (degrees) & (Myr) & (degrees) & & (degrees) & shrinking \\
           \hline
             1b & 75 & 10 & 140 & -(0.92,0.20,0.33) &  -10 & 1.0 & \\
             1a & 75 & 30 & 140 & -(0.92,0.20,0.33) &  -10 & 0.9 & \\
             1b & 60 & 30 & 142 & -(0.97,0.24,0.09) &  0 & 1.1 & \\
             1b & 75 & 0 & 140 & -(0.92,0.20,0.33) &  -15 & 0.9 & preferred model B (Fig.~\ref{fig:eso137-001all_471mom1_mixing1}) \\
             1a & 75 & 40 & 140 & -(0.92,0.20,0.33) &  -10 & 0.8 & \\
             1b & 75 & 40 & 100 & -(0.96,.120,0.24) &  -5 & 0.8 & \\
             1 & 75 & -20 & 140 & -(0.92,0.20,0.33) &  -15 & 0.8 & \\
             1 & 75 & -80 & 140 & -(0.92,0.20,0.33) &  -15 & 1.2 & \\
             1a & 50 & 60 & 182 & -(0.86,0.52,.050) &  -10 & 1.1 & \\
             1 & 75 & -20 & 150 & -(0.90,0.29,0.32) &  -5 & 0.9 & \\
             1a & 75 & -10 & 140 & -(0.92,0.20,0.33) &  -10 & 0.9 & preferred model C (Fig.~\ref{fig:eso137-001all_472mom1_mixing1}) \\
             1a & 75 & -30 & 140 & -(0.92,0.20,0.33) &  -10 & 0.9 & preferred model C \\
             1 & 50 & 70 & 182 & -(0.86,0.52,.050) &  -5 & 1.2 & \\
             1b & 75 & 20 & 140 & -(0.92,0.20,0.33) &  -10 & 1.1 & \\
             1a & 75 & 30 & 140 & -(0.92,0.20,0.33) &  -10 & 1.0 & \\
             1b & 75 & 10 & 130 & -(0.94,0.12,0.33) &  0 & 0.9 & \\
             5 & 60 & -40 & 152 & -(0.95,0.32,0.07) &  5 & 0.9 & highest-ranked model A \\
             1b & 75 & 10 & 140 & -(0.92,0.20,0.33) &  -10 & 0.8 & \\
             1a & 60 & 20 & 132 & -(0.98,0.15,0.10) &  10 & 1.2 & \\
             1b & 50 & 60 & 152 & -(0.95,0.31,0.08) &  -15 & 1.1 & \\
             1b & 75 & 50 & 100 & -(0.96,.120,0.24) &  -10 & 1.1 & \\
             1c & 75 & 0 & 140 & -(0.92,0.20,0.33) &  -10 & 0.9 & \\
             1a & 75 & 70 & 100 & -(0.96,.120,0.24) &  5 & 0.8 & \\
             1 & 75 & -20 & 140 & -(0.92,0.20,0.33) &  -15 & 0.9 & \\
             1c & 75 & -80 & 150 & -(0.90,0.29,0.32) &  -10 & 1.2 & \\
             1b & 75 & 40 & 100 & -(0.96,.120,0.24) &  -10 & 1.2 & \\
             2 & 60 & -10 & 142 & -(0.97,0.24,0.09) &  -5 & 1.0 & \\
             1a & 60 & -20 & 142 & -(0.97,0.24,0.09) &  -5 & 1.2 & preferred model C \\
             1a & 75 & 30 & 140 & -(0.92,0.20,0.33) &  -10 & 0.8 & \\
             1c & 75 & 10 & 140 & -(0.92,0.20,0.33) &  -10 & 1.1 & \\
             5b & 60 & -30 & 132 & -(0.98,0.15,0.10) &  0 & 1.1 & \\
             1a & 75 & -20 & 140 & -(0.92,0.20,0.33) &  -15 & 0.9 & preferred model C \\
             1a & 75 & -10 & 140 & -(0.92,0.20,0.33) &  -15 & 0.9 & preferred model C \\
             1a & 75 & -20 & 130 & -(0.94,0.12,0.33) &  5 & 1.2 & \\
             1b & 75 & 40 & 140 & -(0.92,0.20,0.33) &  -15 & 1.0 & \\
             1 & 75 & -10 & 140 & -(0.92,0.20,0.33) &  -15 & 0.9 & \\
             1b & 60 & 50 & 142 & -(0.97,0.24,0.09) &  0 & 1.1 & \\
             1b & 75 & 30 & 140 & -(0.92,0.20,0.33) &  -15 & 0.8 & \\
             1b & 75 & 20 & 140 & -(0.92,0.20,0.33) &  -15 & 0.8 & \\
             1b & 75 & 10 & 140 & -(0.92,0.20,0.33) &  -10 & 0.9 & \\
             1 & 50 & 60 & 182 & -(0.86,0.52,.050) &  0 & 0.9 & \\
             1b & 75 & 50 & 100 & -(0.96,.120,0.24) &  -5 & 1.0 & \\
             1b & 75 & 40 & 100 & -(0.96,.120,0.24) &  -5 & 0.9 & \\
             1b & 75 & 20 & 140 & -(0.92,0.20,0.33) &  -10 & 1.0 & \\
             1b & 75 & 50 & 100 & -(0.96,.120,0.24) &  -5 & 0.8 & \\
             1b & 50 & 60 & 152 & -(0.95,0.31,0.08) &  -15 & 1.0 & \\
             1a & 50 & 60 & 172 & -(0.89,0.46,0.01) &  -10 & 1.0 & \\
             1b & 75 & 40 & 100 & -(0.96,.120,0.24) &  -10 & 1.1 & \\
             5 & 60 & -40 & 152 & -(0.95,0.32,0.07) &  5 & 1.0 & highest-ranked model A \\
             1 & 50 & 110 & 172 & -(0.89,0.46,0.01) &  -10 & 1.1 & \\
           \hline
        \end{tabular}
         \]
          \begin{list}{}{}
          \item $^{\rm (a)}$ angle between the direction of the galaxy's motion and the galactic disk
          \item $^{\rm (b)}$ time from peak ram pressure
          \item $^{\rm (c)}$ azimuthal viewing angle 
      \end{list}
\end{table*}

\begin{table*}[!ht]
      \caption{Highest-ranked models based on CO, H$\alpha$, and X-ray.}
         \label{tab:molent2}
      \[
         \begin{tabular}{lcrccrcl}
           \hline
          model & $\Theta^{\rm (a)}$ & $\Delta t^{\rm (b)}$ & $az^{\rm (c)}$ & velocity vector & PA rotation & expansion/ & \\
            & (degrees) & (Myr) & (degrees) & & (degrees) & shrinking \\
           \hline
            1b & 75 & 30 & 120 & -(0.95,0.04,0.32) &  15 & 1.0 & \\
            3 & 60 & -20 & 132 & -(0.98,0.15,0.10) &  10 & 1.0 & \\
            5 & 60 & -30 & 142 & -(0.97,0.24,0.09) &  10 & 1.1 & highest-ranked model A \\
            3 & 60 & -30 & 152 & -(0.95,0.32,0.07) &  0 & 0.9 & \\
            5 & 60 & -30 & 142 & -(0.97,0.24,0.09) &  5 & 1.2 & highest-ranked model A \\
            1b & 75 & 30 & 120 & -(0.95,0.04,0.32) &  15 & 0.9 & \\
            3 & 60 & -10 & 122 & -(0.99,0.07,0.09) &  15 & 1.1 & \\
            1b & 75 & 30 & 120 & -(0.95,0.04,0.32) &  15 & 1.1 & \\
            3 & 60 & -10 & 132 & -(0.98,0.15,0.10) &  10 & 1.1 & \\
            3 & 60 & -20 & 132 & -(0.98,0.15,0.10) &  10 & 0.9 & \\
            3 & 60 & -30 & 152 & -(0.95,0.32,0.07) &  0 & 0.8 & \\
            5 & 60 & -30 & 142 & -(0.97,0.24,0.09) &  10 & 1.2 & highest-ranked model A \\
            1b & 75 & 40 & 120 & -(0.95,0.04,0.32) &  15 & 1.1 & \\
            1b & 75 & 20 & 140 & -(0.92,0.20,0.33) &  10 & 1.1 & \\
            3 & 60 & -10 & 132 & -(0.98,0.15,0.10) &  10 & 1.0 & \\
            1b & 60 & 90 & 152 & -(0.95,0.32,0.07) &  0 & 0.9 & \\
            3 & 60 & -20 & 122 & -(0.99,0.07,0.09) &  15 & 1.0 & \\
            5 & 60 & -30 & 142 & -(0.97,0.24,0.09) &  5 & 1.1 & highest-ranked model A \\
            1b & 75 & 40 & 120 & -(0.95,0.04,0.32) &  15 & 1.0 & \\
            5 & 60 & -30 & 132 & -(0.98,0.15,0.10) &  15 & 1.1 & highest-ranked model A \\
            3 & 60 & -30 & 132 & -(0.98,0.15,0.10) &  5 & 0.9 & \\
            5 & 60 & -40 & 142 & -(0.97,0.24,0.09) &  5 & 1.2 & highest-ranked model A \\
            3 & 60 & -20 & 142 & -(0.97,0.24,0.09) &  5 & 1.0 & \\
            3 & 60 & -10 & 122 & -(0.99,0.07,0.09) &  15 & 1.0 & \\
            1b & 60 & 110 & 132 & -(0.98,0.15,0.10) &  10 & 1.1 & \\
            1b & 60 & 100 & 132 & -(0.98,0.15,0.10) &  10 & 1.1 & \\
            1b & 75 & 20 & 140 & -(0.92,0.20,0.33) &  10 & 1.2 & \\
            3 & 60 & -30 & 112 & -(1.00,.020,0.06) &  15 & 0.9 & \\
            1b & 60 & 100 & 132 & -(0.98,0.15,0.10) &  15 & 1.1 & \\
            1b & 75 & 20 & 130 & -(0.94,0.12,0.33) &  15 & 1.2 & \\
            1b & 60 & 110 & 132 & -(0.98,0.15,0.10) &  15 & 1.1 & \\
            1b & 75 & 30 & 110 & -(0.96,.040,0.29) &  15 & 1.0 & \\
            1b & 60 & 110 & 132 & -(0.98,0.15,0.10) &  15 & 1.0 & \\
            1b & 60 & 100 & 132 & -(0.98,0.15,0.10) &  10 & 1.2 & \\
            1b & 75 & 20 & 140 & -(0.92,0.20,0.33) &  10 & 1.0 & \\
            1b & 75 & 20 & 120 & -(0.95,0.04,0.32) &  15 & 1.1 & \\
            3 & 60 & -20 & 122 & -(0.99,0.07,0.09) &  15 & 0.9 & \\
            1b & 75 & 20 & 130 & -(0.94,0.12,0.33) &  15 & 1.1 & \\
            5 & 60 & -30 & 152 & -(0.95,0.32,0.07) &  5 & 1.1 & highest-ranked model A \\
            1b & 60 & 100 & 132 & -(0.98,0.15,0.10) &  15 & 1.0 & \\
            1b & 60 & 100 & 132 & -(0.98,0.15,0.10) &  5 & 1.2 & \\
            1b & 60 & 110 & 132 & -(0.98,0.15,0.10) &  10 & 1.2 & \\
            5 & 60 & -30 & 132 & -(0.98,0.15,0.10) &  15 & 1.2 & highest-ranked model A \\
            1b & 75 & 40 & 110 & -(0.96,.040,0.29) &  15 & 1.0 & \\
            1b & 75 & 30 & 140 & -(0.92,0.20,0.33) &  10 & 1.1 & \\
            1b & 75 & 40 & 120 & -(0.95,0.04,0.32) &  15 & 1.2 & \\
            1b & 60 & 100 & 132 & -(0.98,0.15,0.10) &  5 & 1.1 & \\
            1b & 60 & 90 & 152 & -(0.95,0.32,0.07) &  5 & 1.0 & \\
            1b & 60 & 90 & 152 & -(0.95,0.32,0.07) &  5 & 1.1 & \\
            1b & 75 & 20 & 120 & -(0.95,0.04,0.32) &  15 & 1.2 & \\
            \hline
        \end{tabular}
         \]
         \begin{list}{}{}
          \item $^{\rm (a)}$ angle between the direction of the galaxy's motion and the galactic disk
          \item $^{\rm (b)}$ time from peak ram pressure
          \item $^{\rm (c)}$ azimuthal viewing angle 
      \end{list}
\end{table*}

\newpage

\section{The influence of a higher ICM-ISM mixing rate}

\begin{table*}[!ht]
      \caption{High ICM-ISM mixing rate: highest-ranked models based on CO, H$\alpha$, X-ray, and H$\alpha$ velocity field.}
         \label{tab:moremolent}
      \[
         \begin{tabular}{lcrccrcl}
           \hline
          model & $\Theta^{\rm (a)}$ & $\Delta t^{\rm (b)}$ & $az^{\rm (c)}$ & velocity vector & PA rotation & expansion/ & \\
            & (degrees) & (Myr) & (degrees) & & (degrees) & shrinking \\
           \hline
            1c & 60 & 10 & 132 & -(0.98,0.15,0.10) &  15 & 1.2 & \\
            1b & 50 & 60 & 182 & -(0.86,0.52,.050) &  -15 & 0.9 & \\
            1 & 50 & 70 & 182 & -(0.86,0.52,.050) &  -15 & 0.8 & \\
            1c & 60 & -10 & 142 & -(0.97,0.24,0.09) &  0 & 1.1 & \\
            1c & 60 & 0 & 132 & -(0.98,0.15,0.10) &  15 & 1.1 & \\
            1c & 60 & 10 & 132 & -(0.98,0.15,0.10) &  5 & 1.1 & \\
            1 & 50 & 60 & 182 & -(0.86,0.52,.050) &  -15 & 1.0 & \\
            1b & 50 & 60 & 182 & -(0.86,0.52,.050) &  -15 & 1.0 & \\
            1c & 60 & -10 & 132 & -(0.98,0.15,0.10) &  15 & 1.2 & \\
            1b & 50 & 120 & 152 & -(0.95,0.31,0.08) &  5 & 1.1 & \\
            1 & 50 & 70 & 182 & -(0.86,0.52,.050) &  -15 & 0.9 & \\
            1 & 50 & 60 & 192 & -(0.82,0.56,.110) &  -15 & 0.9 & \\
            1c & 60 & 80 & 142 & -(0.97,0.24,0.09) &  15 & 1.2 & \\
            1 & 50 & 70 & 182 & -(0.86,0.52,.050) &  -15 & 1.0 & \\
            1c & 60 & -10 & 142 & -(0.97,0.24,0.09) &  0 & 1.2 & \\
            1c & 60 & 30 & 132 & -(0.98,0.15,0.10) &  15 & 1.2 & \\
            1c & 60 & 0 & 132 & -(0.98,0.15,0.10) &  15 & 1.2 & \\
            1c & 60 & -10 & 132 & -(0.98,0.15,0.10) &  5 & 1.1 & \\
            1b & 50 & 120 & 152 & -(0.95,0.31,0.08) &  10 & 1.1 & \\
            1c & 60 & 10 & 142 & -(0.97,0.24,0.09) &  0 & 1.1 & \\
            1c & 60 & 90 & 152 & -(0.95,0.32,0.07) &  10 & 1.1 & \\
            5c & 60 & -10 & 132 & -(0.98,0.15,0.10) &  10 & 0.8 & \\
            1b & 50 & 40 & 182 & -(0.86,0.52,.050) &  -15 & 1.2 & \\
            1b & 50 & 40 & 192 & -(0.82,0.56,.110) &  -15 & 1.1 & \\
            1c & 60 & 80 & 132 & -(0.98,0.15,0.10) &  15 & 1.2 & \\
            5c & 60 & 0 & 132 & -(0.98,0.15,0.10) &  10 & 0.8 & \\
            1c & 60 & 70 & 142 & -(0.97,0.24,0.09) &  15 & 1.2 & \\
            1b & 50 & 60 & 182 & -(0.86,0.52,.050) &  -15 & 0.8 & \\
            5c & 60 & 10 & 142 & -(0.97,0.24,0.09) &  0 & 0.8 & \\
            1c & 60 & 0 & 132 & -(0.98,0.15,0.10) &  5 & 1.2 & \\
            1c & 60 & 80 & 132 & -(0.98,0.15,0.10) &  15 & 1.1 & \\
            1c & 60 & 30 & 142 & -(0.97,0.24,0.09) &  5 & 1.2 & \\
            5c & 60 & -10 & 132 & -(0.98,0.15,0.10) &  10 & 0.9 & \\
            1b & 50 & 120 & 152 & -(0.95,0.31,0.08) &  5 & 1.2 & \\
            1 & 50 & 60 & 182 & -(0.86,0.52,.050) &  -15 & 1.1 & \\
            1 & 50 & 70 & 202 & -(0.78,0.60,.180) &  -15 & 1.0 & \\
            5b & 60 & -60 & 132 & -(0.98,0.15,0.10) &  5 & 1.0 & \\
            1c & 60 & 10 & 132 & -(0.98,0.15,0.10) &  15 & 1.1 & \\
            1b & 50 & 120 & 152 & -(0.95,0.31,0.08) &  5 & 1.0 & \\
            1b & 50 & 50 & 192 & -(0.82,0.56,.110) &  -15 & 1.0 & \\
            1b & 50 & 50 & 182 & -(0.86,0.52,.050) &  -15 & 1.1 & \\
            1c & 60 & -20 & 142 & -(0.97,0.24,0.09) &  0 & 1.1 & \\
            1b & 50 & 50 & 182 & -(0.86,0.52,.050) &  -15 & 0.8 & \\
            1 & 50 & 60 & 202 & -(0.78,0.60,.180) &  -15 & 1.0 & \\
            1c & 60 & -20 & 142 & -(0.97,0.24,0.09) &  0 & 1.2 & \\
            1c & 60 & 20 & 142 & -(0.97,0.24,0.09) &  0 & 1.1 & \\
            5c & 60 & 10 & 132 & -(0.98,0.15,0.10) &  15 & 1.1 & \\
            1c & 60 & 10 & 142 & -(0.97,0.24,0.09) &  0 & 1.2 & \\
            5c & 60 & 10 & 142 & -(0.97,0.24,0.09) &  15 & 1.1 & \\
            1b & 50 & 80 & 172 & -(0.89,0.46,0.01) &  -15 & 0.9 & \\
            \hline
        \end{tabular}
         \]
         \begin{list}{}{}
          \item $^{\rm (a)}$ angle between the direction of the galaxy's motion and the galactic disk
          \item $^{\rm (b)}$ time from peak ram pressure
          \item $^{\rm (c)}$ azimuthal viewing angle 
      \end{list}
\end{table*}

\begin{figure*}[!ht]
  \centering
  \resizebox{14cm}{!}{\includegraphics{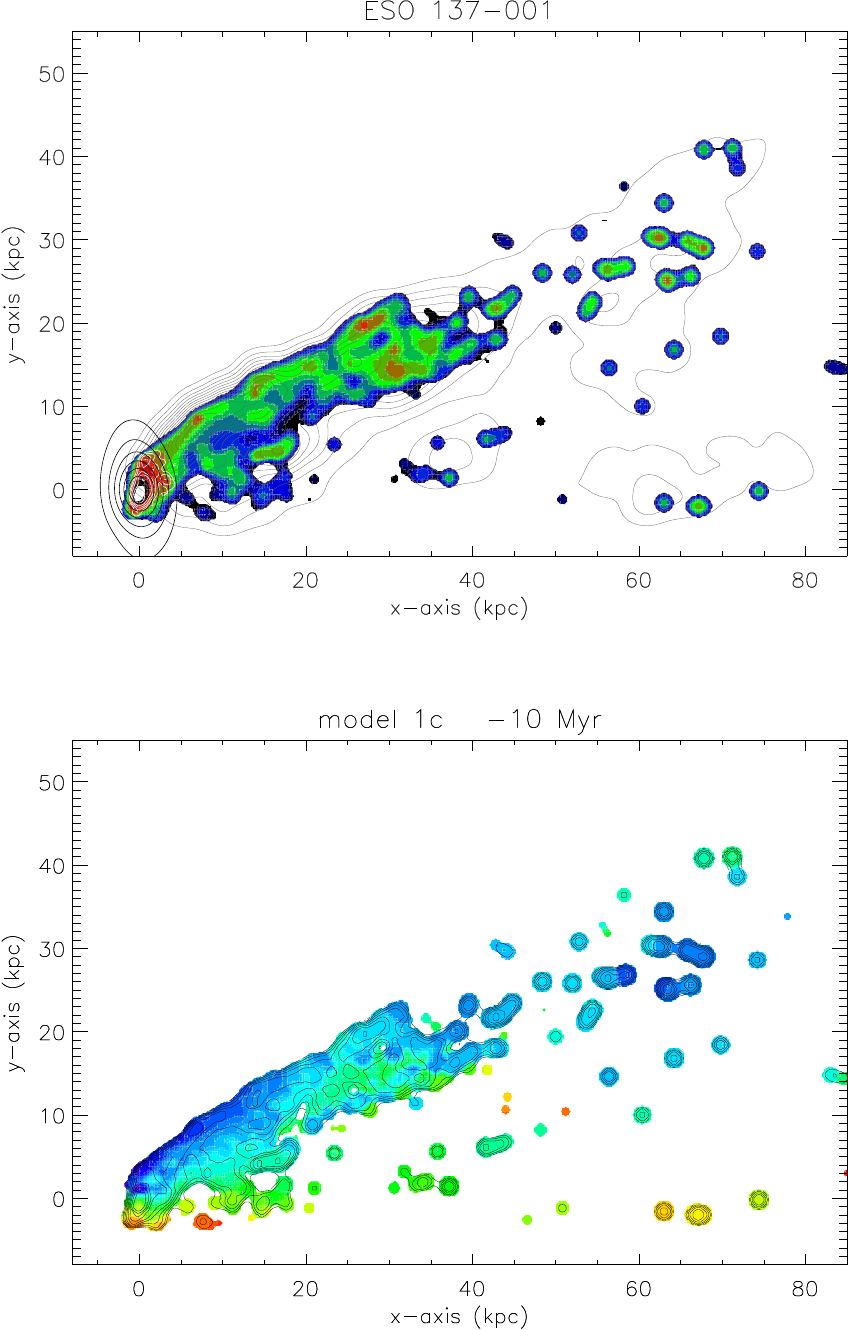}\includegraphics{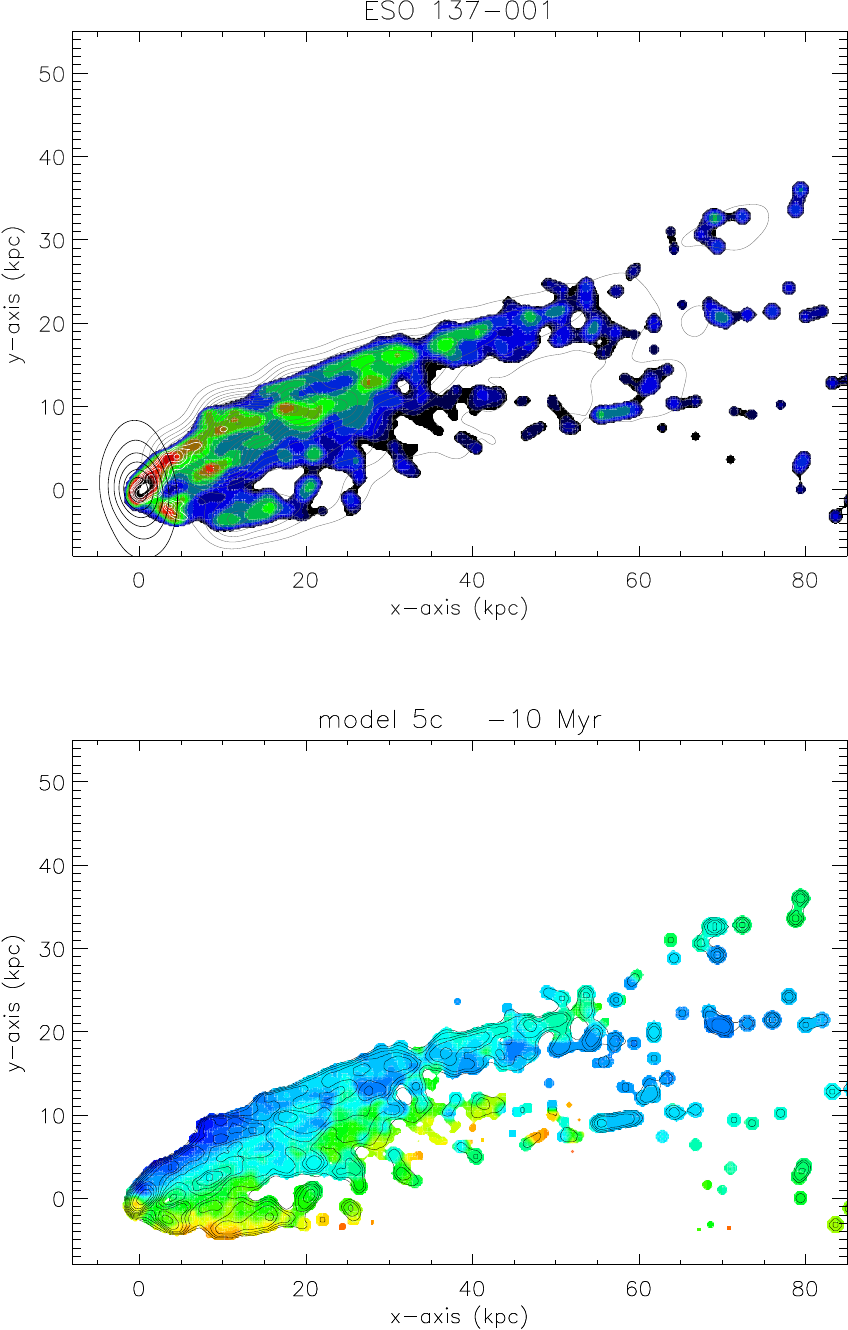}}
  \resizebox{14cm}{!}{\includegraphics{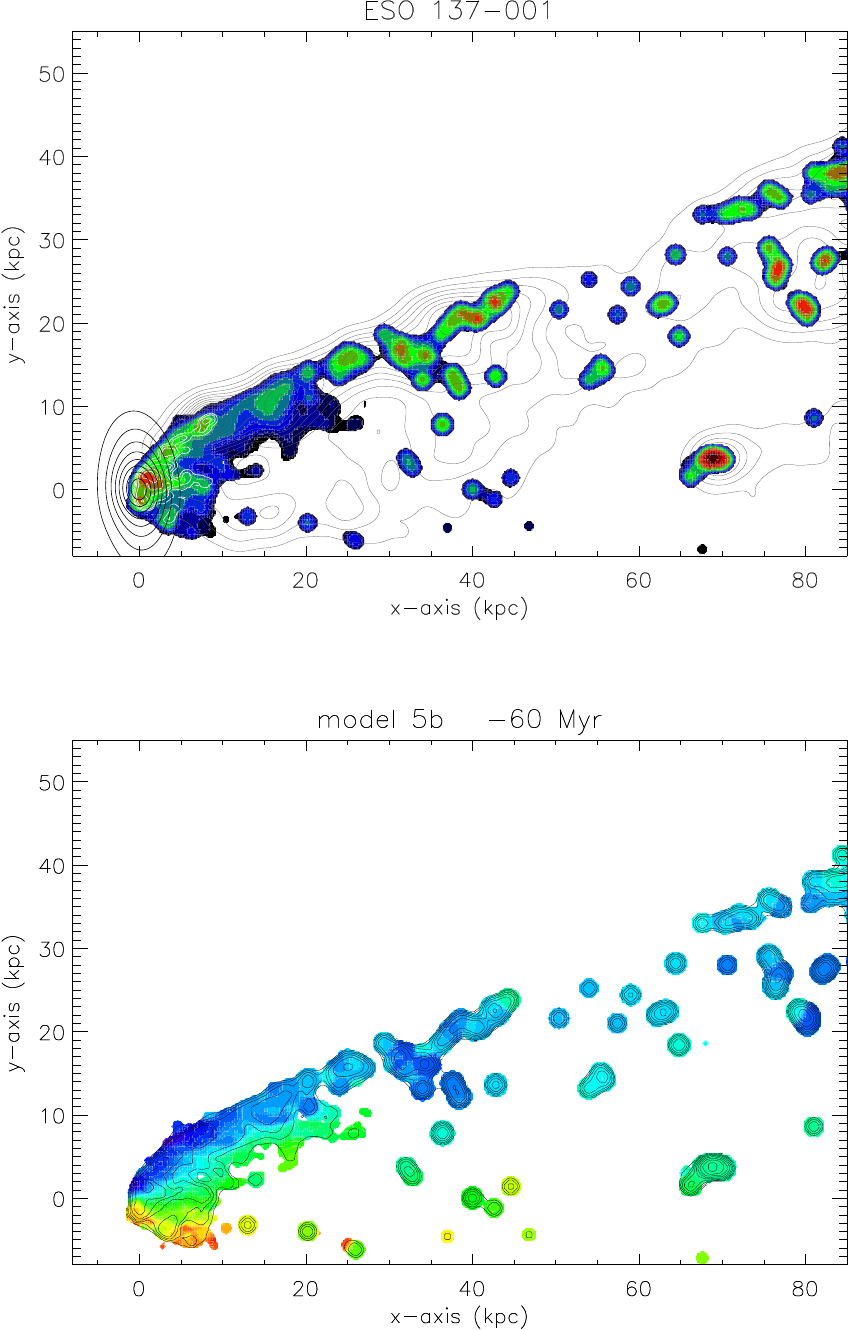}\includegraphics{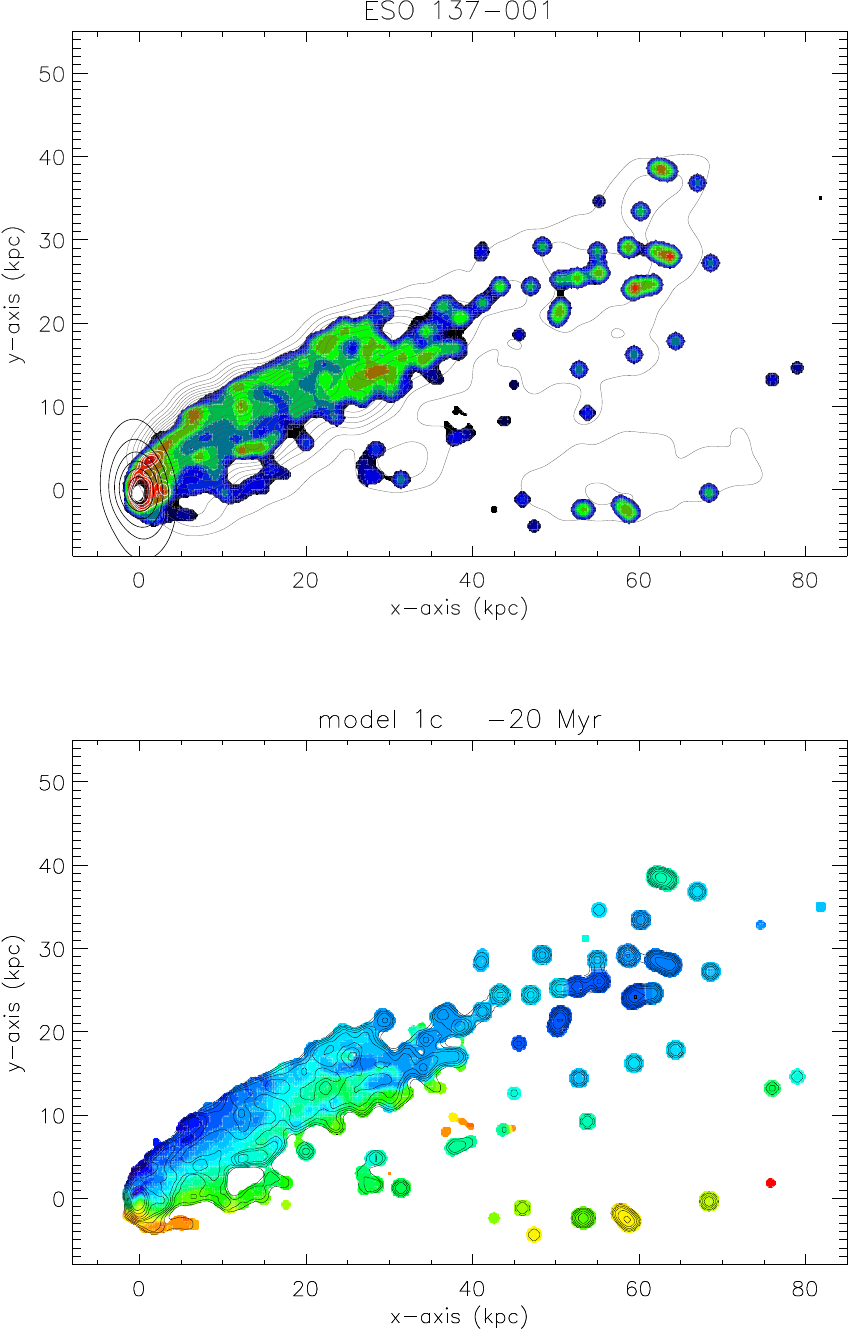}}
  \caption{High ICM-ISM mixing rate: highest-ranked ram pressure pre-peak models of ESO~137-001 based on CO, H$\alpha$, X-ray, and H$\alpha$ velocity field.
    Upper panel: color: H$\alpha$; dark gray contours:
    X-ray; white contours: CO ; black contours: stellar content. Lower panel: H$\alpha$ velocity field.
  \label{fig:moreeso137-001all}}
\end{figure*}

\begin{table*}[!ht]
      \caption{High ICM-ISM mixing rate: highest-ranked models based on the H$\alpha$ velocity field.}
         \label{tab:moremolent1}
      \[
         \begin{tabular}{lcrccrcl}
           \hline
          model & $\Theta^{\rm (a)}$ & $\Delta t^{\rm (b)}$ & $az^{\rm (c)}$ & velocity vector & PA rotation & expansion/ & \\
            & (degrees) & (Myr) & (degrees) & & (degrees) & shrinking \\
           \hline
            1a & 60 & -30 & 142 & -(0.97,0.24,0.09) &  -5 & 1.2 & \\
            5c & 60 & 10 & 142 & -(0.97,0.24,0.09) &  0 & 0.8 & \\
            1a & 75 & 30 & 140 & -(0.92,0.20,0.33) &  -15 & 0.8 & \\
            1 & 75 & -30 & 140 & -(0.92,0.20,0.33) &  -15 & 0.8 & \\
            1c & 60 & 10 & 132 & -(0.98,0.15,0.10) &  15 & 1.2 & \\
            1 & 75 & -20 & 140 & -(0.92,0.20,0.33) &  -10 & 1.0 & \\
            1a & 75 & -20 & 140 & -(0.92,0.20,0.33) &  -10 & 0.9 & \\
            1b & 60 & -50 & 142 & -(0.97,0.24,0.09) &  -5 & 1.2 & \\
            1 & 75 & -30 & 140 & -(0.92,0.20,0.33) &  -10 & 0.8 & \\
            1c & 60 & 0 & 142 & -(0.97,0.24,0.09) &  0 & 1.2 & \\
            1b & 75 & 10 & 140 & -(0.92,0.20,0.33) &  -15 & 0.8 & \\
            1 & 60 & -10 & 132 & -(0.98,0.15,0.10) &  5 & 1.2 & \\
            1 & 60 & -10 & 142 & -(0.97,0.24,0.09) &  -5 & 1.1 & \\
            1 & 60 & -10 & 132 & -(0.98,0.15,0.10) &  15 & 1.0 & \\
            3 & 60 & -70 & 142 & -(0.97,0.24,0.09) &  -5 & 1.2 & \\
            2 & 60 & -10 & 132 & -(0.98,0.15,0.10) &  15 & 1.0 & \\
            3 & 60 & -10 & 132 & -(0.98,0.15,0.10) &  5 & 0.8 & \\
            1a & 75 & -30 & 140 & -(0.92,0.20,0.33) &  -15 & 0.8 & \\
            3 & 60 & -90 & 142 & -(0.97,0.24,0.09) &  -5 & 1.2 & \\
            1b & 50 & 40 & 182 & -(0.86,0.52,.050) &  -15 & 1.2 & \\
            1a & 75 & -20 & 140 & -(0.92,0.20,0.33) &  -15 & 0.9 & \\
            3 & 60 & -90 & 142 & -(0.97,0.24,0.09) &  -5 & 1.1 & \\
            1c & 60 & 90 & 142 & -(0.97,0.24,0.09) &  5 & 0.8 & \\
            1 & 75 & -20 & 140 & -(0.92,0.20,0.33) &  -10 & 0.9 & \\
            3 & 60 & -80 & 142 & -(0.97,0.24,0.09) &  -5 & 1.1 & \\
            5b & 60 & -60 & 142 & -(0.97,0.24,0.09) &  -5 & 1.1 & \\
            1c & 60 & 10 & 132 & -(0.98,0.15,0.10) &  5 & 1.1 & \\
            1b & 75 & 0 & 150 & -(0.90,0.29,0.32) &  -5 & 0.8 & \\
            3 & 60 & -80 & 142 & -(0.97,0.24,0.09) &  -5 & 1.2 & \\
            1c & 75 & -80 & 140 & -(0.92,0.20,0.33) &  -10 & 1.0 & \\
            1 & 50 & 60 & 182 & -(0.86,0.52,.050) &  -15 & 1.0 & \\
            1c & 75 & 10 & 130 & -(0.94,0.12,0.33) &  15 & 1.2 & \\
            1b & 75 & 40 & 100 & -(0.96,.120,0.24) &  -5 & 0.9 & \\
            5a & 60 & -30 & 132 & -(0.98,0.15,0.10) &  5 & 1.0 & \\
            5 & 60 & -20 & 132 & -(0.98,0.15,0.10) &  0 & 0.8 & \\
            1a & 75 & -20 & 140 & -(0.92,0.20,0.33) &  -15 & 1.1 & \\
            7b & 60 & -20 & 142 & -(0.97,0.24,0.09) &  -5 & 0.9 & \\
            5b & 60 & -50 & 132 & -(0.98,0.15,0.10) &  10 & 1.2 & \\
            1c & 60 & 80 & 142 & -(0.97,0.24,0.09) &  15 & 1.2 & \\
            1b & 50 & 40 & 192 & -(0.82,0.56,.110) &  -15 & 1.1 & \\
            1c & 75 & 0 & 140 & -(0.92,0.20,0.33) &  -5 & 0.8 & \\
            1a & 50 & 40 & 192 & -(0.82,0.56,.110) &  -15 & 1.2 & \\
            1b & 75 & 0 & 140 & -(0.92,0.20,0.33) &  -15 & 0.8 & \\
            1b & 75 & 40 & 140 & -(0.92,0.20,0.33) &  -10 & 0.8 & \\
            2 & 60 & -20 & 132 & -(0.98,0.15,0.10) &  15 & 1.0 & \\
            1b & 75 & 50 & 100 & -(0.96,.120,0.24) &  -5 & 0.8 & \\
            1b & 75 & 40 & 140 & -(0.92,0.20,0.33) &  -10 & 0.9 & \\
            1c & 75 & -20 & 140 & -(0.92,0.20,0.33) &  15 & 1.0 & \\
            2 & 60 & -40 & 142 & -(0.97,0.24,0.09) &  -5 & 1.2 & \\
            1b & 75 & 50 & 100 & -(0.96,.120,0.24) &  -5 & 0.9 & \\
            \hline
        \end{tabular}
         \]
         \begin{list}{}{}
          \item $^{\rm (a)}$ angle between the direction of the galaxy's motion and the galactic disk
          \item $^{\rm (b)}$ time from peak ram pressure
          \item $^{\rm (c)}$ azimuthal viewing angle 
      \end{list}
\end{table*}

\begin{figure*}[!ht]
  \centering
  \resizebox{14cm}{!}{\includegraphics{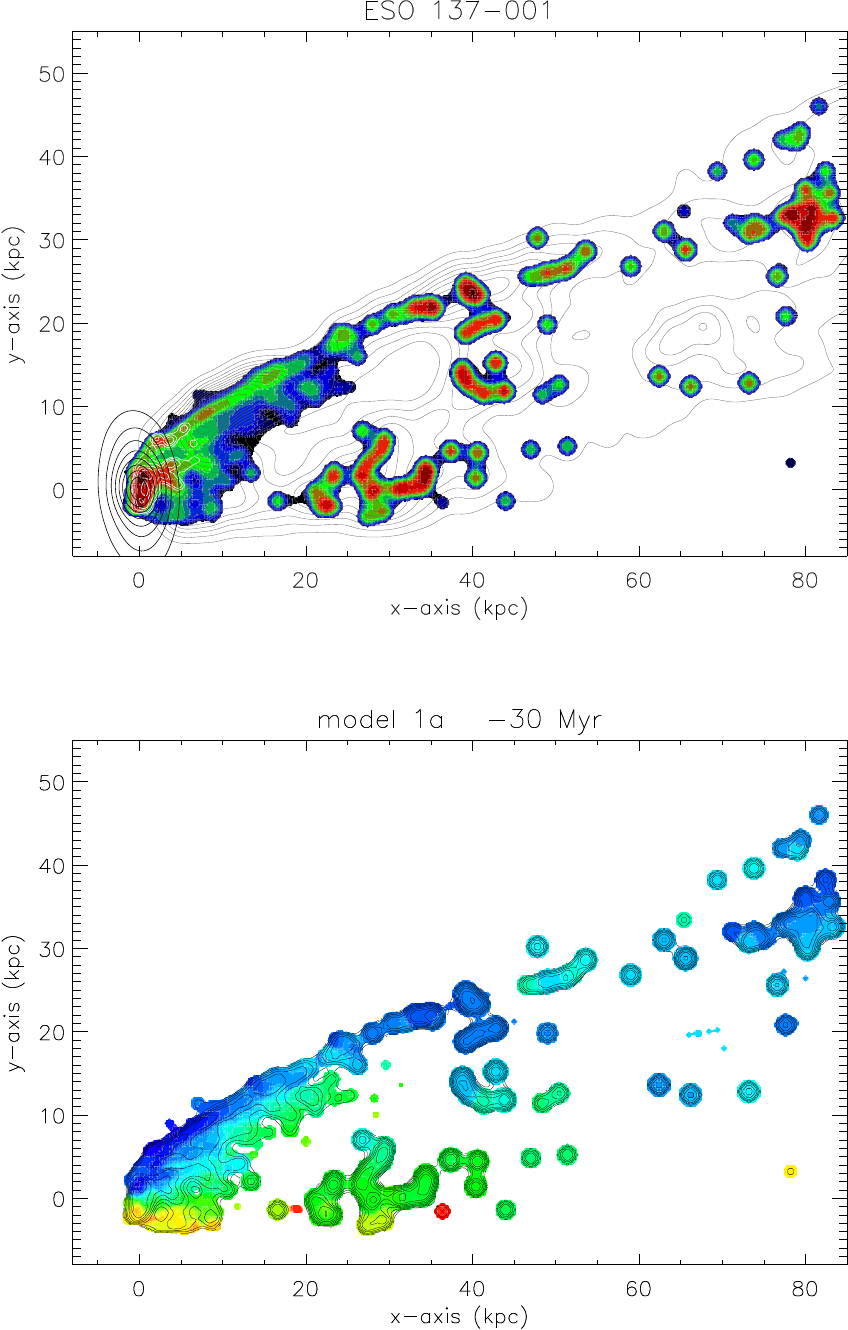}\includegraphics{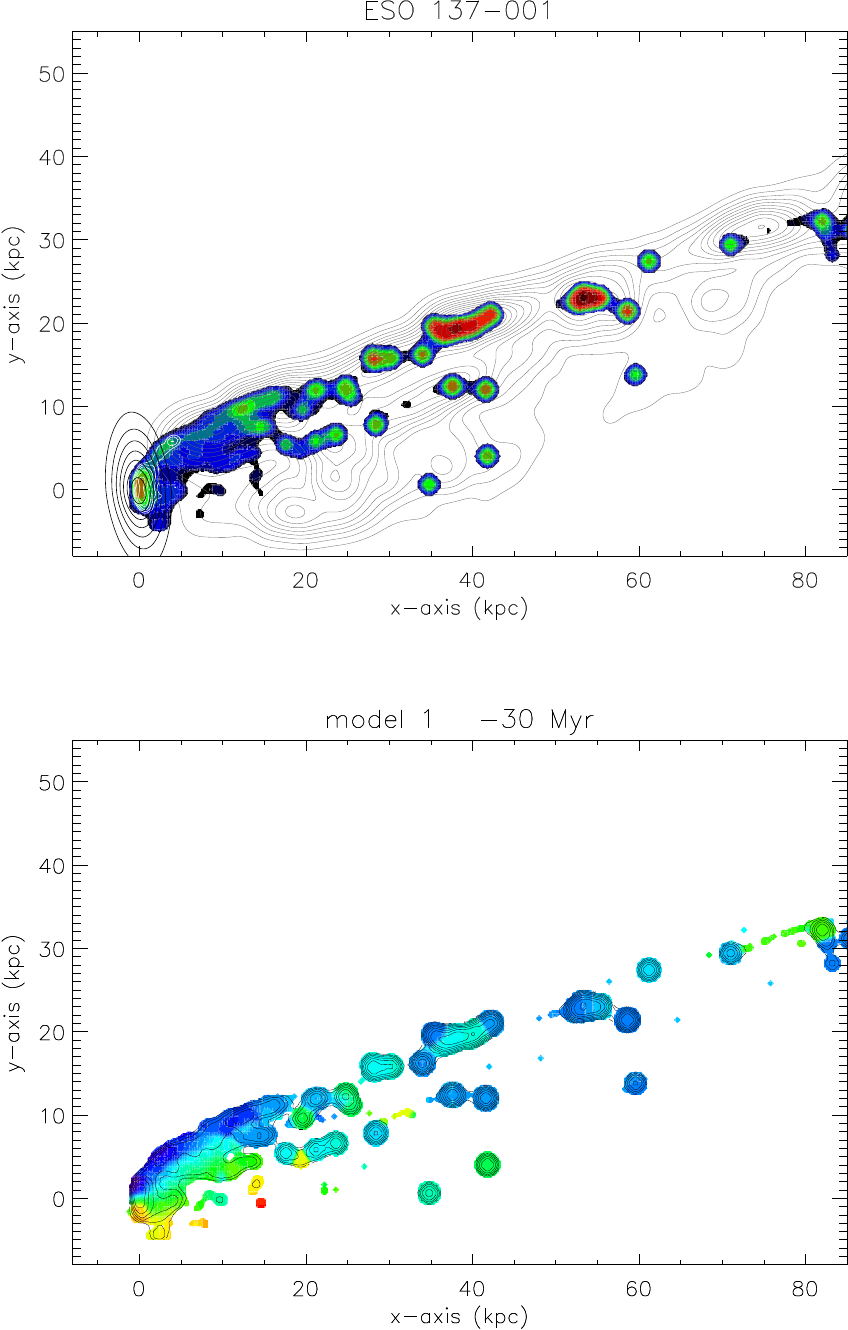}}
  \resizebox{14cm}{!}{\includegraphics{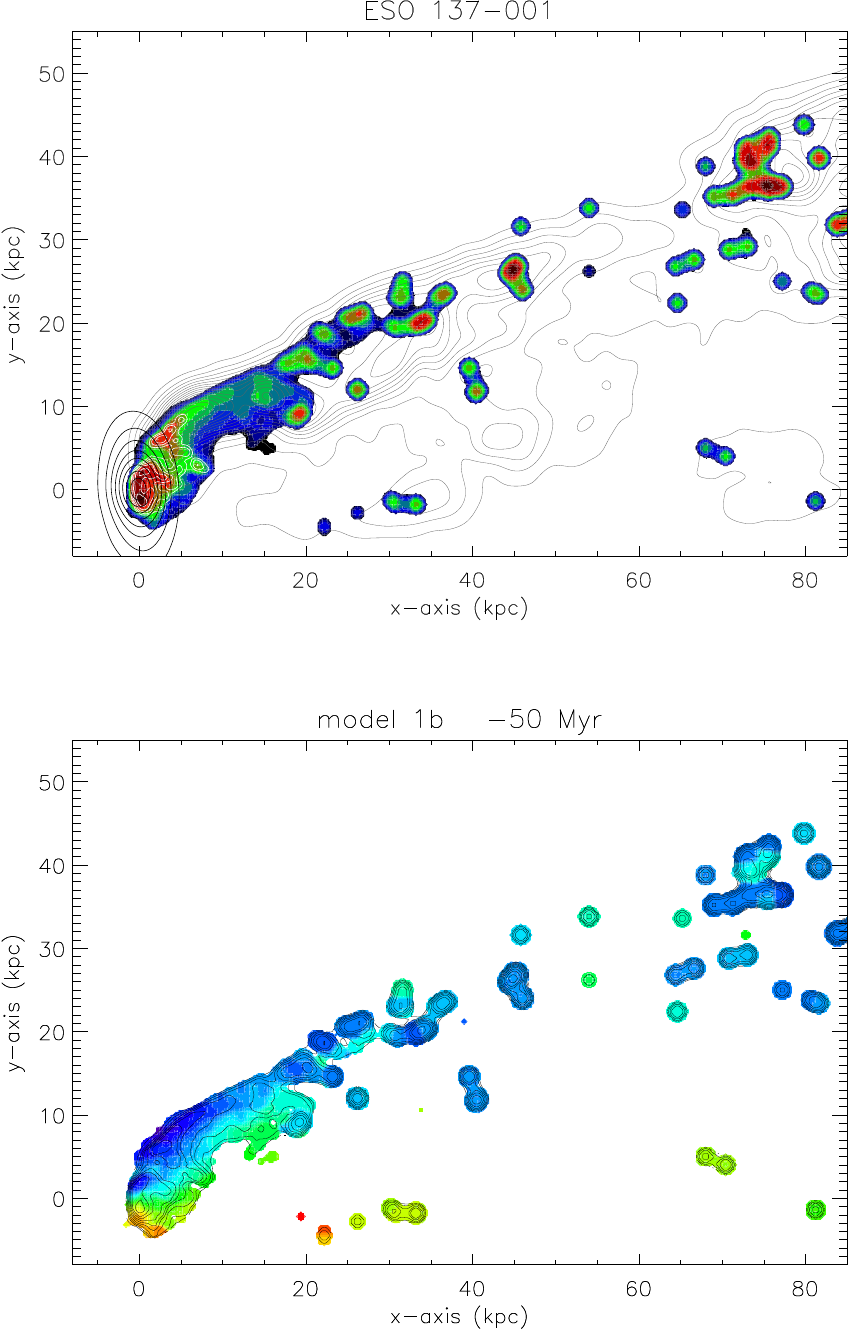}\includegraphics{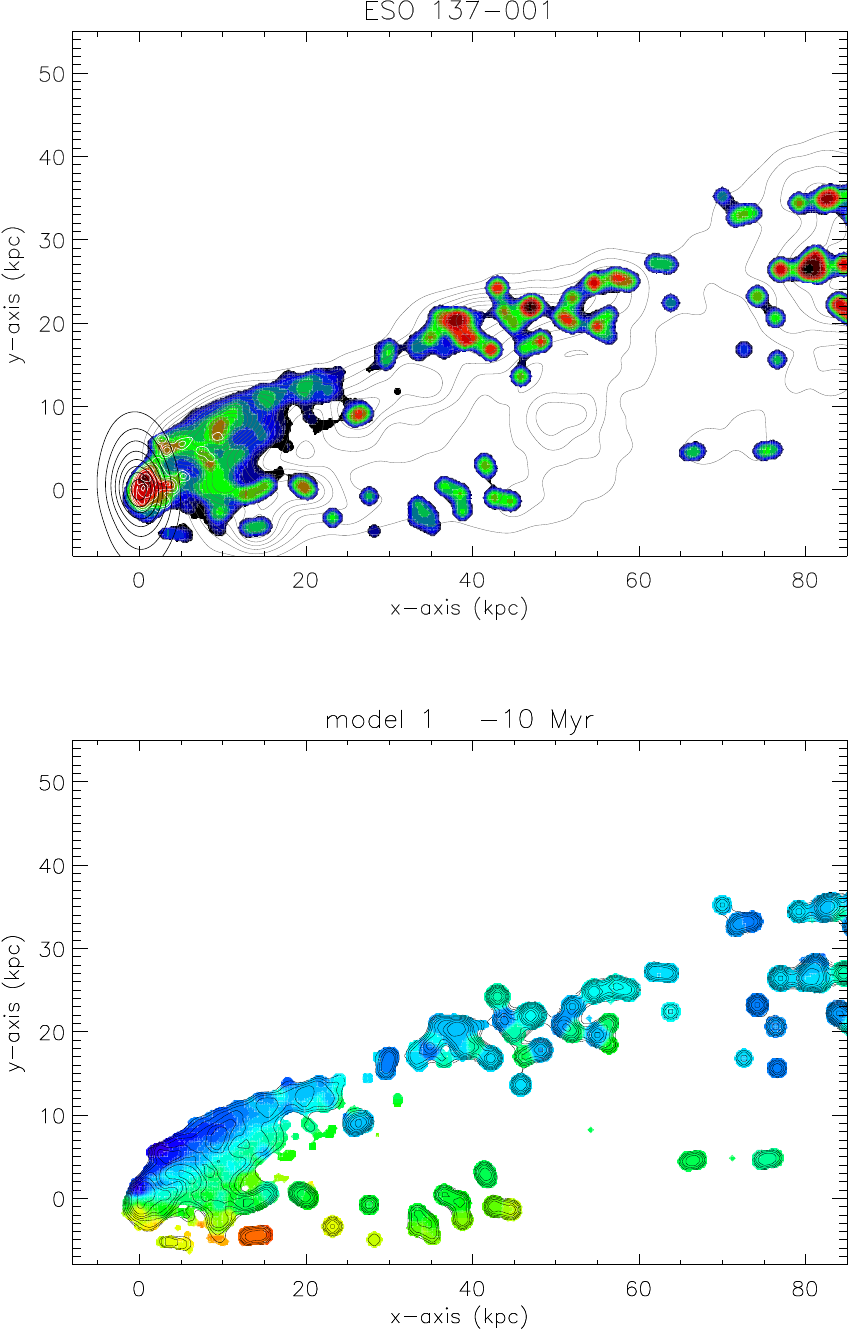}}
  \caption{High ICM-ISM mixing rate: highest-ranked models of ESO~137-001 based on the H$\alpha$ velocity field. Upper panel: color: H$\alpha$; dark gray contours:
    X-ray; white contours: CO ; black contours: stellar content. Lower panel: H$\alpha$ velocity field.
  \label{fig:moreeso137-001allonlyvel}}
\end{figure*}

\begin{table*}[!ht]
      \caption{High ICM-ISM mixing rate: highest-ranked models based on CO, H$\alpha$, and X-ray.}
         \label{tab:moremolent2}
      \[
         \begin{tabular}{lcrccrcl}
           \hline
          model & $\Theta^{\rm (a)}$ & $\Delta t^{\rm (b)}$ & $az^{\rm (c)}$ & velocity vector & PA rotation & expansion/ & \\
            & (degrees) & (Myr) & (degrees) & & (degrees) & shrinking \\
           \hline
            1b & 50 & 50 & 162 & -(0.92,0.39,0.05) &  -10 & 0.9 & \\
            1b & 50 & 60 & 152 & -(0.95,0.31,0.08) &  -5 & 1.0 & \\
            1b & 50 & 50 & 162 & -(0.92,0.39,0.05) &  -10 & 1.0 & \\
            1c & 60 & 10 & 132 & -(0.98,0.15,0.10) &  10 & 1.2 & \\
            5c & 60 & 0 & 132 & -(0.98,0.15,0.10) &  10 & 0.9 & \\
            1b & 50 & 60 & 152 & -(0.95,0.31,0.08) &  0 & 1.1 & \\
            1c & 60 & 10 & 142 & -(0.97,0.24,0.09) &  5 & 1.2 & \\
            1b & 50 & 60 & 152 & -(0.95,0.31,0.08) &  -5 & 1.1 & \\
            1c & 60 & 10 & 132 & -(0.98,0.15,0.10) &  10 & 1.1 & \\
            5c & 60 & 10 & 142 & -(0.97,0.24,0.09) &  10 & 1.0 & \\
            5c & 60 & 10 & 122 & -(0.99,0.07,0.09) &  15 & 1.0 & \\
            5c & 60 & 10 & 142 & -(0.97,0.24,0.09) &  10 & 1.1 & \\
            1c & 60 & -10 & 142 & -(0.97,0.24,0.09) &  5 & 1.1 & \\
            5c & 60 & 0 & 132 & -(0.98,0.15,0.10) &  10 & 0.8 & \\
            1c & 60 & -10 & 142 & -(0.97,0.24,0.09) &  5 & 1.2 & \\
            5c & 60 & 10 & 122 & -(0.99,0.07,0.09) &  15 & 0.9 & \\
            1b & 50 & 50 & 182 & -(0.86,0.52,.050) &  -15 & 0.9 & \\
            5c & 60 & 0 & 152 & -(0.95,0.32,0.07) &  0 & 0.9 & \\
            1c & 60 & 20 & 142 & -(0.97,0.24,0.09) &  5 & 1.2 & \\
            1c & 60 & 20 & 142 & -(0.97,0.24,0.09) &  5 & 1.0 & \\
            5c & 60 & 0 & 122 & -(0.99,0.07,0.09) &  15 & 0.9 & \\
            1c & 60 & 20 & 142 & -(0.97,0.24,0.09) &  5 & 1.1 & \\
            5c & 60 & 10 & 132 & -(0.98,0.15,0.10) &  15 & 0.9 & \\
            1b & 50 & 50 & 152 & -(0.95,0.31,0.08) &  -5 & 1.0 & \\
            1c & 60 & 10 & 122 & -(0.99,0.07,0.09) &  15 & 1.2 & \\
            5c & 60 & 10 & 132 & -(0.98,0.15,0.10) &  15 & 1.0 & \\
            1b & 50 & 50 & 162 & -(0.92,0.39,0.05) &  -5 & 1.0 & \\
            1c & 60 & 10 & 132 & -(0.98,0.15,0.10) &  15 & 1.2 & \\
            1c & 60 & -10 & 132 & -(0.98,0.15,0.10) &  10 & 1.1 & \\
            1c & 60 & 10 & 122 & -(0.99,0.07,0.09) &  15 & 1.1 & \\
            1c & 60 & 20 & 142 & -(0.97,0.24,0.09) &  10 & 1.2 & \\
            1b & 50 & 60 & 162 & -(0.92,0.39,0.05) &  -10 & 1.0 & \\
            5c & 60 & 0 & 152 & -(0.95,0.32,0.07) &  5 & 0.9 & \\
            1b & 50 & 50 & 172 & -(0.89,0.46,0.01) &  -15 & 1.0 & \\
            5c & 60 & 10 & 122 & -(0.99,0.07,0.09) &  15 & 0.8 & \\
            1b & 50 & 60 & 152 & -(0.95,0.31,0.08) &  0 & 1.2 & \\
            1c & 60 & 10 & 142 & -(0.97,0.24,0.09) &  5 & 1.1 & \\
            1b & 50 & 40 & 172 & -(0.89,0.46,0.01) &  -15 & 0.9 & \\
            5c & 60 & 0 & 132 & -(0.98,0.15,0.10) &  10 & 1.0 & \\
            1b & 50 & 60 & 162 & -(0.92,0.39,0.05) &  -10 & 1.1 & \\
            1c & 60 & 0 & 142 & -(0.97,0.24,0.09) &  5 & 1.2 & \\
            5c & 60 & 10 & 132 & -(0.98,0.15,0.10) &  10 & 0.8 & \\
            5c & 60 & 10 & 132 & -(0.98,0.15,0.10) &  10 & 0.9 & \\
            1c & 60 & -10 & 132 & -(0.98,0.15,0.10) &  15 & 1.2 & \\
            1b & 50 & 50 & 172 & -(0.89,0.46,0.01) &  -15 & 0.9 & \\
            1b & 50 & 50 & 152 & -(0.95,0.31,0.08) &  0 & 1.1 & \\
            1c & 60 & -10 & 142 & -(0.97,0.24,0.09) &  10 & 1.2 & \\
            1c & 60 & -10 & 132 & -(0.98,0.15,0.10) &  10 & 1.2 & \\
            1b & 50 & 50 & 162 & -(0.92,0.39,0.05) &  -10 & 1.1 & \\
            1b & 60 & 30 & 142 & -(0.97,0.24,0.09) &  5 & 1.1 & \\
            \hline
        \end{tabular}
         \]
         \begin{list}{}{}
          \item $^{\rm (a)}$ angle between the direction of the galaxy's motion and the galactic disk
          \item $^{\rm (b)}$ time from peak ram pressure
          \item $^{\rm (c)}$ azimuthal viewing angle 
      \end{list}
\end{table*}

\begin{figure*}[!ht]
  \centering
  \resizebox{14cm}{!}{\includegraphics{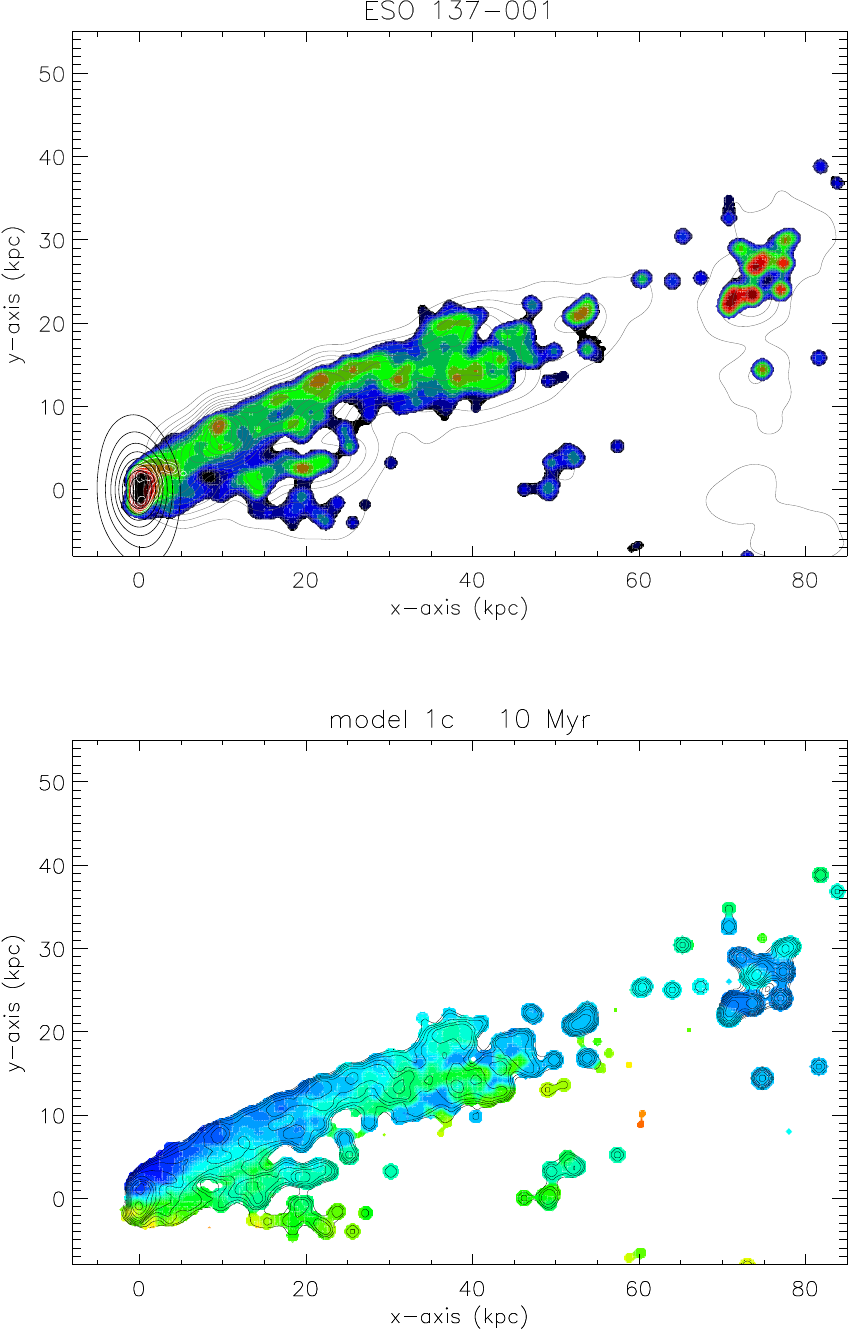}\includegraphics{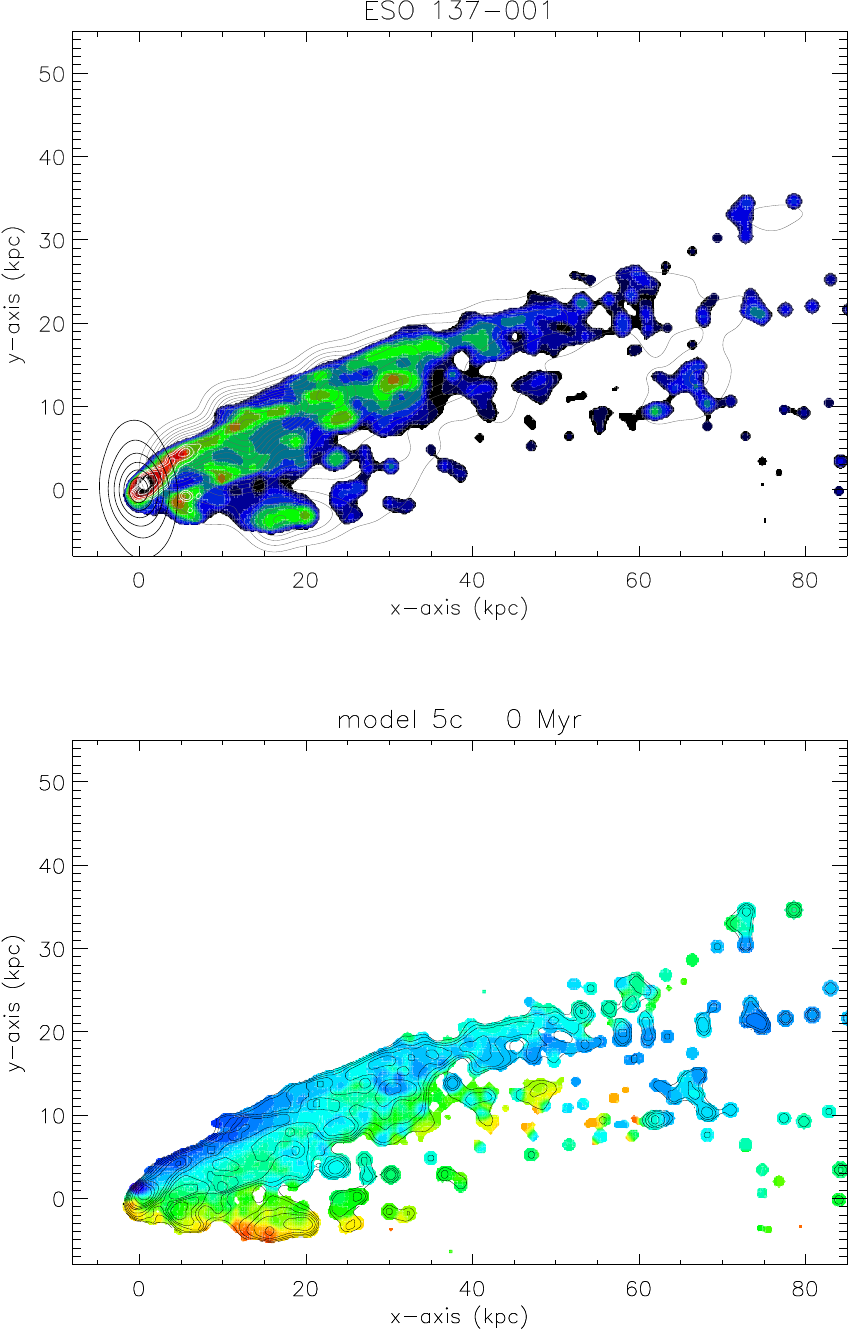}}
  \resizebox{14cm}{!}{\includegraphics{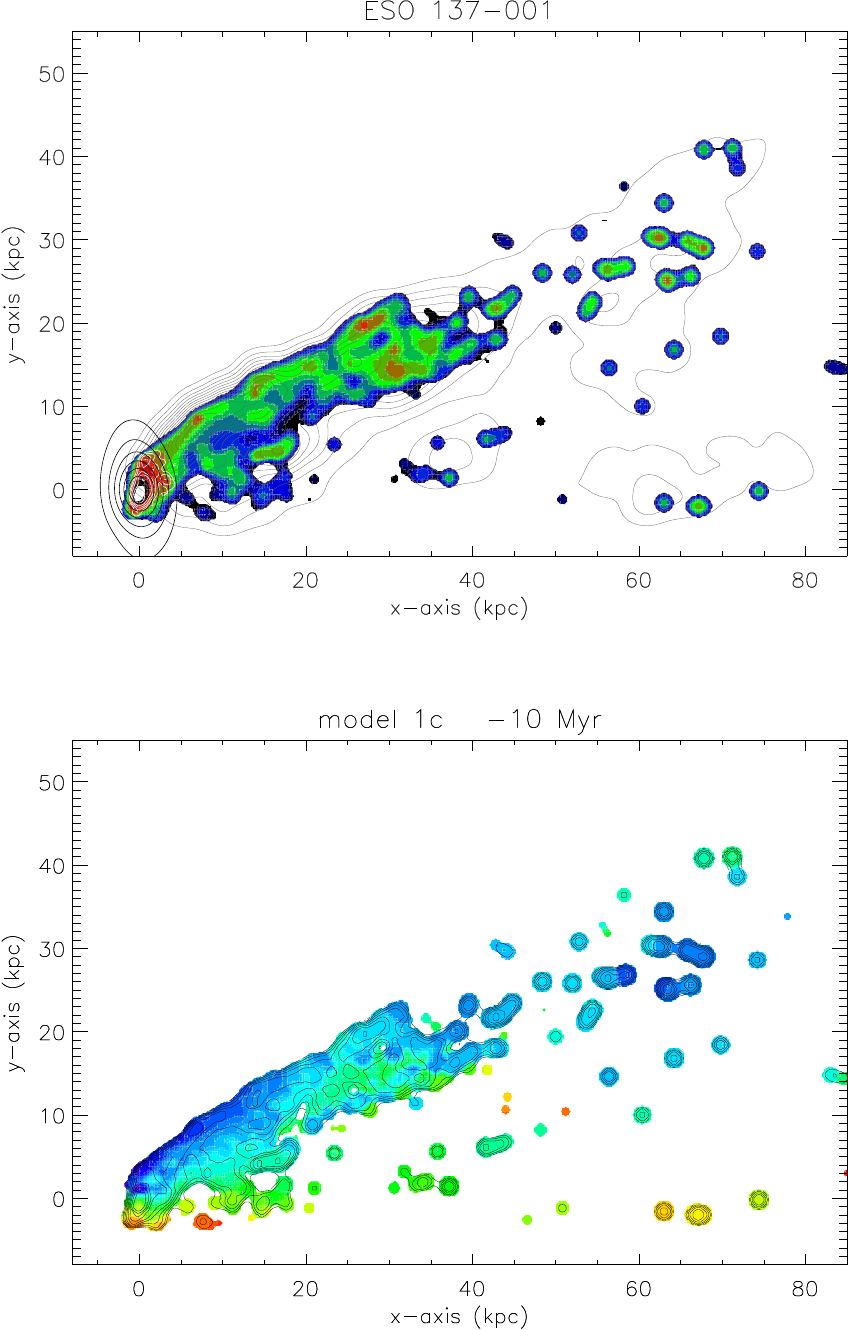}\includegraphics{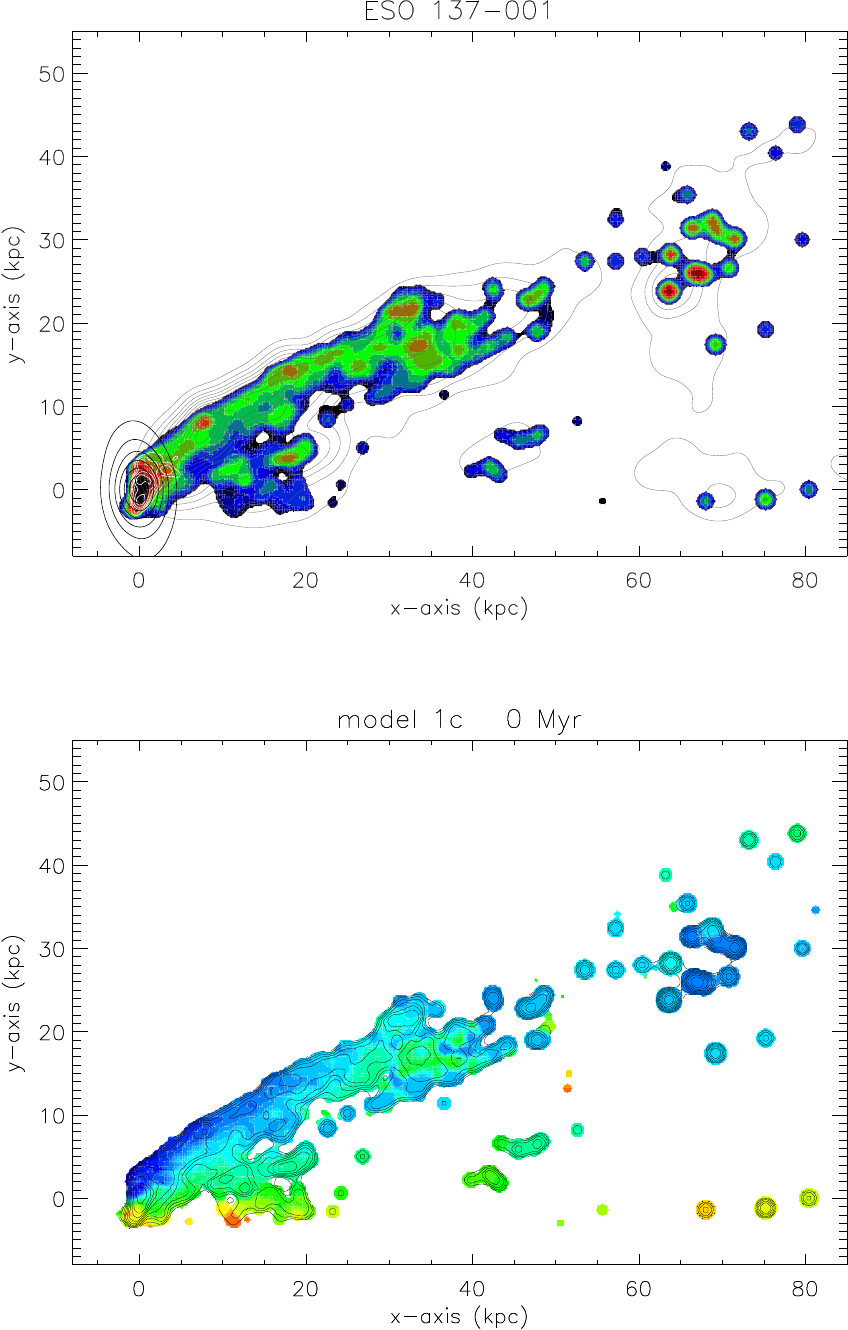}}
  \caption{High ICM-ISM mixing rate: highest-ranked near ram pressure peak models of ESO~137-001 based on CO, H$\alpha$, and X-ray. Upper panel: color: H$\alpha$; dark gray contours:
    X-ray; white contours: CO ; black contours: stellar content. Lower panel: H$\alpha$ velocity field.
  \label{fig:moreeso137-001allnovel}}
\end{figure*}

\newpage

\section{The influence of a lower ICM-ISM mixing rate}

\begin{table*}[!ht]
      \caption{Low ICM-ISM mixing rate: highest-ranked models based on CO, H$\alpha$, X-ray, and H$\alpha$ velocity field.}
         \label{tab:lessmolent}
      \[
         \begin{tabular}{lcrccrcl}
           \hline
          model & $\Theta^{\rm (a)}$ & $\Delta t^{\rm (b)}$ & $az^{\rm (c)}$ & velocity vector & PA rotation & expansion/ & \\
            & (degrees) & (Myr) & (degrees) & & (degrees) & shrinking \\
           \hline
             1c & 60 & 60 & 132 & -(0.98,0.15,0.10) &  10 & 1.1 & \\
             1c & 75 & 0 & 120 & -(0.95,0.04,0.32) &  15 & 1.0 & \\
             1c & 60 & 50 & 142 & -(0.97,0.24,0.09) &  10 & 1.2 & \\
             1c & 60 & 30 & 132 & -(0.98,0.15,0.10) &  10 & 1.1 & \\
             1c & 60 & 50 & 132 & -(0.98,0.15,0.10) &  10 & 1.1 & \\
             1c & 60 & 80 & 132 & -(0.98,0.15,0.10) &  10 & 1.2 & \\
             1a & 60 & 100 & 142 & -(0.97,0.24,0.09) &  10 & 1.1 & \\
             1c & 60 & 30 & 132 & -(0.98,0.15,0.10) &  5 & 1.1 & \\
             1c & 75 & 0 & 120 & -(0.95,0.04,0.32) &  15 & 0.8 & \\
             1c & 60 & 60 & 132 & -(0.98,0.15,0.10) &  5 & 1.0 & \\
             1c & 60 & 20 & 132 & -(0.98,0.15,0.10) &  10 & 1.1 & \\
             1c & 60 & 70 & 132 & -(0.98,0.15,0.10) &  5 & 1.0 & \\
             5c & 60 & 10 & 132 & -(0.98,0.15,0.10) &  15 & 0.9 & \\
             1c & 60 & 20 & 142 & -(0.97,0.24,0.09) &  -5 & 1.0 & \\
             1a & 60 & 110 & 142 & -(0.97,0.24,0.09) &  10 & 1.1 & \\
             1a & 60 & 20 & 132 & -(0.98,0.15,0.10) &  10 & 1.1 & \\
             1a & 60 & 30 & 132 & -(0.98,0.15,0.10) &  10 & 1.1 & \\
             1c & 60 & 20 & 132 & -(0.98,0.15,0.10) &  10 & 1.0 & \\
             1c & 60 & 70 & 132 & -(0.98,0.15,0.10) &  5 & 1.1 & \\
             1c & 75 & 0 & 140 & -(0.92,0.20,0.33) &  -10 & 0.9 & \\
             1c & 60 & 50 & 132 & -(0.98,0.15,0.10) &  10 & 1.2 & \\
             1c & 60 & 70 & 132 & -(0.98,0.15,0.10) &  10 & 1.2 & \\
             1b & 75 & 40 & 100 & -(0.96,.120,0.24) &  -5 & 0.8 & \\
             1c & 75 & 90 & 140 & -(0.92,0.20,0.33) &  0 & 0.9 & \\
             1c & 75 & 10 & 120 & -(0.95,0.04,0.32) &  15 & 1.0 & \\
             1a & 60 & 20 & 132 & -(0.98,0.15,0.10) &  0 & 0.9 & \\
             1a & 75 & 0 & 110 & -(0.96,.040,0.29) &  15 & 1.2 & \\
             1c & 60 & 80 & 132 & -(0.98,0.15,0.10) &  5 & 0.9 & \\
             5c & 60 & 20 & 132 & -(0.98,0.15,0.10) &  15 & 1.0 & \\
             1a & 60 & 50 & 142 & -(0.97,0.24,0.09) &  0 & 0.9 & \\
             1a & 75 & 40 & 140 & -(0.92,0.20,0.33) &  -10 & 0.9 & \\
             1a & 60 & 30 & 132 & -(0.98,0.15,0.10) &  10 & 1.0 & \\
             1c & 60 & 90 & 132 & -(0.98,0.15,0.10) &  5 & 1.0 & \\
             1c & 75 & 60 & 140 & -(0.92,0.20,0.33) &  -5 & 0.8 & \\
             1c & 75 & 0 & 130 & -(0.94,0.12,0.33) &  5 & 1.0 & \\
             3 & 60 & -30 & 132 & -(0.98,0.15,0.10) &  10 & 1.0 & \\
             1b & 75 & 40 & 100 & -(0.96,.120,0.24) &  -5 & 0.9 & \\
             1c & 60 & 90 & 142 & -(0.97,0.24,0.09) &  5 & 0.9 & \\
             1c & 60 & 80 & 132 & -(0.98,0.15,0.10) &  5 & 1.0 & \\
             1c & 60 & 80 & 132 & -(0.98,0.15,0.10) &  10 & 1.1 & \\
             1c & 75 & 80 & 140 & -(0.92,0.20,0.33) &  0 & 0.9 & \\
             1b & 75 & 30 & 140 & -(0.92,0.20,0.33) &  -10 & 0.9 & \\
             1c & 75 & 70 & 140 & -(0.92,0.20,0.33) &  -5 & 0.9 & \\
             1a & 60 & 100 & 132 & -(0.98,0.15,0.10) &  10 & 1.1 & \\
             1c & 60 & 20 & 132 & -(0.98,0.15,0.10) &  5 & 1.1 & \\
             1c & 60 & 90 & 132 & -(0.98,0.15,0.10) &  5 & 0.9 & \\
             5c & 60 & -10 & 132 & -(0.98,0.15,0.10) &  10 & 0.9 & \\
             1c & 60 & 90 & 132 & -(0.98,0.15,0.10) &  10 & 1.2 & \\
             1c & 75 & 0 & 120 & -(0.95,0.04,0.32) &  15 & 0.9 & \\
             1c & 60 & 100 & 142 & -(0.97,0.24,0.09) &  5 & 1.0 & \\
           \hline
        \end{tabular}
         \]
         \begin{list}{}{}
          \item $^{\rm (a)}$ angle between the direction of the galaxy's motion and the galactic disk
          \item $^{\rm (b)}$ time from peak ram pressure
          \item $^{\rm (c)}$ azimuthal viewing angle 
      \end{list}
\end{table*}

\begin{figure*}[!ht]
  \centering
  \resizebox{14cm}{!}{\includegraphics{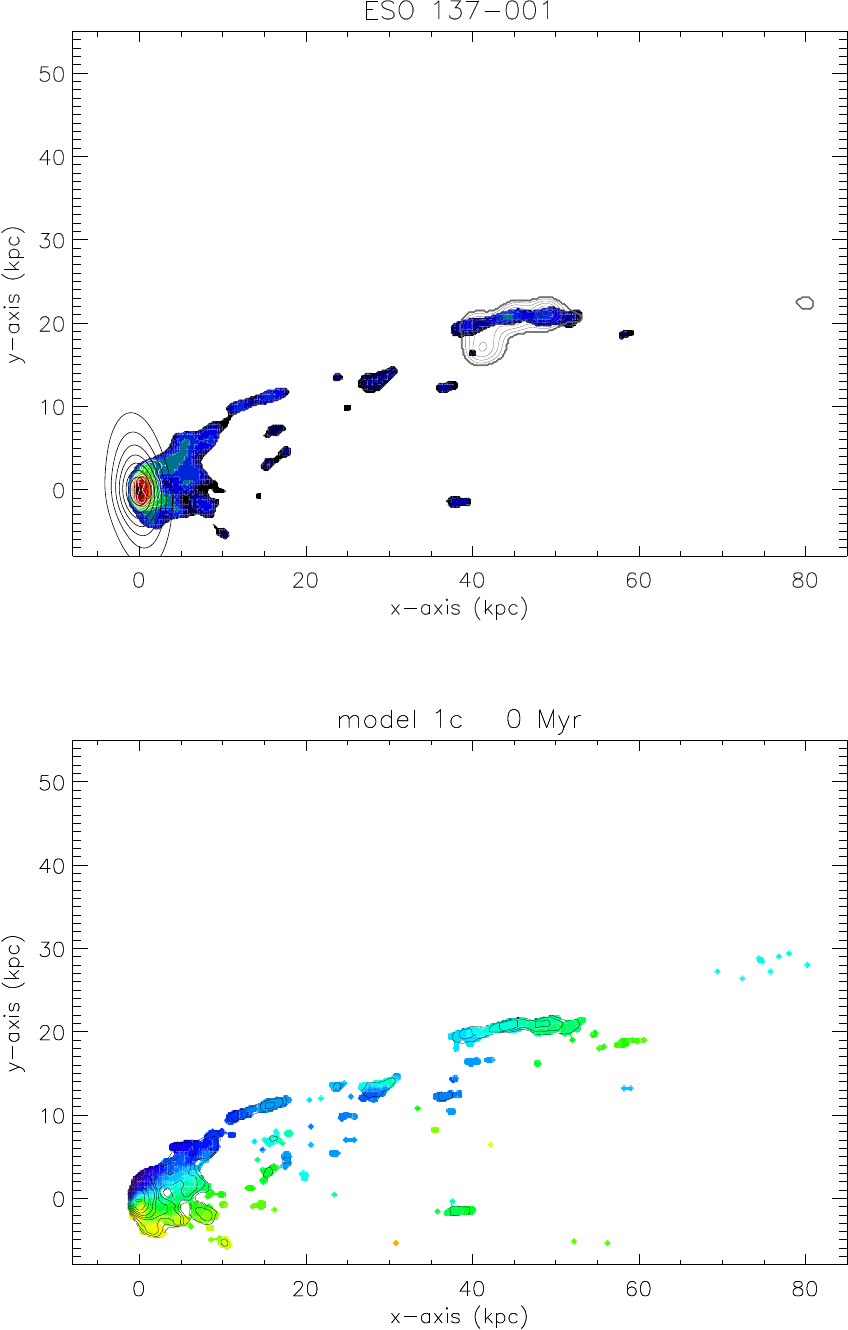}\includegraphics{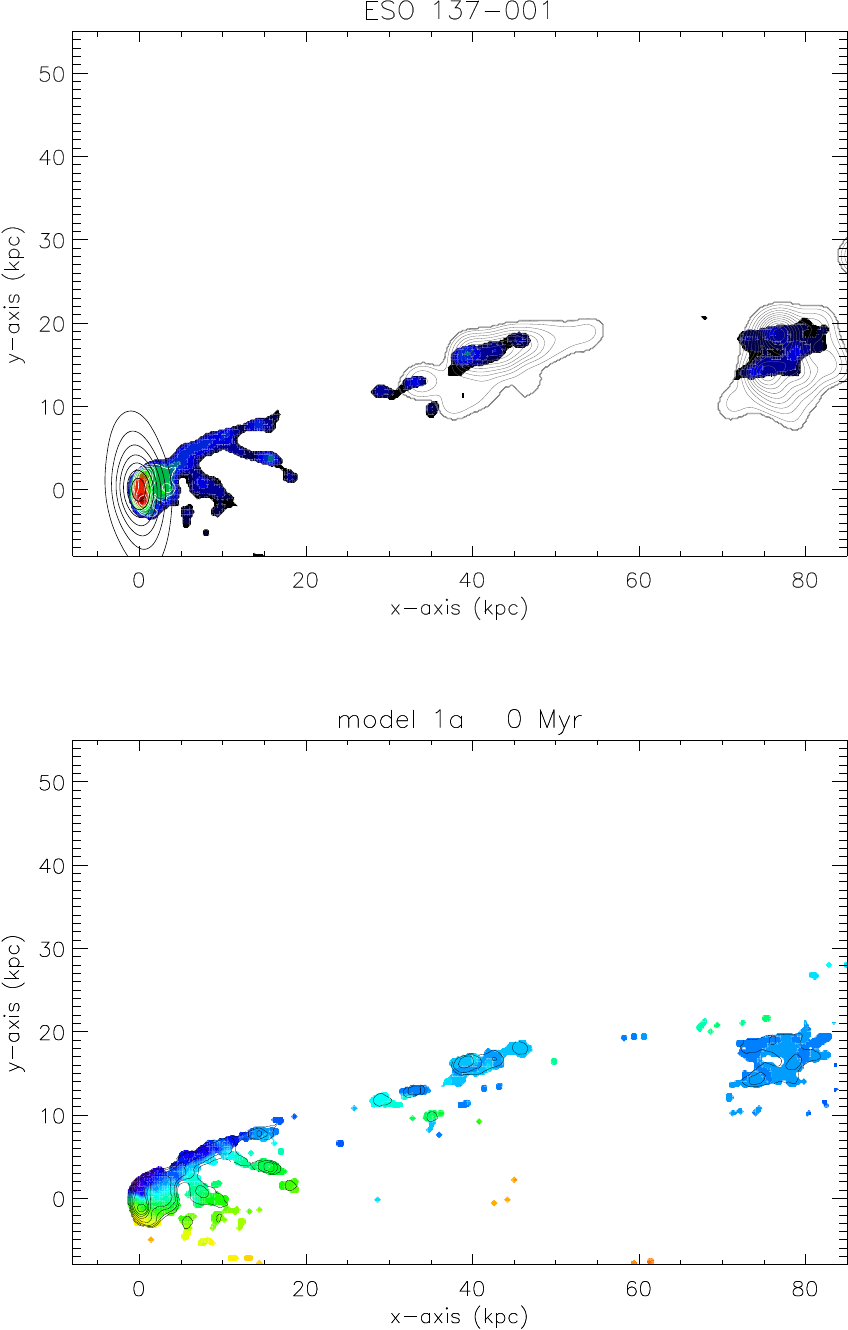}}
  \resizebox{14cm}{!}{\includegraphics{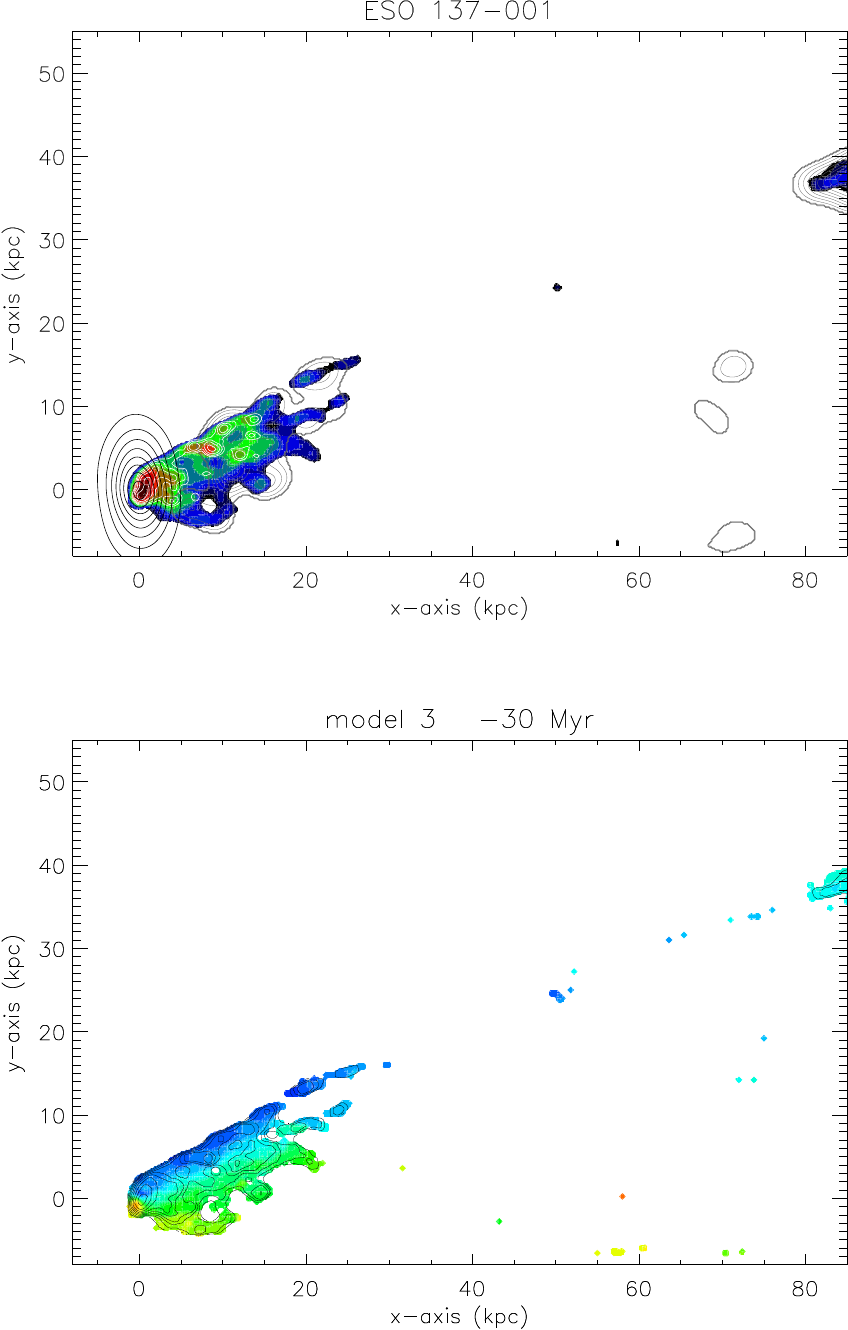}\includegraphics{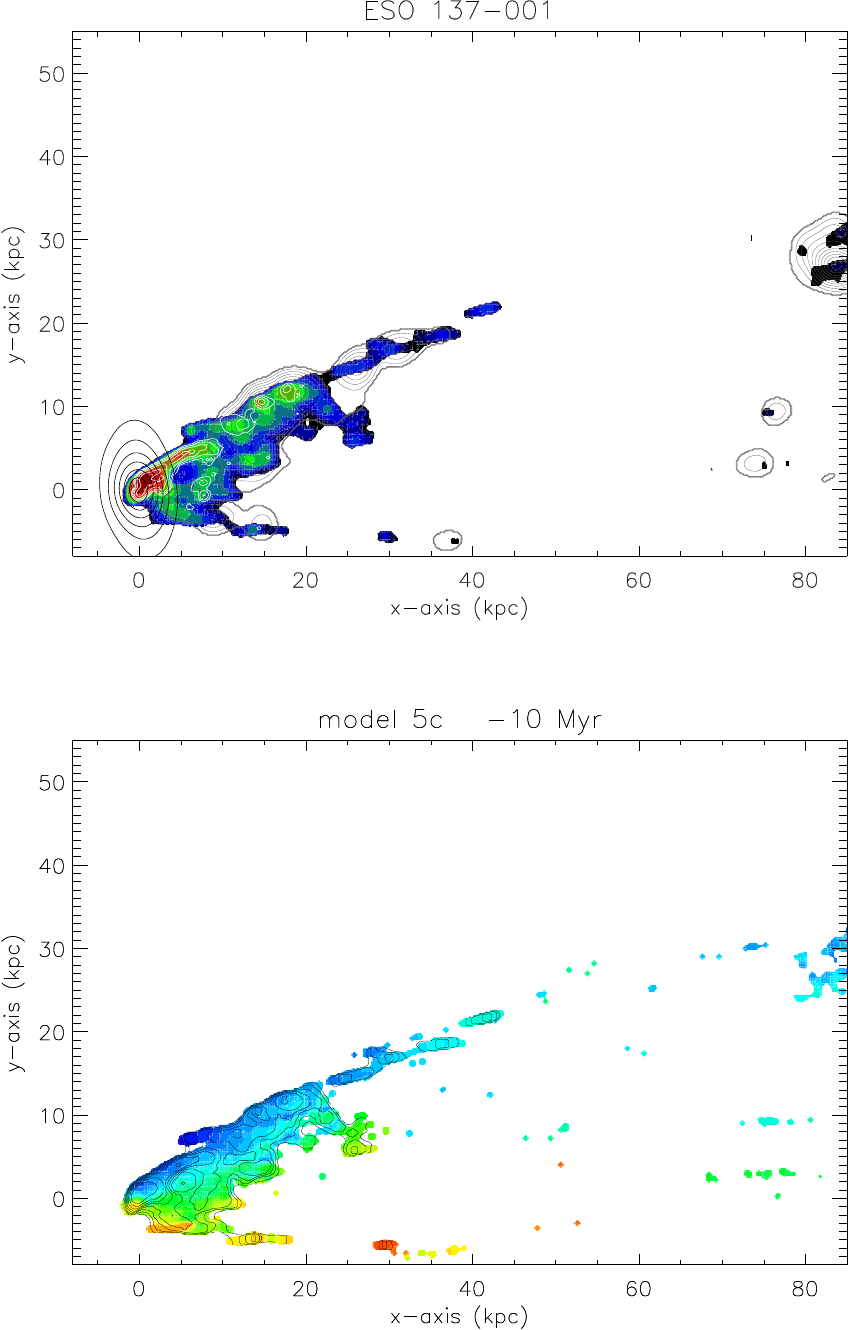}}
  \caption{Low ICM-ISM mixing rate: highest-ranked models of ESO~137-001 based on CO, H$\alpha$, X-ray, and H$\alpha$ velocity field. Upper panel: color: H$\alpha$; dark gray contours:
    X-ray; white contours: CO ; black contours: stellar content. Lower panel: H$\alpha$ velocity field.
  \label{fig:lesseso137-001all}}
\end{figure*}

\begin{table*}[!ht]
      \caption{Low ICM-ISM mixing rate: highest-ranked models based on the H$\alpha$ velocity field.}
         \label{tab:lessmolent1}
      \[
         \begin{tabular}{lcrccrcl}
           \hline
          model & $\Theta^{\rm (a)}$ & $\Delta t^{\rm (b)}$ & $az^{\rm (c)}$ & velocity vector & PA rotation & expansion/ & \\
            & (degrees) & (Myr) & (degrees) & & (degrees) & shrinking \\
           \hline
             1b & 75 & 20 & 130 & -(0.94,0.12,0.33) &  5 & 0.8 & \\
             1 & 75 & -20 & 140 & -(0.92,0.20,0.33) &  -10 & 0.9 & \\
             1b & 75 & 20 & 140 & -(0.92,0.20,0.33) &  -15 & 0.9 & \\
             1c & 75 & 0 & 120 & -(0.95,0.04,0.32) &  15 & 1.0 & \\
             1b & 75 & 30 & 140 & -(0.92,0.20,0.33) &  -15 & 0.8 & \\
             1a & 75 & 30 & 140 & -(0.92,0.20,0.33) &  -15 & 0.8 & \\
             1a & 60 & -10 & 142 & -(0.97,0.24,0.09) &  -5 & 1.2 & \\
             1a & 75 & 40 & 140 & -(0.92,0.20,0.33) &  -10 & 0.9 & \\
             1b & 75 & 30 & 140 & -(0.92,0.20,0.33) &  -10 & 0.8 & \\
             1b & 75 & 30 & 140 & -(0.92,0.20,0.33) &  -10 & 0.9 & \\
             1c & 75 & 10 & 140 & -(0.92,0.20,0.33) &  -10 & 0.8 & \\
             1c & 75 & 10 & 130 & -(0.94,0.12,0.33) &  5 & 1.2 & \\
             1a & 75 & 70 & 100 & -(0.96,.120,0.24) &  0 & 0.9 & \\
             5b & 60 & -50 & 142 & -(0.97,0.24,0.09) &  -5 & 1.1 & \\
             1a & 75 & -30 & 140 & -(0.92,0.20,0.33) &  -10 & 1.0 & \\
             1 & 75 & -10 & 140 & -(0.92,0.20,0.33) &  -10 & 1.0 & \\
             1b & 75 & 10 & 140 & -(0.92,0.20,0.33) &  -10 & 0.9 & \\
             1c & 75 & 10 & 140 & -(0.92,0.20,0.33) &  -10 & 0.9 & \\
             1b & 75 & 20 & 140 & -(0.92,0.20,0.33) &  -15 & 0.8 & \\
             5a & 60 & -20 & 132 & -(0.98,0.15,0.10) &  0 & 0.9 & \\
             1b & 75 & 40 & 100 & -(0.96,.120,0.24) &  -5 & 0.8 & \\
             1c & 75 & 100 & 140 & -(0.92,0.20,0.33) &  0 & 0.8 & \\
             1c & 60 & -60 & 142 & -(0.97,0.24,0.09) &  -10 & 1.2 & \\
             1 & 75 & -10 & 140 & -(0.92,0.20,0.33) &  -10 & 0.8 & \\
             1a & 75 & -30 & 140 & -(0.92,0.20,0.33) &  -10 & 0.9 & \\
             1 & 75 & -10 & 150 & -(0.90,0.29,0.32) &  -5 & 1.0 & \\
             1a & 60 & -10 & 132 & -(0.98,0.15,0.10) &  0 & 1.1 & \\
             1c & 75 & -80 & 140 & -(0.92,0.20,0.33) &  -15 & 1.0 & \\
             1c & 75 & -20 & 140 & -(0.92,0.20,0.33) &  -15 & 0.9 & \\
             1a & 75 & -20 & 140 & -(0.92,0.20,0.33) &  -15 & 1.0 & \\
             1c & 75 & 0 & 120 & -(0.95,0.04,0.32) &  15 & 1.1 & \\
             1 & 75 & -20 & 140 & -(0.92,0.20,0.33) &  -10 & 0.8 & \\
             1b & 75 & 50 & 100 & -(0.96,.120,0.24) &  0 & 0.8 & \\
             1b & 75 & 10 & 140 & -(0.92,0.20,0.33) &  -10 & 1.0 & \\
             5b & 60 & -50 & 132 & -(0.98,0.15,0.10) &  0 & 1.1 & \\
             1c & 75 & 0 & 140 & -(0.92,0.20,0.33) &  -15 & 1.2 & \\
             1c & 75 & 0 & 120 & -(0.95,0.04,0.32) &  15 & 0.8 & \\
             5b & 60 & -30 & 132 & -(0.98,0.15,0.10) &  0 & 1.2 & \\
             1a & 75 & -30 & 140 & -(0.92,0.20,0.33) &  -10 & 0.8 & \\
             1a & 75 & -20 & 140 & -(0.92,0.20,0.33) &  -15 & 0.9 & \\
             1c & 75 & 10 & 140 & -(0.92,0.20,0.33) &  -10 & 1.0 & \\
             1c & 75 & -70 & 140 & -(0.92,0.20,0.33) &  -15 & 1.2 & \\
             1b & 75 & 40 & 140 & -(0.92,0.20,0.33) &  -10 & 1.1 & \\
             1a & 50 & 40 & 172 & -(0.89,0.46,0.01) &  -15 & 1.2 & \\
             1c & 75 & -20 & 120 & -(0.95,0.04,0.32) &  15 & 1.2 & \\
             1b & 75 & 50 & 100 & -(0.96,.120,0.24) &  0 & 0.9 & \\
             1b & 75 & -50 & 140 & -(0.92,0.20,0.33) &  -15 & 1.2 & \\
             1a & 75 & 70 & 100 & -(0.96,.120,0.24) &  0 & 0.8 & \\
             1b & 75 & 20 & 140 & -(0.92,0.20,0.33) &  -15 & 1.0 & \\
             1b & 75 & -50 & 140 & -(0.92,0.20,0.33) &  -15 & 1.1 & \\
           \hline
        \end{tabular}
         \]
         \begin{list}{}{}
          \item $^{\rm (a)}$ angle between the direction of the galaxy's motion and the galactic disk
          \item $^{\rm (b)}$ time from peak ram pressure
          \item $^{\rm (c)}$ azimuthal viewing angle 
      \end{list}
\end{table*}

\begin{figure*}[!ht]
  \centering
  \resizebox{14cm}{!}{\includegraphics{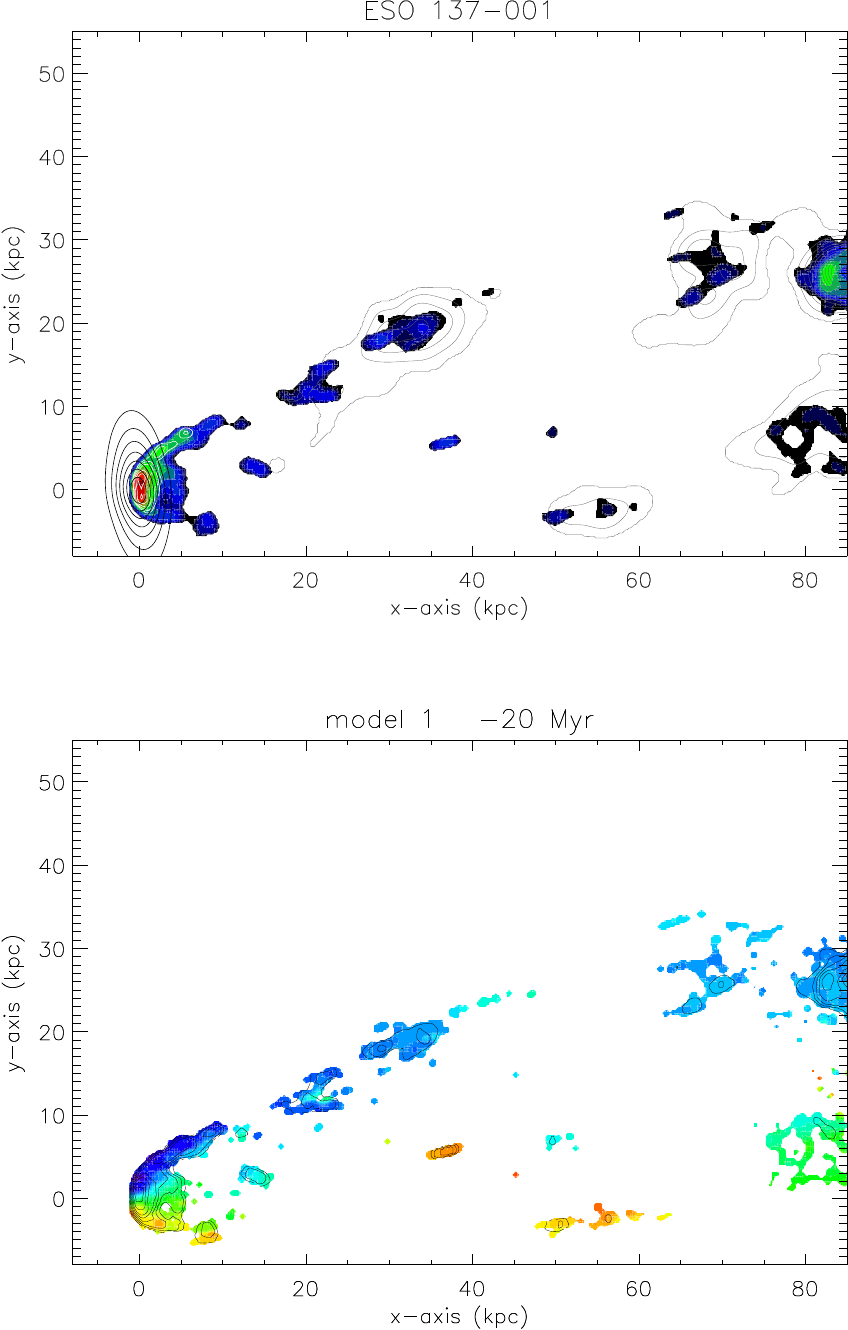}\includegraphics{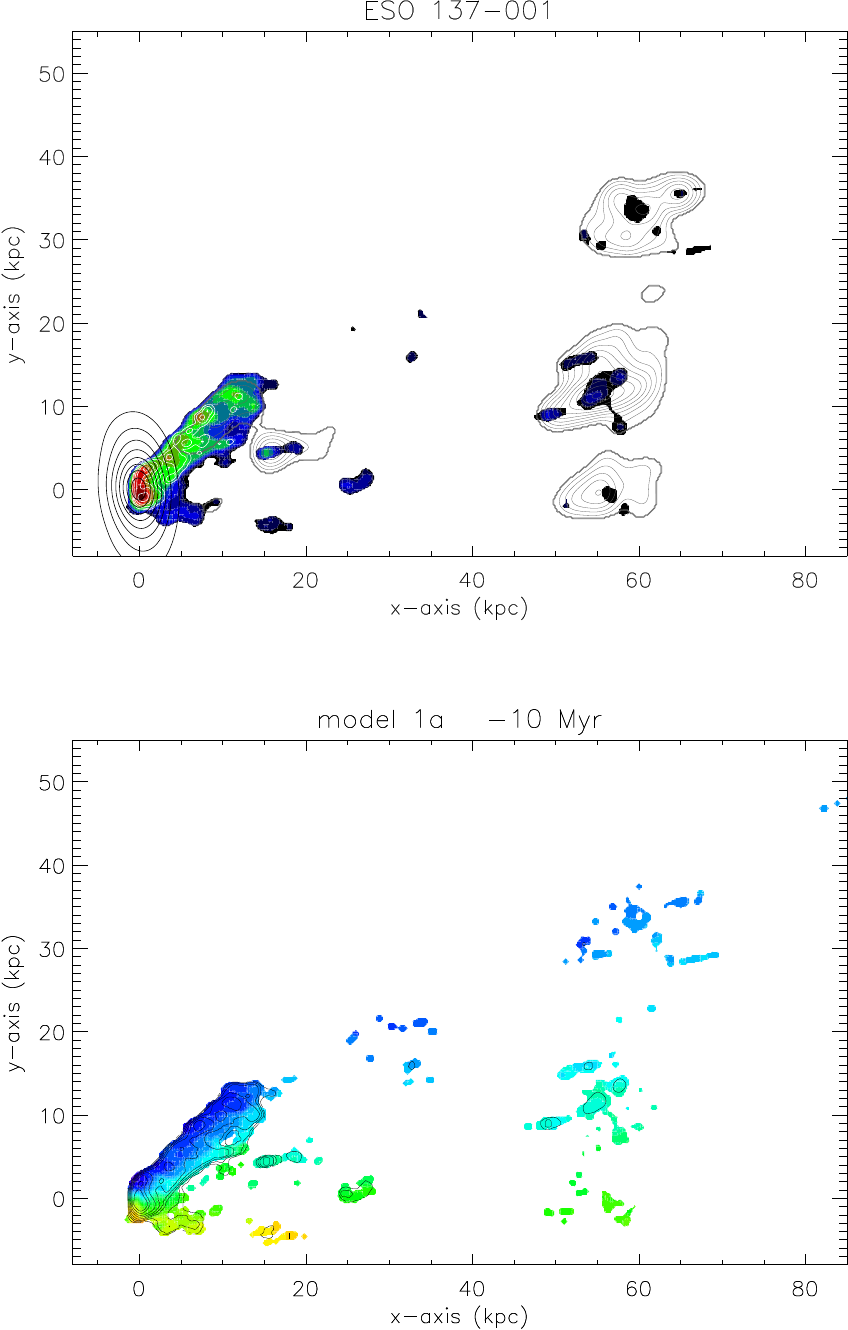}}
  \resizebox{14cm}{!}{\includegraphics{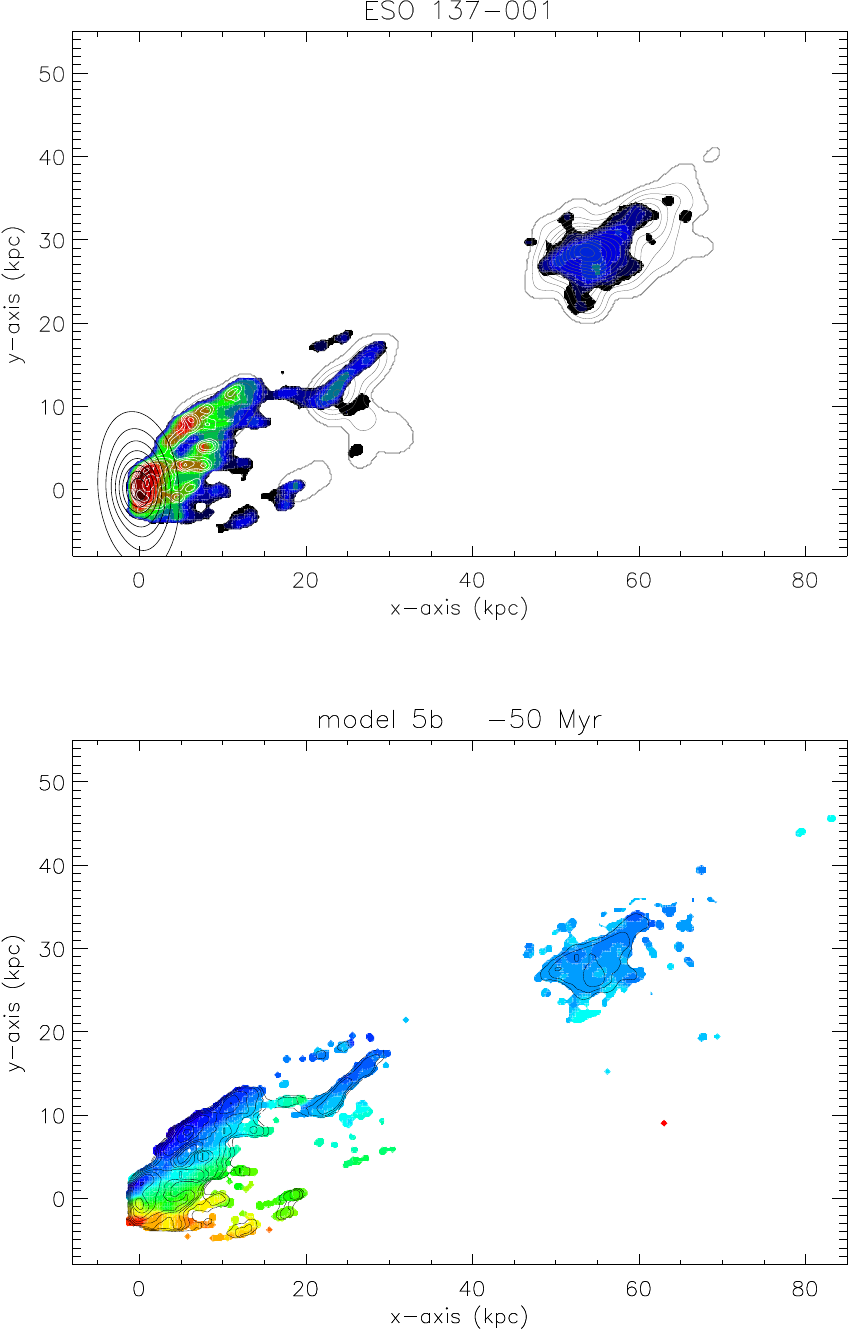}\includegraphics{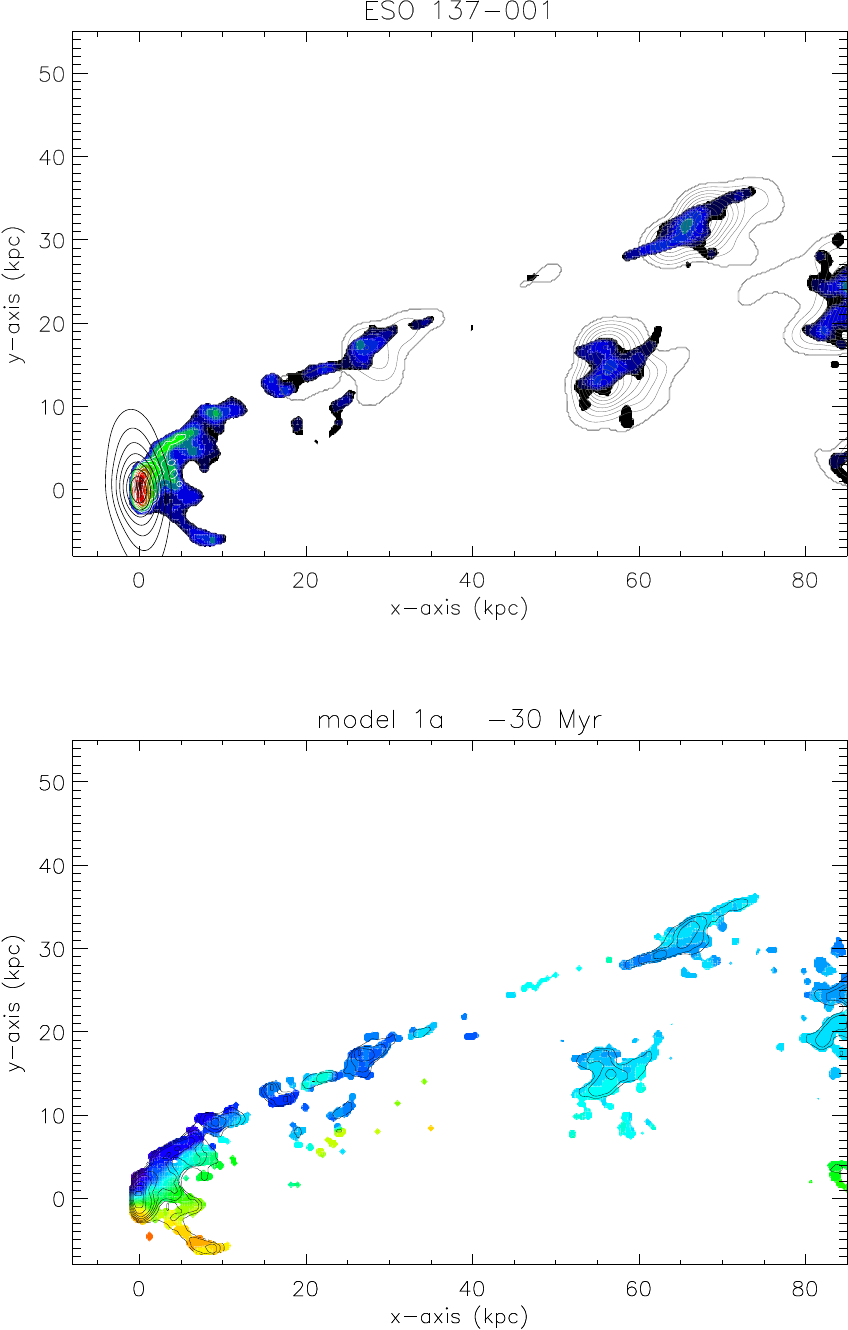}}
  \caption{Low ICM-ISM mixing rate: highest-ranked ram pressure pre-peak models of ESO~137-001 based on the H$\alpha$ velocity field. Upper panel: color: H$\alpha$; dark gray contours:
    X-ray; white contours: CO ; black contours: stellar content. Lower panel: H$\alpha$ velocity field.
  \label{fig:lesseso137-001allonlyvel}}
\end{figure*}

\begin{table*}[!ht]
      \caption{Low ICM-ISM mixing rate: highest-ranked models based on CO, H$\alpha$, and X-ray.}
         \label{tab:lessmolent2}
      \[
         \begin{tabular}{lcrccrcl}
           \hline
          model & $\Theta^{\rm (a)}$ & $\Delta t^{\rm (b)}$ & $az^{\rm (c)}$ & velocity vector & PA rotation & expansion/ & \\
            & (degrees) & (Myr) & (degrees) & & (degrees) & shrinking \\
           \hline
             1c & 60 & 60 & 152 & -(0.95,0.32,0.07) &  0 & 0.9 & \\
             3 & 60 & -30 & 142 & -(0.97,0.24,0.09) &  5 & 1.1 & \\
             3 & 60 & -30 & 132 & -(0.98,0.15,0.10) &  10 & 1.1 & \\
             1c & 60 & 60 & 152 & -(0.95,0.32,0.07) &  5 & 0.9 & \\
             1c & 60 & 60 & 152 & -(0.95,0.32,0.07) &  5 & 1.0 & \\
             1c & 60 & 50 & 152 & -(0.95,0.32,0.07) &  -5 & 1.0 & \\
             3 & 60 & -30 & 132 & -(0.98,0.15,0.10) &  10 & 1.0 & \\
             3 & 60 & -30 & 132 & -(0.98,0.15,0.10) &  15 & 1.1 & \\
             1a & 60 & 30 & 132 & -(0.98,0.15,0.10) &  5 & 1.1 & \\
             1c & 60 & 50 & 152 & -(0.95,0.32,0.07) &  0 & 0.9 & \\
             3 & 60 & -30 & 142 & -(0.97,0.24,0.09) &  5 & 1.2 & \\
             1c & 60 & 60 & 152 & -(0.95,0.32,0.07) &  10 & 1.0 & \\
             1c & 60 & 60 & 152 & -(0.95,0.32,0.07) &  -5 & 0.9 & \\
             1c & 60 & 50 & 152 & -(0.95,0.32,0.07) &  0 & 1.1 & \\
             1c & 60 & 20 & 142 & -(0.97,0.24,0.09) &  0 & 1.1 & \\
             1a & 60 & 30 & 132 & -(0.98,0.15,0.10) &  5 & 1.0 & \\
             1c & 60 & 60 & 152 & -(0.95,0.32,0.07) &  10 & 0.9 & \\
             1c & 60 & 50 & 152 & -(0.95,0.32,0.07) &  -5 & 0.9 & \\
             3 & 60 & -20 & 132 & -(0.98,0.15,0.10) &  10 & 1.1 & \\
             1c & 60 & 60 & 152 & -(0.95,0.32,0.07) &  -5 & 1.0 & \\
             1c & 60 & 60 & 152 & -(0.95,0.32,0.07) &  5 & 0.8 & \\
             1c & 60 & 60 & 152 & -(0.95,0.32,0.07) &  0 & 1.1 & \\
             1c & 60 & 60 & 152 & -(0.95,0.32,0.07) &  0 & 0.8 & \\
             1c & 60 & 60 & 152 & -(0.95,0.32,0.07) &  5 & 1.1 & \\
             1c & 60 & 20 & 152 & -(0.95,0.32,0.07) &  -5 & 1.0 & \\
             3 & 60 & -20 & 132 & -(0.98,0.15,0.10) &  15 & 1.1 & \\
             1a & 60 & 30 & 152 & -(0.95,0.32,0.07) &  0 & 1.1 & \\
             1a & 60 & 110 & 152 & -(0.95,0.32,0.07) &  15 & 0.9 & \\
             1c & 60 & 50 & 152 & -(0.95,0.32,0.07) &  5 & 1.0 & \\
             3 & 60 & -30 & 132 & -(0.98,0.15,0.10) &  10 & 1.2 & \\
             1c & 60 & 40 & 152 & -(0.95,0.32,0.07) &  -5 & 1.0 & \\
             3 & 60 & -30 & 152 & -(0.95,0.32,0.07) &  0 & 1.1 & \\
             1a & 60 & 20 & 152 & -(0.95,0.32,0.07) &  0 & 1.1 & \\
             1c & 75 & -10 & 140 & -(0.92,0.20,0.33) &  5 & 1.1 & \\
             1a & 60 & 30 & 152 & -(0.95,0.32,0.07) &  -5 & 1.0 & \\
             3 & 60 & -30 & 142 & -(0.97,0.24,0.09) &  10 & 1.2 & \\
             1c & 60 & 70 & 152 & -(0.95,0.32,0.07) &  10 & 1.0 & \\
             1c & 60 & 90 & 132 & -(0.98,0.15,0.10) &  15 & 1.0 & \\
             1a & 60 & 30 & 142 & -(0.97,0.24,0.09) &  0 & 1.1 & \\
             1c & 60 & 50 & 152 & -(0.95,0.32,0.07) &  5 & 1.1 & \\
             3 & 60 & -20 & 132 & -(0.98,0.15,0.10) &  10 & 1.2 & \\
             1c & 60 & 60 & 152 & -(0.95,0.32,0.07) &  -5 & 0.8 & \\
             1c & 60 & 30 & 152 & -(0.95,0.32,0.07) &  0 & 1.1 & \\
             1c & 60 & 90 & 112 & -(1.00,.020,0.06) &  15 & 0.9 & \\
             1c & 60 & 90 & 122 & -(0.99,0.07,0.09) &  15 & 0.9 & \\
             1c & 60 & 90 & 132 & -(0.98,0.15,0.10) &  10 & 1.0 & \\
             1a & 60 & 30 & 132 & -(0.98,0.15,0.10) &  10 & 1.1 & \\
             1c & 60 & 60 & 152 & -(0.95,0.32,0.07) &  10 & 1.1 & \\
             1c & 60 & 90 & 122 & -(0.99,0.07,0.09) &  15 & 1.0 & \\
             1c & 60 & 40 & 152 & -(0.95,0.32,0.07) &  5 & 1.1 & \\
           \hline
        \end{tabular}
         \]
         \begin{list}{}{}
          \item $^{\rm (a)}$ angle between the direction of the galaxy's motion and the galactic disk
          \item $^{\rm (b)}$ time from peak ram pressure
          \item $^{\rm (c)}$ azimuthal viewing angle 
      \end{list}
\end{table*}

\begin{figure*}[!ht]
  \centering
  \resizebox{14cm}{!}{\includegraphics{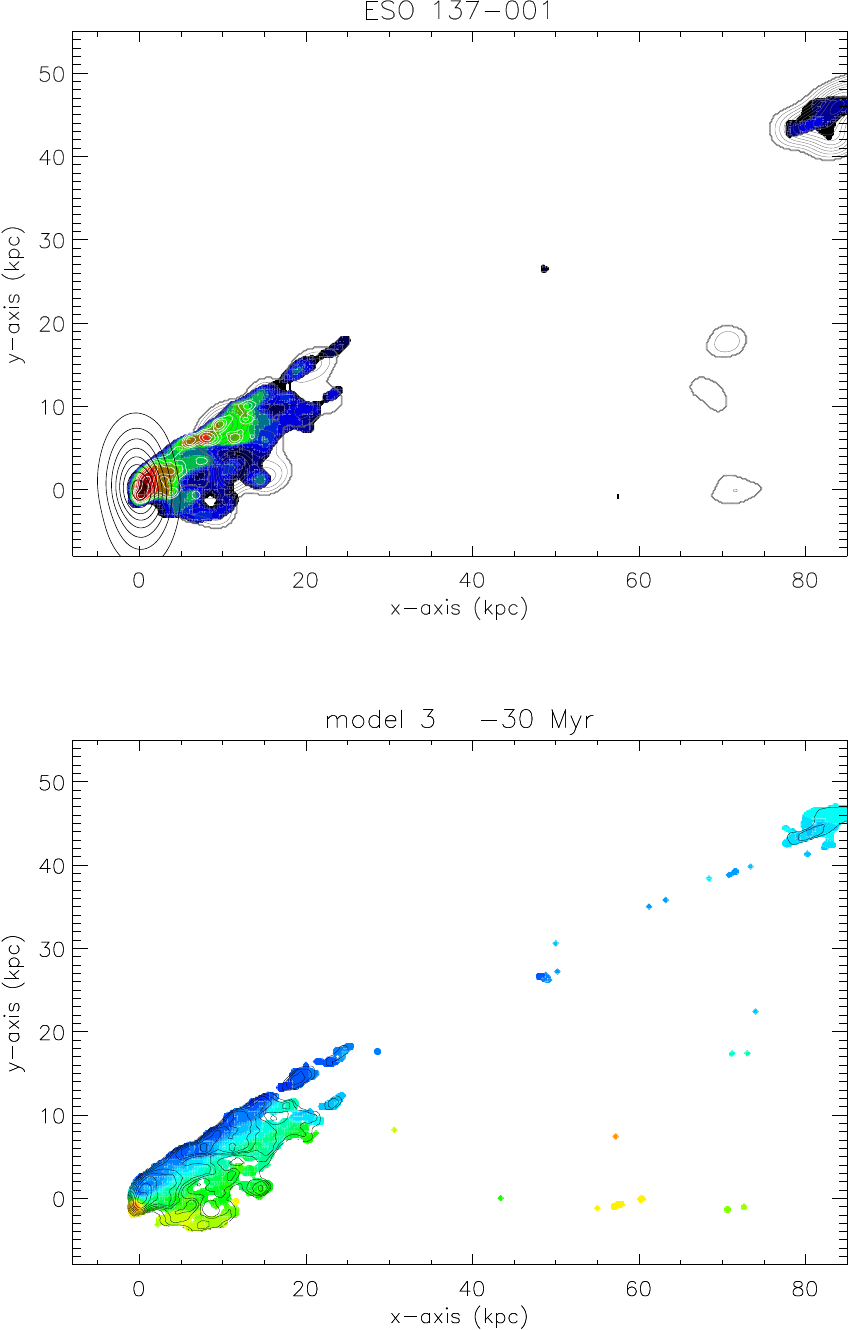}\includegraphics{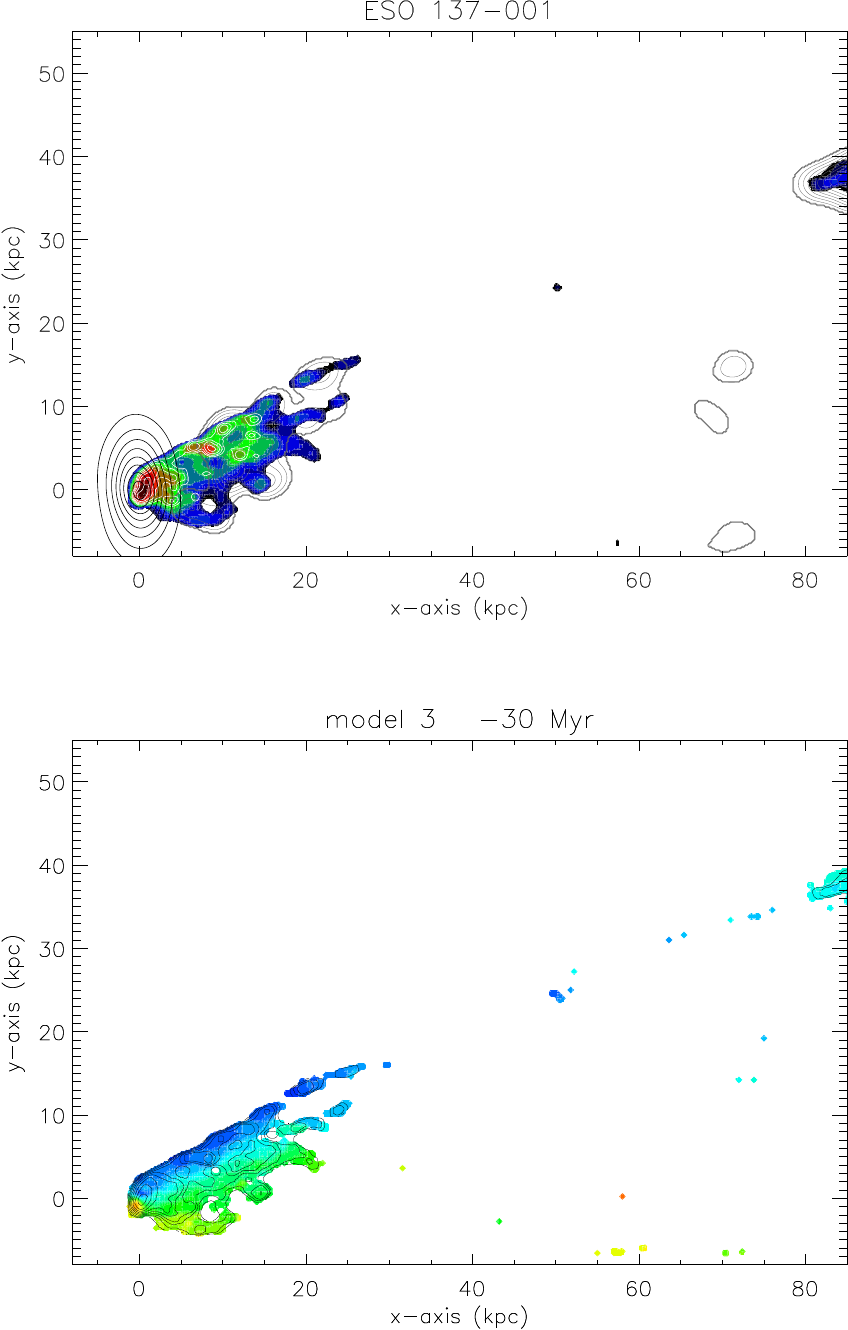}}
  \resizebox{14cm}{!}{\includegraphics{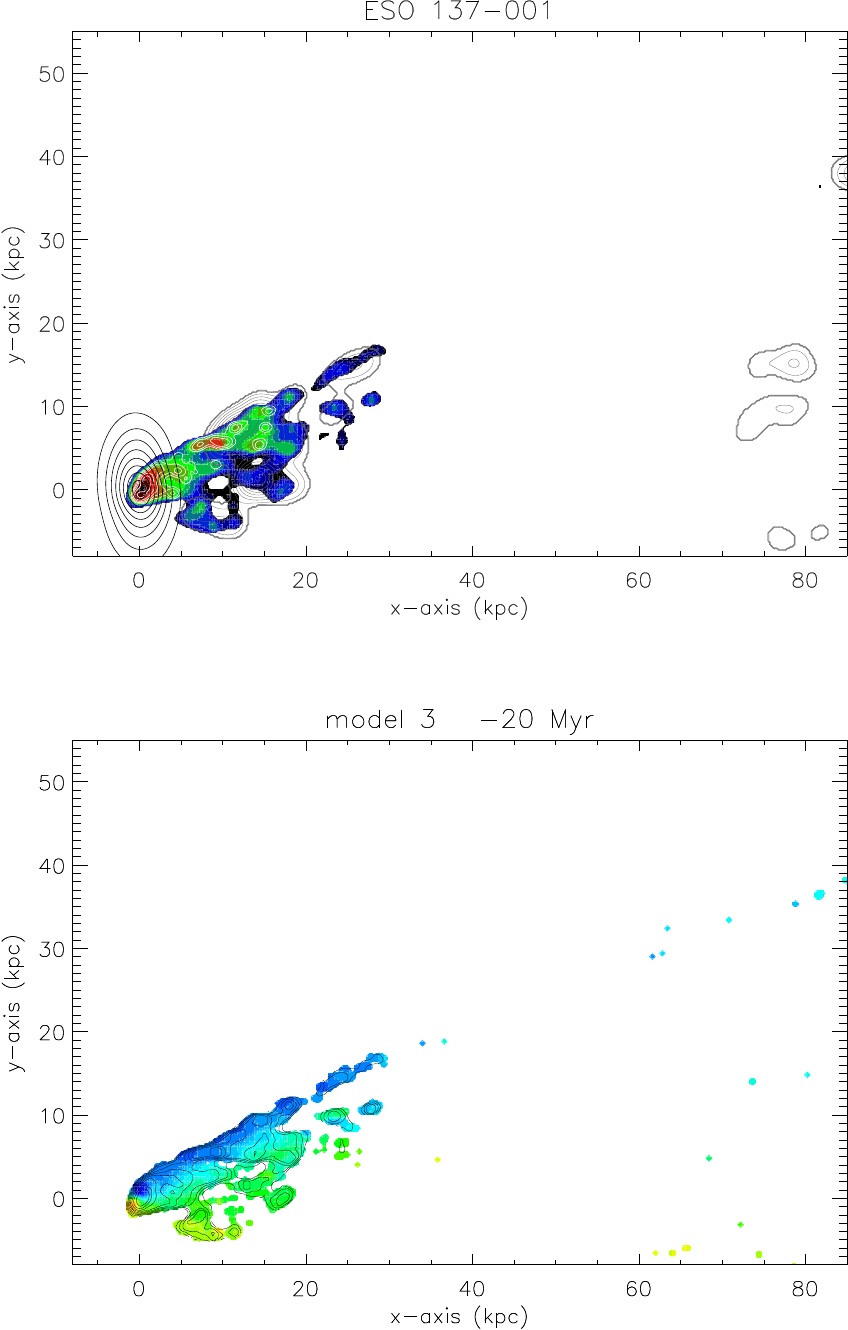}\includegraphics{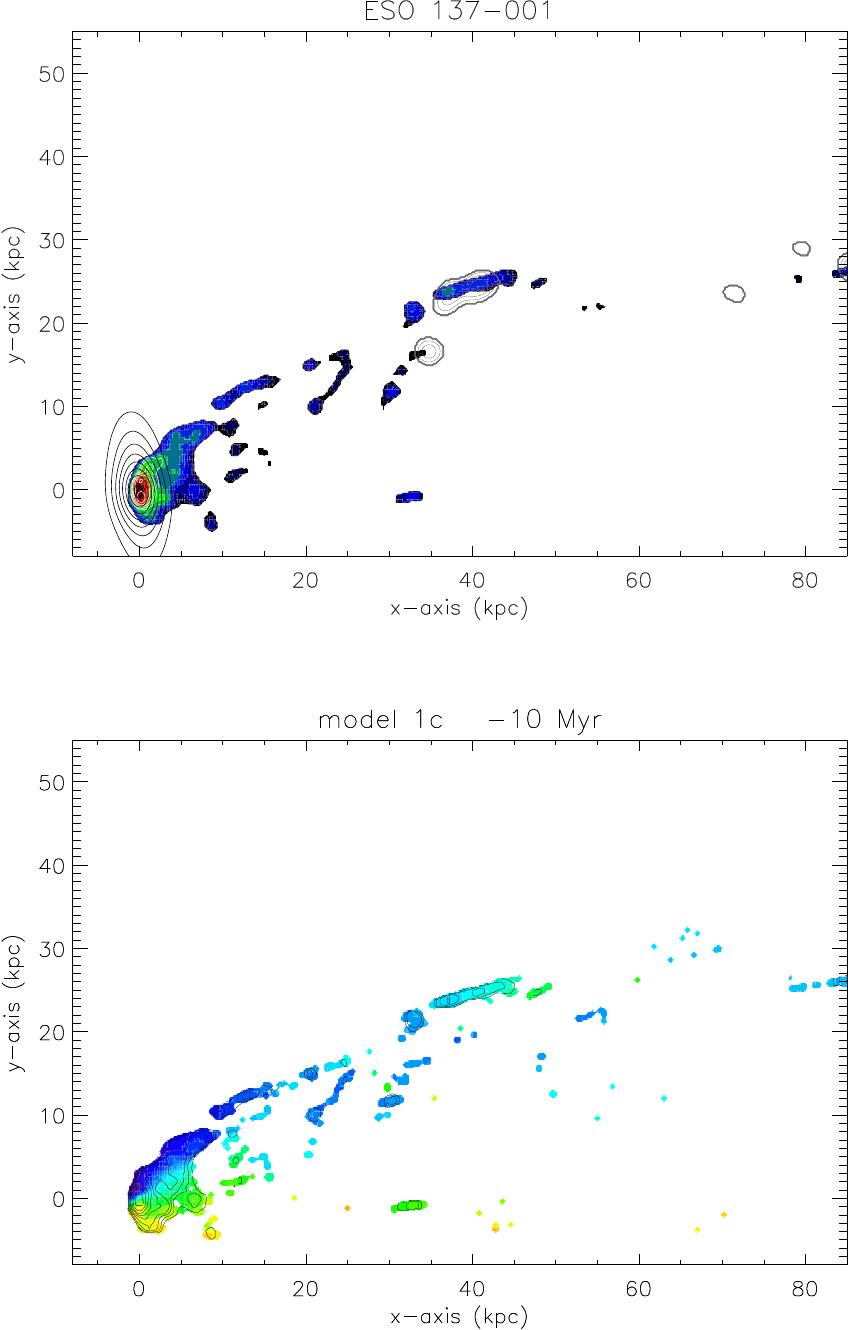}}
  \caption{Low ICM-ISM mixing rate: highest-ranked ram pressure pre-peak models of ESO~137-001 based on CO, H$\alpha$, and X-ray. Upper panel: color: H$\alpha$; dark gray contours:
    X-ray; white contours: CO ; black contours: stellar content. Lower panel: H$\alpha$ velocity field.
  \label{fig:lesseso137-001allnovel}}
\end{figure*}

\newpage

\section{The influence of a lower stripping efficiency of the hot ISM}

\begin{table*}[!ht]
      \caption{Lower stripping efficiency: highest-ranked models based on CO, H$\alpha$, X-ray, and H$\alpha$ velocity field.}
         \label{tab:xpushmolent}
      \[
         \begin{tabular}{lcrccrcl}
           \hline
          model & $\Theta^{\rm (a)}$ & $\Delta t^{\rm (b)}$ & $az^{\rm (c)}$ & velocity vector & PA rotation & expansion/ & \\
            & (degrees) & (Myr) & (degrees) & & (degrees) & shrinking \\
           \hline
             1b & 60 & 110 & 142 & -(0.97,0.24,0.09) &  10 & 1.0 & \\
             1a & 50 & 70 & 172 & -(0.89,0.46,0.01) &  -15 & 0.9 & \\
             1b & 50 & 40 & 182 & -(0.86,0.52,.050) &  -15 & 0.9 & \\
             1b & 60 & 110 & 132 & -(0.98,0.15,0.10) &  10 & 1.1 & \\
             1b & 60 & 120 & 132 & -(0.98,0.15,0.10) &  10 & 1.1 & \\
             1a & 50 & 70 & 152 & -(0.95,0.31,0.08) &  -15 & 0.8 & \\
             1a & 50 & 100 & 182 & -(0.86,0.52,.050) &  -10 & 1.0 & \\
             5a & 60 & 80 & 142 & -(0.97,0.24,0.09) &  15 & 0.9 & \\
             1c & 60 & 80 & 132 & -(0.98,0.15,0.10) &  5 & 1.0 & \\
             1c & 60 & 80 & 132 & -(0.98,0.15,0.10) &  5 & 1.1 & \\
             1a & 50 & 70 & 172 & -(0.89,0.46,0.01) &  -15 & 0.8 & \\
             1a & 50 & 90 & 152 & -(0.95,0.31,0.08) &  -10 & 0.9 & \\
             1a & 50 & 70 & 172 & -(0.89,0.46,0.01) &  -10 & 0.9 & \\
             1c & 60 & 70 & 132 & -(0.98,0.15,0.10) &  10 & 1.0 & \\
             1c & 60 & 80 & 132 & -(0.98,0.15,0.10) &  10 & 0.8 & \\
             1c & 60 & 80 & 132 & -(0.98,0.15,0.10) &  10 & 0.9 & \\
             1b & 60 & 60 & 142 & -(0.97,0.24,0.09) &  0 & 1.1 & \\
             1c & 60 & 90 & 132 & -(0.98,0.15,0.10) &  5 & 1.0 & \\
             5a & 60 & 80 & 132 & -(0.98,0.15,0.10) &  15 & 0.9 & \\
             1b & 60 & 110 & 152 & -(0.95,0.32,0.07) &  10 & 1.1 & \\
             1c & 60 & 80 & 142 & -(0.97,0.24,0.09) &  0 & 0.9 & \\
             1b & 60 & 110 & 142 & -(0.97,0.24,0.09) &  5 & 0.9 & \\
             1b & 50 & 50 & 152 & -(0.95,0.31,0.08) &  -15 & 0.9 & \\
             1b & 50 & 60 & 152 & -(0.95,0.31,0.08) &  -15 & 0.9 & \\
             1b & 60 & 110 & 152 & -(0.95,0.32,0.07) &  10 & 1.0 & \\
             1a & 50 & 70 & 152 & -(0.95,0.31,0.08) &  -15 & 0.9 & \\
             1b & 60 & 110 & 132 & -(0.98,0.15,0.10) &  10 & 1.0 & \\
             1a & 60 & 160 & 142 & -(0.97,0.24,0.09) &  10 & 1.2 & \\
             1a & 50 & 60 & 172 & -(0.89,0.46,0.01) &  -15 & 0.9 & \\
             1b & 60 & 110 & 152 & -(0.95,0.32,0.07) &  10 & 1.2 & \\
             1c & 60 & 80 & 132 & -(0.98,0.15,0.10) &  10 & 1.0 & \\
             1c & 60 & 90 & 132 & -(0.98,0.15,0.10) &  5 & 1.1 & \\
             1a & 50 & 80 & 182 & -(0.86,0.52,.050) &  -15 & 0.8 & \\
             1b & 60 & 100 & 152 & -(0.95,0.32,0.07) &  10 & 1.1 & \\
             1c & 60 & 90 & 152 & -(0.95,0.32,0.07) &  10 & 1.1 & \\
             1a & 50 & 80 & 152 & -(0.95,0.31,0.08) &  -10 & 0.8 & \\
             1c & 60 & 90 & 132 & -(0.98,0.15,0.10) &  5 & 0.9 & \\
             7c & 60 & 100 & 152 & -(0.95,0.32,0.07) &  10 & 0.8 & \\
             1c & 60 & 80 & 132 & -(0.98,0.15,0.10) &  5 & 1.2 & \\
             1 & 50 & 80 & 192 & -(0.82,0.56,.110) &  -10 & 0.9 & \\
             1 & 50 & 90 & 182 & -(0.86,0.52,.050) &  -10 & 0.8 & \\
             2 & 60 & -20 & 132 & -(0.98,0.15,0.10) &  10 & 1.0 & \\
             1a & 60 & 150 & 142 & -(0.97,0.24,0.09) &  10 & 1.1 & \\
             1b & 50 & 40 & 182 & -(0.86,0.52,.050) &  -15 & 1.0 & \\
             1 & 50 & 100 & 172 & -(0.89,0.46,0.01) &  -15 & 0.8 & \\
             1b & 50 & 50 & 172 & -(0.89,0.46,0.01) &  -15 & 0.9 & \\
             1a & 60 & 150 & 142 & -(0.97,0.24,0.09) &  10 & 1.2 & \\
             2 & 60 & -10 & 132 & -(0.98,0.15,0.10) &  10 & 0.8 & \\
             1b & 50 & 60 & 152 & -(0.95,0.31,0.08) &  -15 & 0.8 & \\
             1b & 60 & 120 & 142 & -(0.97,0.24,0.09) &  10 & 1.0 & \\
           \hline
        \end{tabular}
         \]
         \begin{list}{}{}
          \item $^{\rm (a)}$ angle between the direction of the galaxy's motion and the galactic disk
          \item $^{\rm (b)}$ time from peak ram pressure
          \item $^{\rm (c)}$ azimuthal viewing angle 
      \end{list}
\end{table*}

\begin{figure*}[!ht]
  \centering
  \resizebox{14cm}{!}{\includegraphics{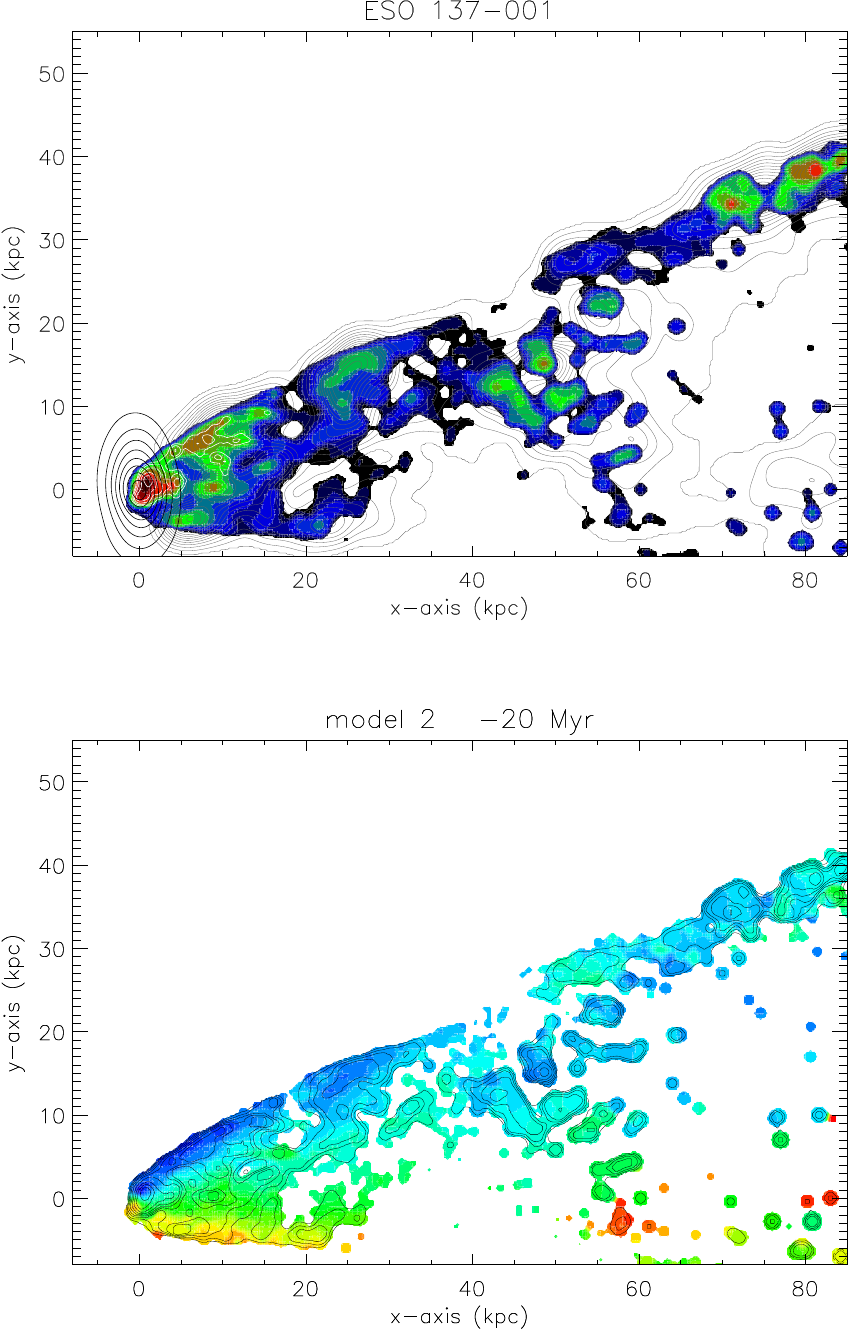}\includegraphics{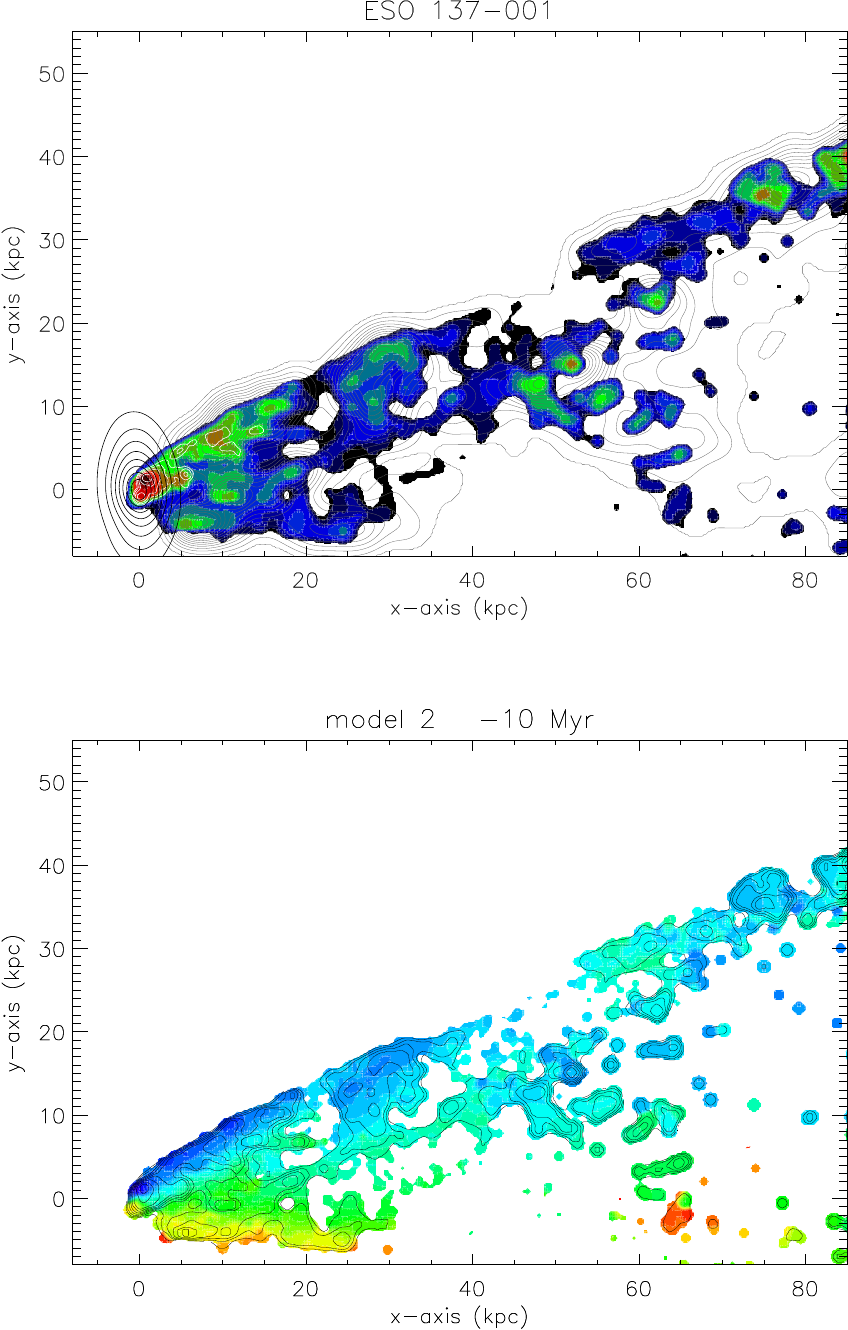}}
  \caption{Lower stripping efficiency: highest-ranked ram pressure pre-peak models of ESO~137-001 based on CO, H$\alpha$, X-ray, and H$\alpha$ velocity field. Upper panel: color: H$\alpha$; dark gray contours:
    X-ray; white contours: CO ; black contours: stellar content. Lower panel: H$\alpha$ velocity field.
  \label{fig:xpusheso137-001all}}
\end{figure*}

\begin{table*}[!ht]
      \caption{Lower stripping efficiency: highest-ranked models based on the H$\alpha$ velocity field.}
         \label{tab:xpushmolent1}
      \[
         \begin{tabular}{lcrccrcl}
           \hline
          model & $\Theta^{\rm (a)}$ & $\Delta t^{\rm (b)}$ & $az^{\rm (c)}$ & velocity vector & PA rotation & expansion/ & \\
            & (degrees) & (Myr) & (degrees) & & (degrees) & shrinking \\
           \hline
             1a & 75 & 30 & 140 & -(0.92,0.20,0.33) &  -10 & 0.8 & \\
             1c & 60 & 80 & 142 & -(0.97,0.24,0.09) &  10 & 0.8 & \\
             1a & 75 & -30 & 140 & -(0.92,0.20,0.33) &  -5 & 0.9 & \\
             1a & 75 & 40 & 140 & -(0.92,0.20,0.33) &  -15 & 0.8 & \\
             1 & 50 & 90 & 192 & -(0.82,0.56,.110) &  -10 & 1.1 & \\
             1b & 75 & 10 & 140 & -(0.92,0.20,0.33) &  -10 & 0.9 & \\
             1 & 50 & 90 & 192 & -(0.82,0.56,.110) &  -10 & 1.0 & \\
             1c & 60 & 80 & 142 & -(0.97,0.24,0.09) &  0 & 0.9 & \\
             1c & 60 & 80 & 132 & -(0.98,0.15,0.10) &  10 & 0.8 & \\
             7a & 60 & 20 & 132 & -(0.98,0.15,0.10) &  15 & 0.8 & \\
             1 & 50 & 100 & 172 & -(0.89,0.46,0.01) &  -15 & 0.8 & \\
             1 & 75 & -10 & 140 & -(0.92,0.20,0.33) &  -10 & 0.8 & \\
             1a & 50 & 30 & 182 & -(0.86,0.52,.050) &  -15 & 1.0 & \\
             1b & 75 & -50 & 140 & -(0.92,0.20,0.33) &  -15 & 1.0 & \\
             1b & 75 & 20 & 130 & -(0.94,0.12,0.33) &  10 & 1.0 & \\
             1b & 60 & 60 & 142 & -(0.97,0.24,0.09) &  0 & 1.1 & \\
             3 & 60 & -90 & 132 & -(0.98,0.15,0.10) &  10 & 1.2 & \\
             1 & 50 & 90 & 172 & -(0.89,0.46,0.01) &  15 & 1.0 & \\
             1b & 75 & -50 & 140 & -(0.92,0.20,0.33) &  -15 & 0.9 & \\
             1b & 75 & -50 & 140 & -(0.92,0.20,0.33) &  -15 & 1.2 & \\
             1a & 50 & 70 & 172 & -(0.89,0.46,0.01) &  -15 & 1.2 & \\
             1a & 75 & 30 & 140 & -(0.92,0.20,0.33) &  -15 & 0.8 & \\
             1b & 75 & 40 & 100 & -(0.96,.120,0.24) &  -5 & 0.9 & \\
             1b & 75 & 10 & 140 & -(0.92,0.20,0.33) &  -10 & 0.8 & \\
             1c & 75 & 30 & 100 & -(0.96,.120,0.24) &  -5 & 0.8 & \\
             1a & 75 & 20 & 140 & -(0.92,0.20,0.33) &  -10 & 0.8 & \\
             1a & 50 & 70 & 172 & -(0.89,0.46,0.01) &  -15 & 0.9 & \\
             3 & 60 & -100 & 152 & -(0.95,0.32,0.07) &  5 & 1.2 & \\
             1a & 75 & -40 & 140 & -(0.92,0.20,0.33) &  -10 & 1.0 & \\
             1a & 50 & 80 & 172 & -(0.89,0.46,0.01) &  -15 & 1.1 & \\
             1a & 50 & 60 & 172 & -(0.89,0.46,0.01) &  -15 & 1.1 & \\
             1 & 75 & -30 & 140 & -(0.92,0.20,0.33) &  -15 & 1.1 & \\
             1a & 75 & -20 & 140 & -(0.92,0.20,0.33) &  -10 & 1.1 & right panel of Fig.~\ref{fig:eso137-001all_mixing1_Xpushonlyvel_04} \\
             1 & 75 & -20 & 140 & -(0.92,0.20,0.33) &  -10 & 0.8 & \\
             1b & 75 & 20 & 130 & -(0.94,0.12,0.33) &  10 & 0.9 & \\
             1a & 75 & 40 & 140 & -(0.92,0.20,0.33) &  -15 & 0.9 & \\
             1b & 75 & 20 & 130 & -(0.94,0.12,0.33) &  5 & 0.8 & \\
             1a & 75 & 40 & 130 & -(0.94,0.12,0.33) &  10 & 1.0 & \\
             1b & 50 & 40 & 182 & -(0.86,0.52,.050) &  -15 & 0.9 & \\
             1a & 50 & 50 & 172 & -(0.89,0.46,0.01) &  -15 & 1.2 & \\
             1b & 75 & 0 & 140 & -(0.92,0.20,0.33) &  10 & 0.8 & left panel of Fig.~\ref{fig:eso137-001all_mixing1_Xpushonlyvel_04} \\
             1c & 75 & 0 & 140 & -(0.92,0.20,0.33) &  -15 & 0.9 & \\
             1c & 75 & 0 & 140 & -(0.92,0.20,0.33) &  -10 & 1.1 & \\
             1 & 50 & 90 & 182 & -(0.86,0.52,.050) &  -10 & 0.8 & \\
             1b & 75 & 30 & 140 & -(0.92,0.20,0.33) &  -15 & 0.9 & \\
             1b & 75 & -50 & 140 & -(0.92,0.20,0.33) &  -15 & 1.1 & \\
             1a & 75 & 50 & 100 & -(0.96,.120,0.24) &  -5 & 1.0 & \\
             1c & 75 & 10 & 140 & -(0.92,0.20,0.33) &  -10 & 1.0 & \\
             1c & 75 & 10 & 130 & -(0.94,0.12,0.33) &  5 & 1.1 & \\
             1b & 75 & 20 & 130 & -(0.94,0.12,0.33) &  5 & 1.2 & \\
           \hline
        \end{tabular}
         \]
         \begin{list}{}{}
          \item $^{\rm (a)}$ angle between the direction of the galaxy's motion and the galactic disk
          \item $^{\rm (b)}$ time from peak ram pressure
          \item $^{\rm (c)}$ azimuthal viewing angle 
      \end{list}
\end{table*}

\begin{figure*}[!ht]
  \centering
  \resizebox{14cm}{!}{\includegraphics{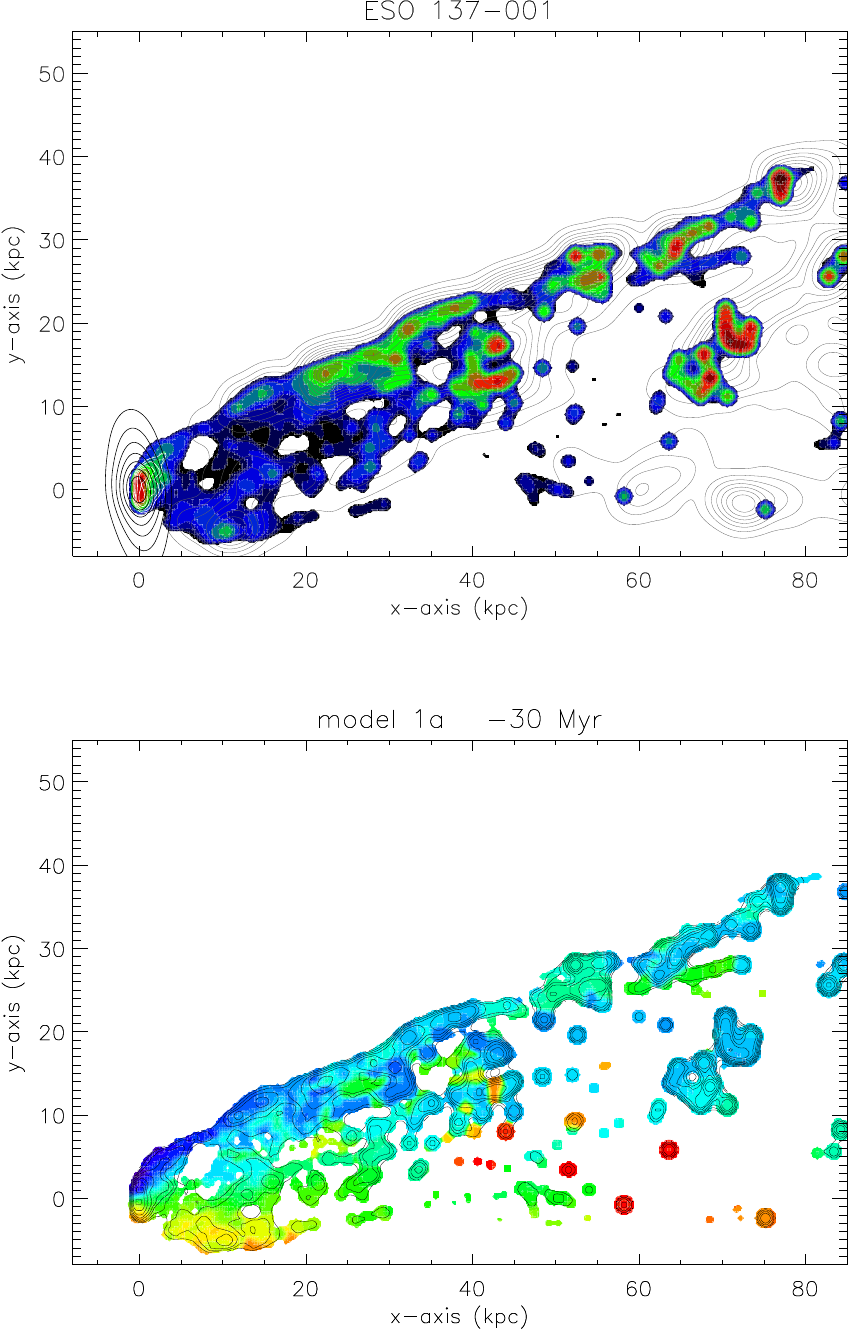}\includegraphics{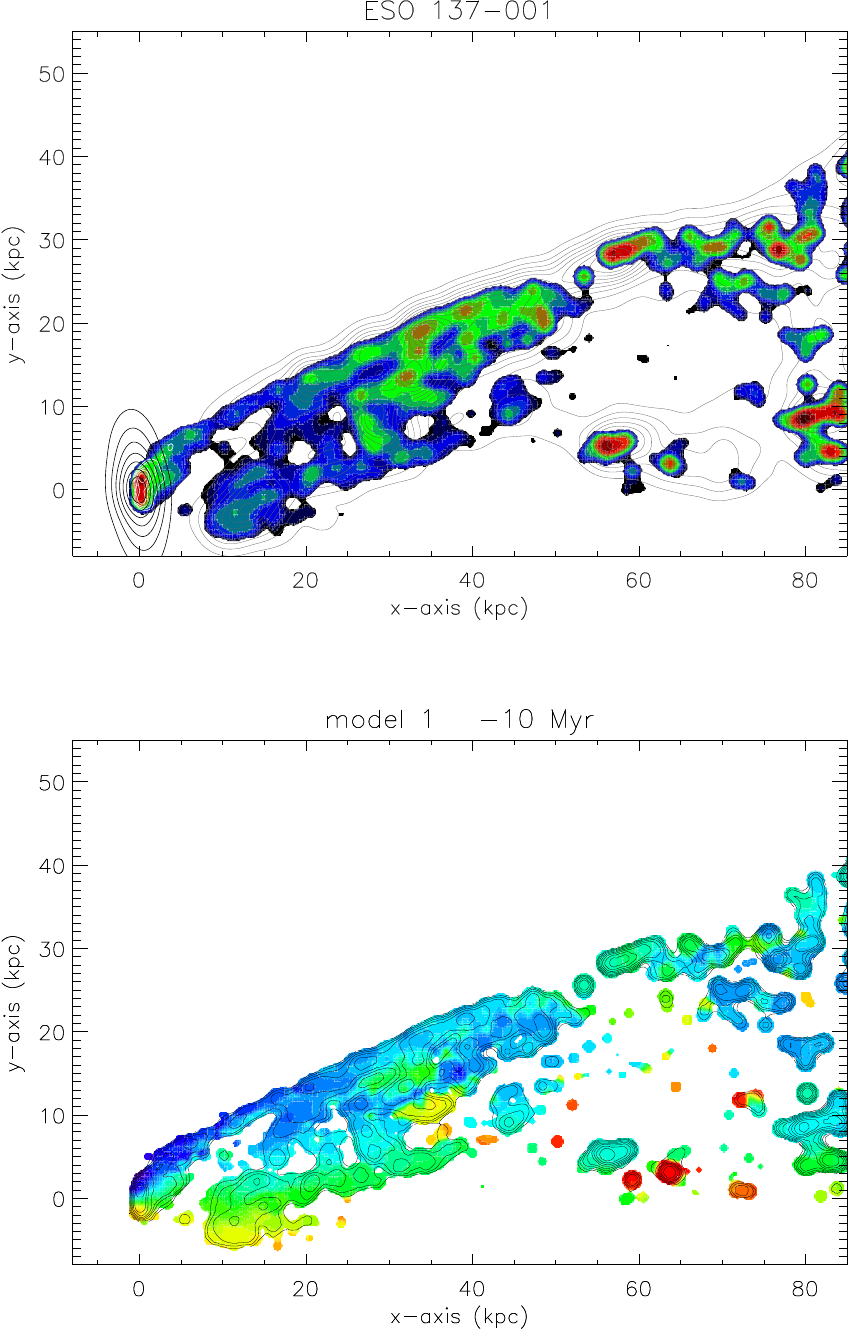}}
  \resizebox{14cm}{!}{\includegraphics{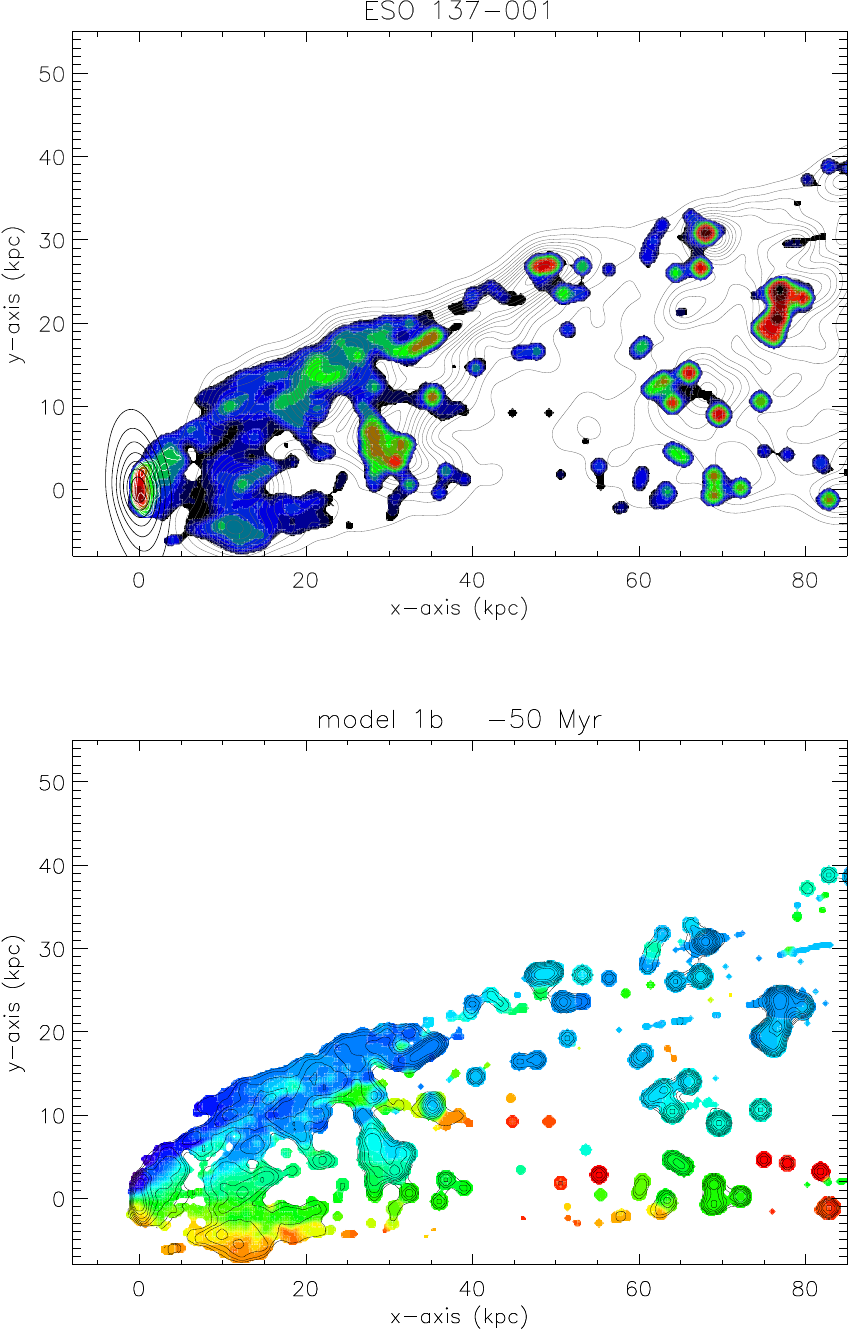}\includegraphics{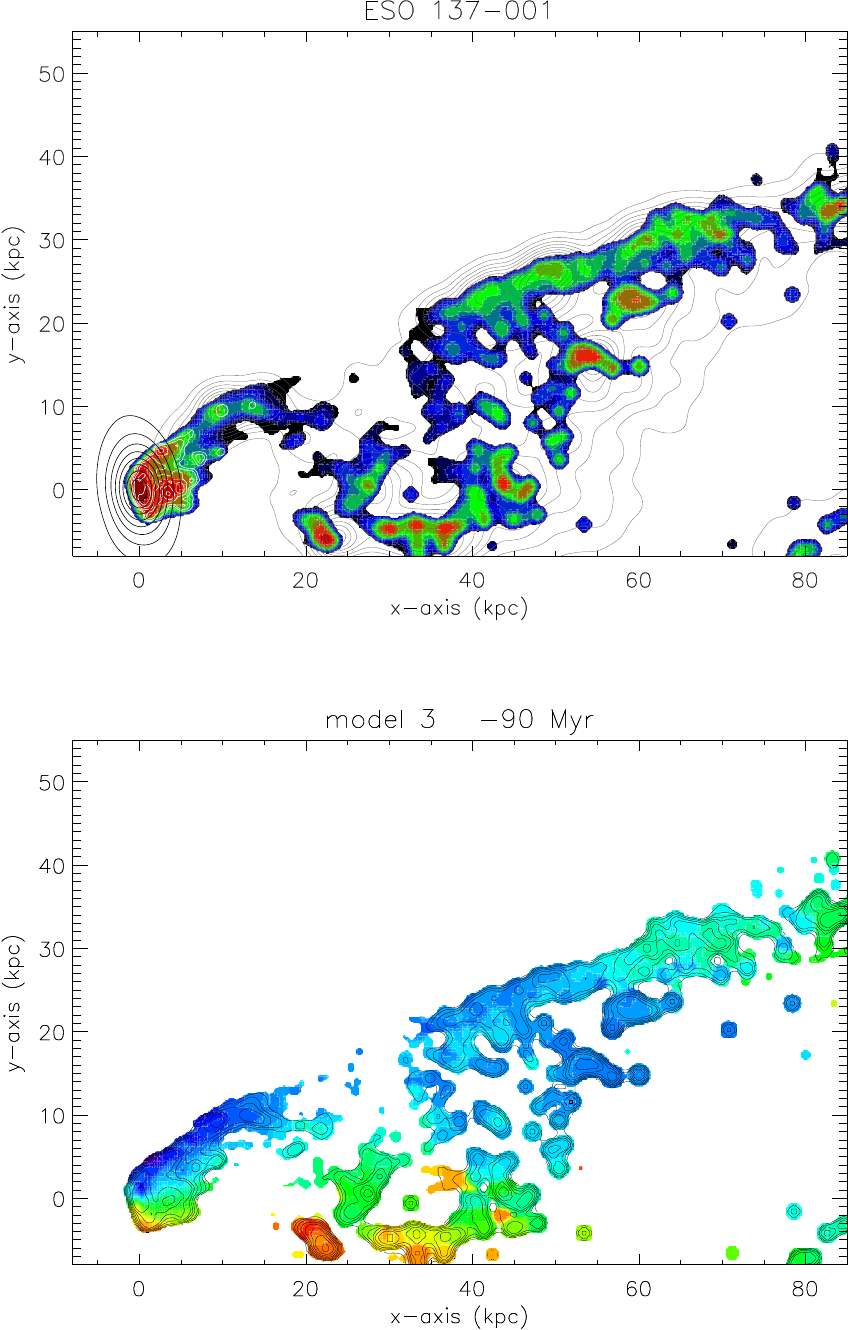}}
  \caption{Lower stripping efficiency: highest-ranked ram pressure pre-peak models of ESO~137-001 based on the H$\alpha$ velocity field. Upper panel: color: H$\alpha$; dark gray contours:
    X-ray; white contours: CO ; black contours: stellar content. Lower panel: H$\alpha$ velocity field.
  \label{fig:xpusheso137-001allonlyvel}}
\end{figure*}

\begin{table*}[!ht]
      \caption{Lower stripping efficiency: highest-ranked models based on CO, H$\alpha$, and X-ray.}
         \label{tab:xpushmolent2}
      \[
         \begin{tabular}{lcrccrcl}
           \hline
          model & $\Theta^{\rm (a)}$ & $\Delta t^{\rm (b)}$ & $az^{\rm (c)}$ & velocity vector & PA rotation & expansion/ & \\
            & (degrees) & (Myr) & (degrees) & & (degrees) & shrinking \\
           \hline
             1b & 60 & 110 & 152 & -(0.95,0.32,0.07) &  10 & 1.0 & \\
             1b & 60 & 110 & 152 & -(0.95,0.32,0.07) &  10 & 1.1 & \\
             1b & 60 & 110 & 152 & -(0.95,0.32,0.07) &  10 & 0.9 & \\
             1b & 60 & 110 & 152 & -(0.95,0.32,0.07) &  5 & 1.0 & \\
             1b & 60 & 110 & 152 & -(0.95,0.32,0.07) &  5 & 1.1 & \\
             2 & 60 & -20 & 132 & -(0.98,0.15,0.10) &  15 & 1.2 & \\
             1b & 60 & 110 & 132 & -(0.98,0.15,0.10) &  15 & 1.1 & \\
             1b & 60 & 110 & 152 & -(0.95,0.32,0.07) &  5 & 0.9 & \\
             1b & 60 & 110 & 142 & -(0.97,0.24,0.09) &  10 & 1.1 & \\
             3 & 60 & -10 & 122 & -(0.99,0.07,0.09) &  15 & 1.0 & \\
             3 & 60 & -10 & 132 & -(0.98,0.15,0.10) &  10 & 1.0 & \\
             1b & 60 & 110 & 142 & -(0.97,0.24,0.09) &  10 & 1.0 & \\
             1c & 50 & 40 & 152 & -(0.95,0.31,0.08) &  -5 & 1.0 & \\
             1b & 60 & 110 & 142 & -(0.97,0.24,0.09) &  15 & 1.0 & \\
             7b & 60 & 110 & 152 & -(0.95,0.32,0.07) &  10 & 0.8 & \\
             7b & 60 & 110 & 142 & -(0.97,0.24,0.09) &  15 & 0.8 & \\
             1b & 60 & 110 & 152 & -(0.95,0.32,0.07) &  10 & 1.2 & \\
             1b & 60 & 110 & 152 & -(0.95,0.32,0.07) &  5 & 1.2 & \\
             7b & 60 & 100 & 152 & -(0.95,0.32,0.07) &  10 & 0.8 & \\
             7b & 60 & 110 & 152 & -(0.95,0.32,0.07) &  15 & 0.8 & \\
             2 & 60 & -20 & 132 & -(0.98,0.15,0.10) &  15 & 1.1 & \\
             1b & 60 & 110 & 132 & -(0.98,0.15,0.10) &  15 & 1.2 & \\
             7b & 60 & 100 & 152 & -(0.95,0.32,0.07) &  15 & 0.8 & \\
             7b & 60 & 100 & 142 & -(0.97,0.24,0.09) &  15 & 0.8 & \\
             1b & 60 & 110 & 142 & -(0.97,0.24,0.09) &  15 & 1.1 & \\
             1b & 60 & 120 & 132 & -(0.98,0.15,0.10) &  15 & 1.0 & \\
             1c & 50 & 40 & 152 & -(0.95,0.31,0.08) &  0 & 1.1 & \\
             1b & 60 & 100 & 152 & -(0.95,0.32,0.07) &  5 & 1.2 & \\
             1c & 50 & 40 & 152 & -(0.95,0.31,0.08) &  -5 & 1.1 & \\
             1b & 60 & 110 & 132 & -(0.98,0.15,0.10) &  15 & 1.0 & \\
             1c & 50 & 40 & 152 & -(0.95,0.31,0.08) &  -10 & 1.0 & \\
             1c & 50 & 30 & 152 & -(0.95,0.31,0.08) &  -5 & 1.0 & \\
             1b & 60 & 110 & 152 & -(0.95,0.32,0.07) &  10 & 0.8 & \\
             7b & 60 & 110 & 142 & -(0.97,0.24,0.09) &  10 & 0.8 & \\
             7b & 60 & 100 & 132 & -(0.98,0.15,0.10) &  15 & 0.8 & \\
             1b & 60 & 120 & 132 & -(0.98,0.15,0.10) &  15 & 1.1 & \\
             1b & 60 & 110 & 142 & -(0.97,0.24,0.09) &  10 & 1.2 & \\
             1b & 60 & 110 & 132 & -(0.98,0.15,0.10) &  10 & 1.1 & \\
             1b & 60 & 110 & 152 & -(0.95,0.32,0.07) &  15 & 0.9 & \\
             7b & 60 & 100 & 142 & -(0.97,0.24,0.09) &  10 & 0.8 & \\
             1b & 60 & 100 & 152 & -(0.95,0.32,0.07) &  10 & 1.1 & \\
             7b & 60 & 110 & 132 & -(0.98,0.15,0.10) &  15 & 0.8 & \\
             1b & 60 & 110 & 142 & -(0.97,0.24,0.09) &  10 & 0.9 & \\
             1b & 60 & 110 & 152 & -(0.95,0.32,0.07) &  15 & 1.0 & \\
             1a & 50 & 80 & 152 & -(0.95,0.31,0.08) &  -5 & 1.0 & \\
             1b & 60 & 110 & 142 & -(0.97,0.24,0.09) &  15 & 0.9 & \\
             1c & 50 & 30 & 152 & -(0.95,0.31,0.08) &  -10 & 1.0 & \\
             1a & 60 & 150 & 142 & -(0.97,0.24,0.09) &  15 & 1.1 & \\
             1a & 60 & 160 & 142 & -(0.97,0.24,0.09) &  15 & 1.1 & \\
             7b & 60 & 100 & 152 & -(0.95,0.32,0.07) &  10 & 0.9 & \\
           \hline
        \end{tabular}
         \]
         \begin{list}{}{}
          \item $^{\rm (a)}$ angle between the direction of the galaxy's motion and the galactic disk
          \item $^{\rm (b)}$ time from peak ram pressure
          \item $^{\rm (c)}$ azimuthal viewing angle 
      \end{list}
\end{table*}

\begin{figure*}[!ht]
  \centering
  \resizebox{14cm}{!}{\includegraphics{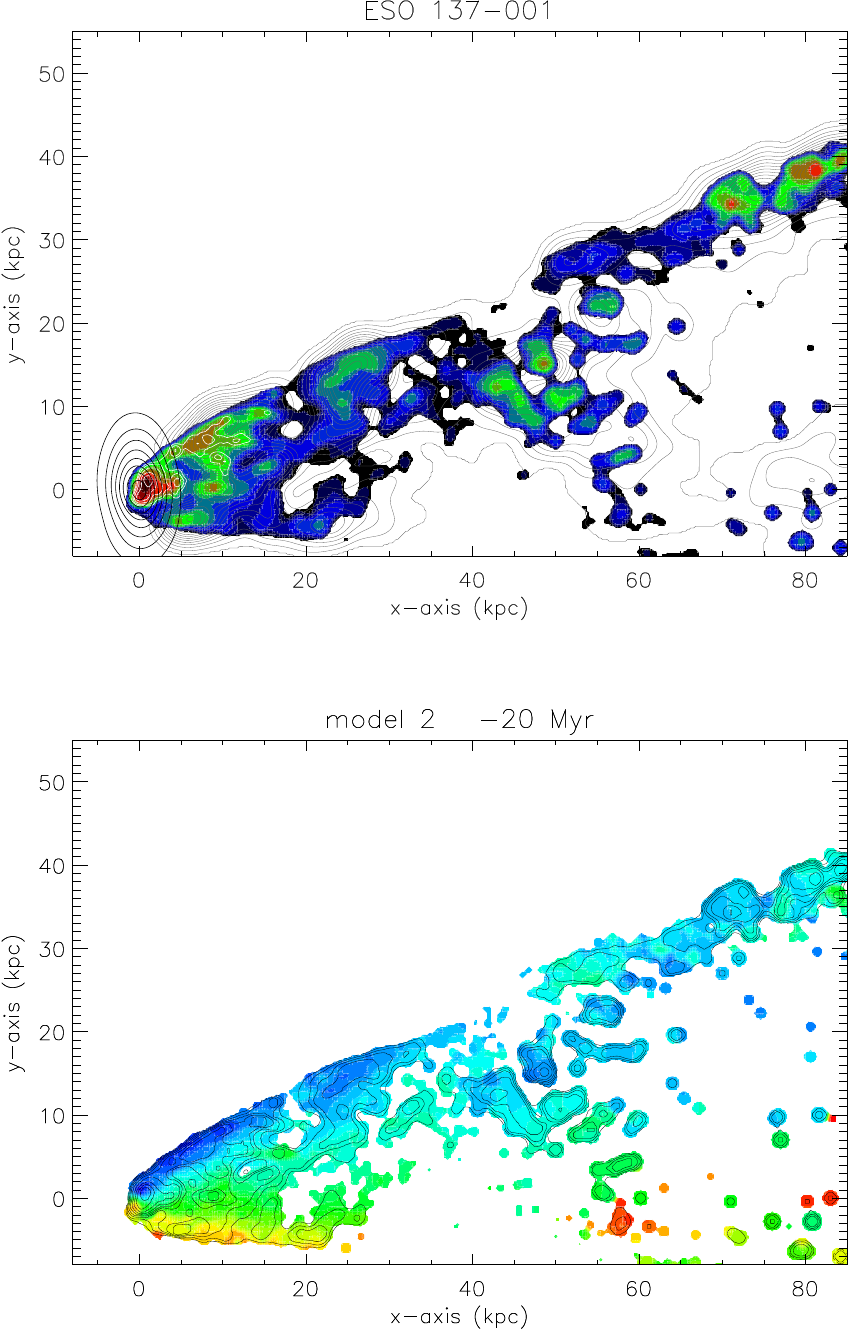}\includegraphics{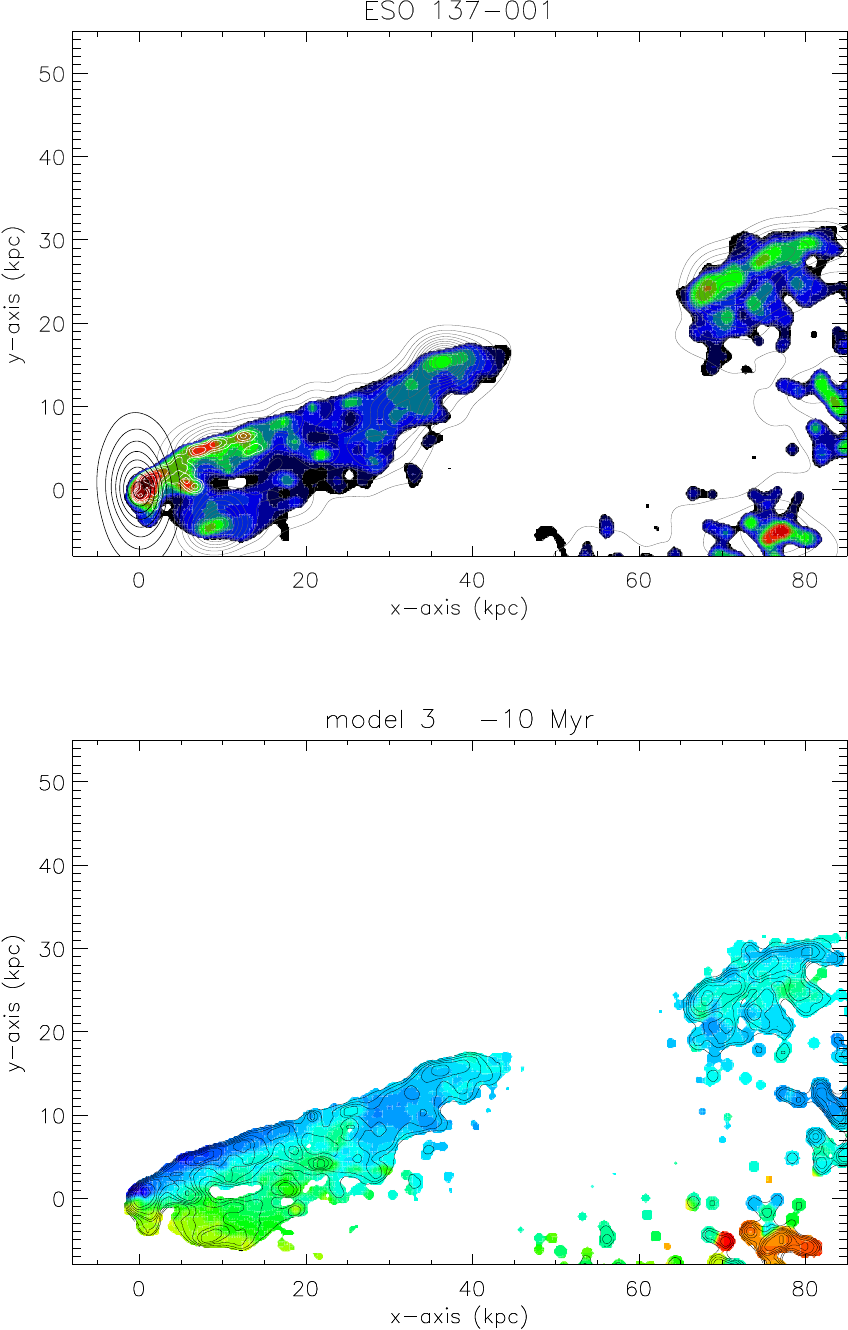}}
  \caption{Lower stripping efficiency: highest-ranked ram pressure pre-peak models of ESO~137-001 based on CO, H$\alpha$, and X-ray. Upper panel: color: H$\alpha$; dark gray contours:
    X-ray; white contours: CO ; black contours: stellar content. Lower panel: H$\alpha$ velocity field.
  \label{fig:xpusheso137-001allnovel}}
\end{figure*}

\end{appendix}
  
\end{document}